\documentclass[journal,letterpaper,11pt,onecolumn]{IEEEtran}

\input{definition}
    \setlist{nosep,leftmargin=*}
\usepackage{etoolbox}
\usepackage[export]{adjustbox}
\usepackage{silence}
\usepackage[colorlinks,allcolors=blue]{hyperref}

\newcommand\IndexInf{\Index_\infty}
\newcommand\IndexPos{\Index_\infty^+}

\newcommand\alt{\mathrm{alt}}
\newcommand\local{}
\newcommand\Global{\mathrm G}
\newcommand\Network{\mathrm N}

\DeclareMathOperator\spec{spec}
\DeclareMathOperator\co{co}
\DeclareMathOperator\vol{vol}
\DeclareMathOperator\HProd{\circ}

\title{\LARGE\bf Topological entropy of switched nonlinear and \\%
	interconnected systems}
\author{Guosong~Yang, Daniel~Liberzon, and Jo\~{a}o~P.~Hespanha%
    \thanks{This work was supported in part by the National Science Foundation grant CMMI-2106043 and in part by the National Science Foundation grant EPCN-1608880.}%
    \thanks{G.~Yang is with the Department of Electrical and Computer Engineering, Rutgers University, Piscataway, NJ 08854 USA (e-mail: guosong.yang@rutgers.edu).}%
    \thanks{D.~Liberzon is with the Coordinated Science Laboratory, University of Illinois Urbana-Champaign, Urbana, IL 61801 USA (e-mail: liberzon@illinois.edu).}%
    \thanks{J.~P.~Hespanha is with the Center for Control, Dynamical Systems, and Computation, University of California, Santa Barbara, Santa Barbara, CA 93106 USA (e-mail: hespanha@ucsb.edu).}%
}

\begin{document}

\maketitle

\begin{abstract}
A general upper bound for topological entropy of switched nonlinear systems is constructed, using an asymptotic average of upper limits of the matrix measures of Jacobian matrices of strongly persistent individual modes, weighted by their active rates.
A general lower bound is constructed as well, using a similar weighted average of lower limits of the traces of these Jacobian matrices.
In a case of interconnected structure, the general upper bound is readily applied to derive upper bounds for entropy that depend only on ``network-level'' information.
In a case of block-diagonal structure, less conservative upper and lower bounds for entropy are constructed.
In each case, upper bounds for entropy that require less information about the switching signal are also derived. 
The upper bounds for entropy and their relations are illustrated by numerical examples of a switched Lotka--Volterra ecosystem model.
\end{abstract}

\section{Introduction}\label{sec:intro}
Topological entropy is a fundamental concept in dynamical systems theory. Roughly speaking, it describes the rate at which uncertainty about the state of a system grows as time evolves. More concretely, one can think of it as the exponential growth rate of the number of trajectories that are separable with a finite precision, or in terms of the exponential growth rate of the size of the reachable set. Adler et al. first defined topological entropy as an extension to Kolmogorov's metric entropy \cite{Kolmogorov1958}, quantifying the expansion of a function using the minimal cardinality of a subcover refinement \cite{AdlerKonheimMcAndrew1965}. A different definition using the number of separable trajectories was later proposed by Bowen \cite{Bowen1971} and independently by Dinaburg \cite{Dinaburg1970}. An equivalence between these two definitions was established in \cite{Bowen1971b}. Most existing results on this topic considered only time-invariant systems, as time-varying dynamics introduce complexities that require new methods to understand \cite{KolyadaSnoha1996,KawanLatushkin2016}. The results on topological entropy of switched systems in the current paper contributes to the understanding of these complexities.

Entropy notions also play a prominent role in control theory. The first such result was a notion of topological feedback entropy for discrete-time systems defined by Nair et al. \cite{NairEvansMareelsMoran2004}, which extended the formulation in \cite{AdlerKonheimMcAndrew1965} and described the growth rate of control complexity for achieving set invariance. A notion of invariance entropy for continuous-time systems was later introduced by Colonius and Kawan \cite{ColoniusKawan2009}, which was closer in concept to the trajectory-counting formulation in \cite{Bowen1971,Dinaburg1970}. An equivalence between these two definitions was established in \cite{ColoniusKawanNair2013}. Entropy notions were also proposed for stabilization \cite{Colonius2012}, state estimation \cite{Savkin2006,MatveevPogromsky2016,LiberzonMitra2018} and model detection \cite{LiberzonMitra2018}.

In this paper, we study topological entropy of continuous-time switched nonlinear systems. Switched systems have been a popular topic in recent years (see, e.g., \cite{Liberzon2003Book,ShortenWirthMasonWulffKing2007} and references therein). In general, a switched system does not inherit stability properties of its individual modes. For example, a switched linear system with two stable modes may still be unstable (see, e.g., \cite[p.~19]{Liberzon2003Book}). On the other hand, it is well known that a switched linear system generated by a finite family of pairwise commuting Hurwitz matrices is globally exponentially stable under arbitrary switching (see, e.g., \cite[Th.~2.5, p.~31]{Liberzon2003Book}). This result has been extended to global uniform asymptotic stability for switching nonlinear systems with pairwise commuting, globally asymptotically stable modes \cite{Mancilla2000,VuLiberzon2005}. A simplest case of pairwise commuting modes is when the modes are simultaneously diagonalizable, which motivates us to consider cases with diagonal and block-diagonal structures in addition to the general case.

We are also interested in switched systems with interconnected structures. In systems and control theory, it is a tremendously important paradigm to represent a large complex system as an interconnection of smaller and simpler subsystems, and establish properties of the overall system by analyzing its individual components. A classical example of this paradigm is the use of small-gain theorems for establishing stability of interconnected linear, nonlinear, and switched systems (see, e.g., \cite{DesoerVidyasagar2009,LiuJiangHill2014,YangLiberzon2015SCL}).

Our interest in topological entropy of switched systems is strongly motivated by its relations to data-rate requirements in control problems. For a continuous-time linear time-invariant control system, the minimal data rate for feedback stabilization is given by the sum of the positive real parts of eigenvalues of the system matrix \cite{HespanhaOrtegaVasudevan2002} (or, in the discrete-time case, the sum of their logarithms \cite{HespanhaOrtegaVasudevan2002,NairEvans2003,TatikondaMitter2004}), which is equal to the topological entropy in open-loop \cite{Bowen1971,ColoniusKawan2009}. Data-rate requirements and entropy notions for nonlinear and interconnected time-invariant control systems have been extensively studied (see, e.g., \cite{NairEvansMareelsMoran2004,LiberzonHespanha2005,Colonius2012,KawanDelvenne2016,MatveevProskurnikovPogromskyFridman2019,TomarZamani2020}). For switched systems, however, neither data-rate requirements in control problems nor topological entropy are completely understood. Sufficient data rates for feedback stabilization of switched linear systems were established in \cite{Liberzon2014,YangLiberzon2018}. Similar data-rate conditions were constructed in \cite{SibaiMitra2017} by extending a notion of estimation entropy from \cite{LiberzonMitra2018}. In \cite{YangSchmidtLiberzon2018,YangSchmidtLiberzonHespanha2020}, formulae and bounds for topological entropy of switched linear systems were constructed using the active rates of individual modes, an approach that is also adopted in the current paper. Relations between topological entropy and stability of switched linear systems were studied in \cite{YangHespanhaLiberzon2019,YangSchmidtLiberzonHespanha2020}. For discrete-time switched linear systems, the topological entropy under worse-case switching sequences was obtained based on the notion of joint spectral radius \cite{BergerJungers2020}, while a formula for estimation entropy was constructed under additional regularity conditions \cite{VicinansaLiberzon2019}.

For switched nonlinear systems, topological entropy has not been closely explored so far. The main objective of this paper is to construct upper and lower bounds for topological entropy of switched nonlinear systems with general and interconnected structures. We start by introducing in Section~\ref{sec:pre} the necessary preliminaries to understand our bounds for entropy, including the definitions of topological entropy for switched systems, and some switching-related quantities such as the active rates of individual modes, and persistent and strongly persistent modes.

In Section~\ref{sec:ent}, we construct a general upper bound for topological entropy of switched nonlinear systems, using an asymptotic average of upper limits of the matrix measures of Jacobian matrices of strongly persistent modes, weighted by their active rates. A general lower bound is constructed as well, using a similar weighted average of lower limits of the traces of these Jacobian matrices. These bounds, as well as all bounds for entropy in this paper, depend on the values of Jacobian matrices of only the strongly persistent modes and over only an $ \omega $-limit set. Moreover, the bounds constructed using matrix measures hold regardless of the induced norm with which the matrix measures are computed, as long as the same one is used for all modes.

In Section~\ref{sec:ent-inter}, we consider switched systems with interconnected structures. We show in Subsection~\ref{ssec:ent-inter} that the general upper bound from Section~\ref{sec:ent} can be readily applied to derive upper bounds for entropy of switched interconnected systems that depend only on ``network-level'' information. In Subsection~\ref{ssec:ent-blk-diag}, we consider an interconnected case where the dynamics of subsystems are independent (i.e., a case of block-diagonal structure), and construct less conservative upper and lower bounds for entropy. The upper and lower bounds for entropy from Section~\ref{sec:ent} can be seen as a special case of the bounds here, as any switched system can be seen as a switched block-diagonal system with a single block.

In each general and interconnected case, we also derive in the corresponding section upper bounds for entropy that are more conservative but require less information about the switching signal, with their relations illustrated by numerical examples of a switched Lotka--Volterra ecosystem model in Section~\ref{sec:eg}. The main proofs, along with the necessary technical preliminaries, are presented in Section~\ref{sec:proof}. Section~\ref{sec:end} concludes the paper with a brief summary and some remarks on future research directions.

A preliminary version of some of the results in current paper was presented in the paper \cite{YangLiberzonHespanha2021}. The current paper improves upon \cite{YangLiberzonHespanha2021} by introducing persistent and strongly persistent modes, considering cases with interconnected and block-diagonal structures, and providing more extensive numerical examples, analysis details, and explanatory remarks.

\emph{Notations:}
Denote by $ \mathbf{1}_n $ the vector of ones in $ \dR^n $, and by $ I_n $ and $ \mathbf{1}_{n \times n} $ the identity matrix and matrix of ones in $ \dR^{n \times n} $, respectively; the subscript is omitted when the dimension is clear from context.
For a complex number $ c $, denote by $ \re(c) $ its real part.
For $ k \geq 2 $ vectors $ v_1 \in \dR^{n_1}, \ldots, v_k \in \dR^{n_k} $, denote by $ (v_1, \ldots, v_k) := [v_1^\top \cdots\, v_k^\top]^\top \in \dR^{n_1 + \cdots + n_k} $ their concatenation.
For a matrix $ A \in \dR^{n \times n} $, denote by $ \tr(A) $, $ \det(A) $, and $ \spec(A) $ its trace, determinant, and spectrum (as a multiset in which each eigenvalue has a number of instances equal to its algebraic multiplicity), respectively.
For a finite set $ E $, denote by $ \#E $ its cardinality.
For a set $ S \subset \dR^n $, denote by $ \co(S) $ and $ \vol(S) $ its convex hull and volume (Lebesgue measure), respectively.
Denote by $ |v|_\infty := \max_{1 \leq i \leq n} |v_i| $ the $ \infty $-norm of a vector $ v = (v_1, \ldots, v_n) \in \dR^n $, and by $ \|A\|_\infty = \max_{1 \leq i \leq n} \sum_{j=1}^{n} |a_{ij}| $ the induced $ \infty $-norm of a matrix $ A = [a_{ij}] \in \dR^{n \times n} $.
By default, all logarithms are natural logarithms (in order to avoid generating an extra multiplicative factor $ \ln 2 $ when computing topological entropy).

\section{Preliminaries}\label{sec:pre}
Consider a \emph{finite} family of continuous-time nonlinear dynamical systems
\begin{equation}\label{eq:ti-set}
    \dot x = f_p(x), \qquad p \in \Index
\end{equation}
with state $ x \in \dR^n $, in which each system is labeled with an index $ p $, and $ \Index $ is the corresponding finite set of indices. We assume that the functions $ f_p $ are continuously differentiable, and the systems in \eqref{eq:ti-set} are forward complete. We are interested in the corresponding \emph{switched system} defined by
\begin{equation}\label{eq:sw}
    \dot x = f_\sigma(x)
\end{equation}
with a right-continuous, piecewise-constant \emph{switching signal} $ \sigma: \dR_{\geq 0} \to \Index $. We call the system with index $ p $ in \eqref{eq:ti-set} \emph{mode} $ p $ of the switched system \eqref{eq:sw}, and $ \sigma(t) $ the \emph{active mode} at time $ t $. We denote by $ \xi_\sigma(t, x) $ the solution to \eqref{eq:sw} with switching signal $ \sigma $ at time $ t \geq 0 $ with initial state $ x $ (at time $ 0 $). Under the assumptions above, $ \xi_\sigma(t, x) $ is unique, absolutely continuous in $ t $, continuously differentiable in $ x $, and satisfies the differential equation \eqref{eq:sw} away from discontinuities of $ \sigma $, which are called \emph{switching times}, or simply \emph{switches}.
We assume that there is at most one switch at each time, and finitely many switches in each finite time interval (i.e., the set of switches contains no accumulation point).
For brevity, we denote by
\begin{equation*}
	J_x f_p(v) := J_x f_p(x)|_{x = v}, \qquad J_x \xi_\sigma(t, v) := J_x \xi_\sigma(t, x)|_{x = v}
\end{equation*}
the Jacobian matrix of function $ f_p(x) $ in mode $ p $ of \eqref{eq:sw} at state $ x = v $, and that of solution $ \xi_\sigma(t, x) $ to \eqref{eq:sw} with respect to initial state $ x $ at time $ t $ and $ x = v $, respectively.

\subsection{Entropy definitions}\label{ssec:pre-ent}
We now define the topological entropy of the switched system \eqref{eq:sw} with switching signal $ \sigma $ and initial states drawn from a compact set with nonempty interior $ K \subset \dR^n $ called the \emph{initial set}. Let $ |\cdot| $ be a norm on $ \dR^n $ and $ \|\cdot\| $ the corresponding induced norm on $ \dR^{n \times n} $. Given a time horizon $ T \geq 0 $ and a radius $ \varepsilon > 0 $, we define the following open ball in $ \dR^n $ with center $ x $:
\begin{equation}\label{eq:ball-dfn}
     B_{f_\sigma}(x, \varepsilon, T) := \Big\{ \bar x \in \dR^n: \max_{t \in [0, T]} |\xi_\sigma(t, \bar x) - \xi_\sigma(t, x)| < \varepsilon \Big\}.
\end{equation}
We say that a finite set $ E \subset K $ is \emph{$ (T, \varepsilon) $-spanning} if
\begin{equation}\label{eq:span-dfn}
    K \subset \bigcup_{x \in E} B_{f_\sigma}(x, \varepsilon, T),
\end{equation}
or equivalently, for each $ \bar x \in K $, there exists an $ x \in E $ such that $ |\xi_\sigma(t, \bar x) - \xi_\sigma(t, x)| < \varepsilon $ for all $ t \in [0, T] $. We denote by $ S(f_\sigma, \varepsilon, T, K) \geq 1 $ the minimal cardinality of a $ (T, \varepsilon) $-spanning set, or equivalently, the cardinality of a minimal $ (T, \varepsilon) $-spanning set, which is nondecreasing in $ T $ and nonincreasing in $ \varepsilon $. The \emph{topological entropy} of the switched system \eqref{eq:sw} with switching signal $ \sigma $ and initial set $ K $ is defined in terms of the exponential growth rate of $ S(f_\sigma, \varepsilon, T, K) $ by
\begin{equation}\label{eq:ent-dfn}
    h(f_\sigma, K) := \lim_{\varepsilon \searrow 0} \limsup_{T \to \infty} \frac{1}{T} \log S(f_\sigma, \varepsilon, T, K) \geq 0.
\end{equation}
For brevity, we at times refer to $ h(f_\sigma, K) $ simply as the (topological) entropy of \eqref{eq:sw}.

\begin{rmk}\label{rmk:norm}
In view of the equivalence of norms on a finite-dimensional vector space, the value of $ h(f_\sigma, K) $ is the same for every norm $ |\cdot| $ on $ \dR^n $. In particular, it is invariant under a change of basis. (More generally, one can define topological entropy on a metric space $ (X, d) $ instead of the normed space $ (\dR^n, |\cdot|) $, in which case its value depends on the given metric, but is a topological invariant for an initial set contained in a compact positively invariant set; see \cite[Prop.~3.1.2, p.\,109]{KatokHasselblatt1995} and \cite[p.\,1703]{ColoniusKawan2009} for more details.)
\end{rmk}

We also provide an alternative definition for the entropy of \eqref{eq:sw}. For $ T \geq 0 $ and $ \varepsilon > 0 $ given as before, we say that a finite set $ E \subset K $ is \emph{$ (T, \varepsilon) $-separated} if
\begin{equation}\label{eq:sep-dfn}
    \bar x \notin B_{f_\sigma}(x, \varepsilon, T) \qquad \forall x, \bar x \in E: \bar x \neq x,
\end{equation}
or equivalently, for each pair of distinct $ x, \bar x \in E $, there exists a time $ t \in [0, T] $ such that $ |\xi_\sigma(t, \bar x) - \xi_\sigma(t, x)| \geq \varepsilon $. We denote by $ N(f_\sigma, \varepsilon, T, K) \geq 1 $ the maximal cardinality of a $ (T, \varepsilon) $-separated set, or equivalently, the cardinality of a maximal $ (T, \varepsilon) $-separated set, which is also nondecreasing in $ T $ and nonincreasing in $ \varepsilon $. As stated in the following result, the entropy of \eqref{eq:sw} can be equivalently defined in terms of the exponential growth rate of $ N(f_\sigma, \varepsilon, T, K) $; the proof is along the lines of \cite[p.\,110]{KatokHasselblatt1995} and thus omitted here.
\begin{lem}\label{lem:ent-dfn-alt}
The topological entropy of the switched system \eqref{eq:sw} satisfies
\begin{equation}\label{eq:ent-dfn-alt}
    h(f_\sigma, K) = \lim_{\varepsilon \searrow 0} \limsup_{T \to \infty} \dfrac{1}{T} \log N(f_\sigma, \varepsilon, T, K).
\end{equation}
\end{lem}

\begin{rmk}\label{rmk:ent-dfn-liminf}
Following \cite[pp.\,109--110]{KatokHasselblatt1995}, for a time-invariant system $ \dot x = f(x) $ and an initial set $ K $ contained in a compact positively invariant set, the value of $ h(f, K) $ remains the same when the upper limits in \eqref{eq:ent-dfn} and \eqref{eq:ent-dfn-alt} are replaced with lower limits. However, this is not necessarily true for the switched system \eqref{eq:sw}, since the subadditivity required in the proof of \cite[Lemma~3.1.5, p.\,109]{KatokHasselblatt1995} may not hold.
\end{rmk}

\subsection{Active times, active rates, and persistent modes}\label{ssec:pre-act-time}
Here we introduce some switching-related quantities that will be useful for constructing bounds for the topological entropy of the switched system \eqref{eq:sw}. Given a switching signal $ \sigma $, we define the \emph{active time} of mode $ p $ of \eqref{eq:sw} over interval $ [0, t] $ by
\begin{equation}\label{eq:act-time}
    \tau_p(t) := \int_{0}^{t} \indFcn_p(\sigma(s)) \d s, \qquad p \in \Index
\end{equation}
with the indicator function
\begin{equation*}
    \indFcn_p(\sigma(s)) :=
    \begin{cases}
        1, &\sigma(s) = p, \\
        0, &\sigma(s) \neq p.
    \end{cases}
\end{equation*}
We also define the \emph{active rate} of mode $ p $ over $ [0, t] $ by
\begin{equation}\label{eq:act-rate}
    \rho_p(t) := \frac{\tau_p(t)}{t}, \qquad p \in \Index
\end{equation}
with $ \rho_p(0) := \indFcn_p(\sigma(0)) $, and the \emph{asymptotic active rate} of mode $ p $ by
\begin{equation}\label{eq:act-rate-sup}
    \hat\rho_p := \limsup_{t \to \infty} \rho_p(t), \qquad p \in \Index.
\end{equation}
Clearly, the active times $ \tau_p(t) \geq 0 $ are nondecreasing and satisfy $ \sum_{p \in \Index} \tau_p(t) = t $ for all $ t \geq 0 $; the active rates $ \rho_p(t) \in [0, 1] $ satisfy $ \sum_{p \in \Index} \rho_p(t) = 1 $ for all $ t \geq 0 $. In contrast, due to the upper limit in \eqref{eq:act-rate-sup}, it is possible that $ \sum_{p \in \Index} \hat\rho_p > 1 $ for the asymptotic active rates $ \hat\rho_p \geq 0 $ (for a numerical example, see \cite[Example~1]{YangSchmidtLiberzonHespanha2020}). Moreover, we define
\begin{equation}\label{eq:index-inf-pos}
	\IndexInf := \{p \in \Index: \sup\{t \geq 0: \sigma(t) = p\} = \infty\}, \qquad \IndexPos := \{p \in \Index: \hat\rho_p > 0\},
\end{equation}
and call them the sets of \emph{persistent} and \emph{strongly persistent} modes, respectively. Clearly, they satisfy $ \IndexPos \subset \IndexInf $.

\section{Entropy of general switched nonlinear systems}\label{sec:ent}
In this section, we construct general upper and lower bounds for the topological entropy of the switched system \eqref{eq:sw}. The upper bounds will be constructed based on a notion of matrix measure. For an induced norm $ \|\cdot\| $ on $ \dR^{n \times n} $, its one-sided directional derivative at $ I $ in direction $ A \in \dR^{n \times n} $ is called the \emph{matrix measure} of $ A $ and denoted by $ \mu(A) $, that is,
\begin{equation}\label{eq:mx-meas-dfn}
    \mu(A) := \lim_{t \searrow 0} \frac{\|I + t A\| - 1}{t}.
\end{equation}
The matrix measure $ \mu(A) $ can also be seen as the right derivative of the functions $ t \mapsto \|e^{A t}\| $ and $ t \mapsto \log\|e^{t A}\| $ at $ t = 0 $ \cite[Fact~11.15.7, p.\,690]{Bernstein2009}. A summary of properties of matrix measure can be found in \cite[Th.~2.8.5, p.\,31]{DesoerVidyasagar2009}. In particular, $ \mu $ is a convex function and satisfies
\begin{equation}\label{eq:mx-meas-norm-eig}
    {-\mu}(-A) \leq \re(\lambda) \leq \mu(A) \leq \|A\| \qquad \forall A \in \dR^{n \times n},\, \lambda \in \spec(A)
\end{equation}
and
\begin{equation}\label{eq:mx-meas-vec}
	\mu(A + B) \leq \mu(A) + \mu(B), \quad \mu(cA) = c \mu(A) \qquad \forall A, B \in \dR^{n \times n},\, c \geq 0.
\end{equation}
For standard induced norms, there are explicit formulae for the matrix measure \cite[Th.~2.8.24, p.\,33]{DesoerVidyasagar2009}; for example, for the induced $ \infty $-norm, the matrix measure of $ A = [a_{ij}] $ is given by
\begin{equation}\label{eq:mx-meas-inf}
    \mu(A) = \max_{1 \leq i \leq n} \bigg( a_{ii} + \sum_{j \neq i} |a_{ij}| \bigg).
\end{equation}
We denote by $ \xi_\sigma(t, K) := \{\xi_\sigma(t, x): x \in K\} $ the \emph{reachable set} of \eqref{eq:sw} with switching signal $ \sigma $ at time $ t $ from initial set $ K $.

The following result provides upper and lower bounds for the entropy of \eqref{eq:sw}; they are derived based on later results in this paper, which will be explained in the proof in Subsection~\ref{ssec:proof-sw-ent}.
\begin{thm}\label{thm:sw-ent}
Given a switching signal $ \sigma $, consider the active rates $ \rho_p(t) $ defined by \eqref{eq:act-rate} and the sets of persistent and strongly persistent modes $ \IndexInf $ and $ \IndexPos $ defined by \eqref{eq:index-inf-pos}.
\begin{enumerate}
	\item Let
		\begin{equation}\label{eq:sw-meas-sup}
			\hat\mu_p := \limsup_{t \to \infty:\, \sigma(t) = p} \max_{v \in \xi_\sigma(t, \co(K))} \mu(J_x f_p(v)), \qquad p \in \Index.
		\end{equation}
		If the constants $ \hat\mu_p $ are finite for all $ p \in \IndexInf $, then the topological entropy of the switched system \eqref{eq:sw} is upper bounded by
		\begin{equation}\label{eq:sw-ent-upper}
		    h(f_\sigma, K) \leq \max \bigg\{ \limsup_{t \to \infty} \sum_{p \in \IndexPos} n \hat\mu_p \rho_p(t),\, 0 \bigg\}.
		\end{equation}
	\item Let
		\begin{equation}\label{eq:sw-tr-inf}
			\check\chi_p := \liminf_{t \to \infty:\, \sigma(t) = p} \min_{v \in \xi_\sigma(t, K)} \tr(J_x f_p(v)), \qquad p \in \Index.
		\end{equation}
		If the constants $ \check\mu_p $ are finite for all $ p \in \IndexInf $, then the topological entropy of the switched system \eqref{eq:sw} is lower bounded by
		\begin{equation}\label{eq:sw-ent-lower}
		    h(f_\sigma, K) \geq \max \bigg\{ \limsup_{t \to \infty} \sum_{p \in \IndexPos} \check\chi_p \rho_p(t),\, 0 \bigg\}.
		\end{equation}
\end{enumerate}
\end{thm}

Note that:
\begin{enumerate*}[1)]
	\item The upper bound \eqref{eq:sw-ent-upper} holds regardless of the induced norm with which the matrix measures in \eqref{eq:sw-meas-sup} are computed, as long as the same one is used for all modes.
	\item The bounds \eqref{eq:sw-ent-upper} and \eqref{eq:sw-ent-lower} are formulated in terms of asymptotic weighted averages of the constants $ \hat\mu_p $ and $ \check\chi_p $ defined by \eqref{eq:sw-meas-sup} and \eqref{eq:sw-tr-inf}, respectively, with the active rates $ \rho_p(t) $ as weights.
	\item Due to the upper limit in \eqref{eq:sw-meas-sup}, the upper bound \eqref{eq:sw-ent-upper} depends on the values of Jacobian matrices $ J_x f_p(x) $ over only the $ \omega $-limit set from the convex hull of initial set, $ \co(K) $, instead of all reachable points from $ \co(K) $. In particular, \eqref{eq:sw-ent-upper} is able to yield a finite bound in the case of unbounded Jacobian matrices but a compact $ \omega $-limit set. This property is obtained using a result on asymptotic weighted averages established in Lemma~\ref{lem:sw-diag-avg-max-sup} in Subsection~\ref{ssec:pre-avg}, and similar properties hold for all bounds for entropy in this paper.
	\item Provided that the constants $ \hat\mu_p $ and $ \check\chi_p $ are finite for all persistent modes $ p \in \IndexInf $, the bounds \eqref{eq:sw-ent-upper} and \eqref{eq:sw-ent-lower} depend on their values over only the strongly persistent modes $ p \in \IndexPos $ (as the rest disappears after taking the asymptotic weighted average; see the proof of Lemma~\ref{lem:sw-diag-avg-max-sup} for more details). However, if there exists a mode $ p \in \IndexInf\backslash\IndexPos $ such that $ \hat\mu_p $ or $ \check\chi_p $ is infinite, then the corresponding bound may not hold.
\end{enumerate*}

\begin{rmk}\label{rmk:sw-ent-inf}
The bounds \eqref{eq:sw-ent-upper} and \eqref{eq:sw-ent-lower} can be extended to some cases where the constants $ \mu_p $ and $ \check\chi_p $ are infinite. For example, \eqref{eq:sw-ent-lower} still holds if $ \check\chi_p \in \dR \cup \{+\infty\} $ for all $ p \in \IndexInf\backslash\IndexPos $ and $ \check\chi_p \in \dR $ for all $ p \in \IndexPos $. Moreover, if $ \check\chi_p \in \dR \cup \{+\infty\} $ for all $ p \in \IndexInf $, and there exists a mode $ q \in \IndexPos $ such that $ \check\chi_q = +\infty $, then the entropy of \eqref{eq:sw} satisfies $ h(f_\sigma, K) = +\infty $. The proofs are along the lines of that of Theorem~\ref{thm:sw-ent} and thus omitted here (see also Remark~\ref{rmk:sw-diag-avg-max-sup} in Subsection~\ref{ssec:pre-avg}). Similar extensions can be made for the upper bound \eqref{eq:sw-ent-upper} as well as all bounds for entropy of switched systems in this paper.
\end{rmk}

\begin{rmk}\label{rmk:ent-simp}
In many scenarios, we can construct simpler but more conservative bounds for entropy based on those in Theorem~\ref{thm:sw-ent}. For example, let $ S \subset \dR^n $ be a set such that one of the following holds:
\begin{enumerate}
    \item\label{rmk:ent-simp-Rn} $ S = \dR^n $;
    \item\label{rmk:ent-simp-Inv} $ S $ is compact and positively invariant for \eqref{eq:sw} and contains the convex hull of initial set, $ \co(K) $; or
    \item\label{rmk:ent-simp-Omega} $ S $ is compact and contains the $ \omega $-limit set from $ \co(K) $.
\end{enumerate}
Suppose that there exists a family of constants $ \hat\mu_p^* $ for $ p \in \IndexPos $ such that $ \hat\mu_p^* \geq \mu(J_x f_p(v)) $ for all $ p \in \IndexPos $ and $ v \in S $. Then the upper bound \eqref{eq:sw-ent-upper} holds with $ \hat\mu_p^* $ in place of $ \hat\mu_p $. Similar variants hold for all bounds for entropy in this paper. In the numerical examples in Section~\ref{sec:eg}, the computation will be simplified based on the third scenario here.
\end{rmk}

Thinking of the non-switched case as a switched system with constant switched signal, Theorem~\ref{thm:sw-ent} implies the following upper and lower bounds for the entropy of a time-invariant system.
\begin{cor}\label{cor:ti-ent-bnd}
The topological entropy of a time-invariant system $ \dot x = f(x) $ with state $ x \in \dR^n $ satisfies
\begin{equation}\label{eq:ti-ent-bnd}
    \max\{\check\chi,\, 0\} \leq h(f, K) \leq \max\{n \hat\mu,\, 0\}
\end{equation}
with
\begin{equation*}
	\check\chi := \liminf_{t \to \infty} \min_{v \in \xi(t, K)} \tr(J_x f(v)), \qquad \hat\mu := \limsup_{t \to \infty} \max_{v \in \xi(t, \co(K))} n \mu(J_x f(v)).
\end{equation*}
\end{cor}

Based on the upper bound \eqref{eq:sw-ent-upper}, we construct additional upper bounds for the entropy of \eqref{eq:sw} that require less information about the switching signal.
\begin{cor}\label{cor:sw-ent-upper-sum-max}
Given a switching signal $ \sigma $, consider the asymptotic active rates $ \hat\rho_p $ defined by \eqref{eq:act-rate-sup} and the sets of persistent and strongly persistent modes $ \IndexInf $ and $ \IndexPos $ defined by \eqref{eq:index-inf-pos}. If the constants $ \hat\mu_p $ defined by \eqref{eq:sw-meas-sup} are finite for all $ p \in \IndexInf $, then the topological entropy of the switched system \eqref{eq:sw} is upper bounded by
\begin{equation}\label{eq:sw-ent-upper-sum}
    h(f_\sigma, K) \leq \sum_{p \in \IndexPos} \max\{n \hat\mu_p,\, 0\} \hat\rho_p
\end{equation}
and also by
\begin{equation}\label{eq:sw-ent-upper-max}
    h(f_\sigma, K) \leq \max_{p \in \IndexPos} \max\{n \hat\mu_p,\, 0\}.
\end{equation}
\end{cor}

Note that the upper bound \eqref{eq:sw-ent-upper-sum} can be seen as an average of the results of applying the second inequality in \eqref{eq:ti-ent-bnd} to each strongly persistent mode of \eqref{eq:sw} while omitting its inactive time, weighted by the corresponding asymptotic active rate $ \hat\rho_p $; the upper bound \eqref{eq:sw-ent-upper-max} can be seen as a maximum of these results over strongly persistent modes. Similar properties hold for the corresponding bounds in the cases with interconnected and block-diagonal structures in Subsections~\ref{ssec:ent-inter} and~\ref{ssec:eg-sw-blk-diag}, respectively.

\begin{proof}[Proof of Corollary~\ref{cor:sw-ent-upper-sum-max}]
First, as the upper limit is a subadditive function, the upper bound \eqref{eq:sw-ent-upper} implies that
\begin{equation*}
\begin{aligned}
    h(f_\sigma, K) &\leq \max \bigg\{ \sum_{p \in \IndexPos} \limsup_{t \to \infty} n \hat\mu_p \rho_p(t),\, 0 \bigg\} \\
    &\leq \sum_{p \in \IndexPos} \max\{n \hat\mu_p,\, 0\} \limsup_{t \to \infty} \rho_p(t) = \sum_{p \in \IndexPos} \max\{n \hat\mu_p,\, 0\} \hat\rho_p.
\end{aligned}
\end{equation*}
Second, \eqref{eq:sw-ent-upper} also implies that
\begin{align*}
    h(f_\sigma, K) &\leq \max \bigg\{ \limsup_{t \to \infty} \Big( \max_{p \in \IndexPos} n \hat\mu_p \Big) \sum_{p \in \IndexPos} \rho_p(t),\, 0 \bigg\} \\
    &\leq \max \Big\{ \max_{p \in \IndexPos} n \hat\mu_p,\, 0 \Big\} = \max_{p \in \IndexPos} \max\{n \hat\mu_p,\, 0\}. \qedhere
\end{align*}
\end{proof}

The results in Theorem~\ref{thm:sw-ent} and Corollary~\ref{cor:sw-ent-upper-sum-max} are compared in Remark~\ref{rmk:sw-ent-compare} below; for a numerical example, see Example~\ref{eg:sw} in Subsection~\ref{ssec:eg-sw}.
\begin{rmk}\label{rmk:sw-ent-compare}
\begin{enumerate}
	\item\label{rmk:sw-ent-compare-order} The upper bound \eqref{eq:sw-ent-upper} is less conservative than or equivalent to the upper bounds \eqref{eq:sw-ent-upper-sum} and \eqref{eq:sw-ent-upper-max}, while \eqref{eq:sw-ent-upper-sum} and \eqref{eq:sw-ent-upper-max} are both useful in the sense that neither is less conservative, as it is possible that $ \sum_{p \in \Index} \hat\rho_p > 1 $ for the asymptotic active rates $ \hat\rho_p $.
	\item\label{rmk:sw-ent-compare-info} For a fixed family of modes, the upper bounds \eqref{eq:sw-ent-upper-sum} and \eqref{eq:sw-ent-upper-max} require less information about the switching signal than the upper bound \eqref{eq:sw-ent-upper}, as the former two depend on the asymptotic active rates $ \hat\rho_p $ instead of the active rates $ \rho_p(t) $. Moreover, if for each $ p \in \IndexPos $, a constant $ \hat\mu_p^* $ as in the first two scenarios in Remark~\ref{rmk:ent-simp} is used in place of $ \hat\mu_p $, then \eqref{eq:sw-ent-upper-max} yields the same upper bound for all switching signals such that all modes are strongly persistent (i.e., $ \IndexPos = \IndexInf = \Index $).
\end{enumerate}
\end{rmk}

Consider the case where the modes of the switched system \eqref{eq:sw} are all linear, that is, there is a family of matrices $ A_p \in \dR^{n \times n} $ for $ p \in \Index $ such that
\begin{equation*}
    f_p(x) = A_p x \qquad \forall p \in \Index,\, x \in \dR^n.
\end{equation*}
Then the constants $ \hat\mu_p $ and $ \check\chi_p $ defined by \eqref{eq:sw-meas-sup} and \eqref{eq:sw-tr-inf} satisfy
\begin{equation*}
    \hat\mu_p = \mu(A_p), \quad \check\chi_p = \tr(A_p) \qquad \forall p \in \IndexInf.
\end{equation*}
Therefore, Theorem~\ref{thm:sw-ent} and Corollary~\ref{cor:sw-ent-upper-sum-max} extend \cite[Th.~1 and Remark~5]{YangSchmidtLiberzonHespanha2020} for the case of linear modes to the case of nonlinear modes, respectively, and improve them by considering persistent and strongly persistent modes.

\section{Entropy of switched interconnected systems}\label{sec:ent-inter}
\subsection{General switched interconnected systems}\label{ssec:ent-inter}
Consider the case where the switched system \eqref{eq:sw} is a network of $ m \geq 2 $ interconnected subsystems. For $ i \in \{1, \ldots, m\} $, we denote by $ x_i \in \dR^{n_i} $ the state of the $ i $-th subsystem, and by $ f_p^i $ the corresponding components of the function $ f_p $ in mode $ p $ of \eqref{eq:sw} for $ p \in \Index $. Then \eqref{eq:sw} can be written as the \emph{switched interconnected system}
\begin{equation}\label{eq:sw-inter}
	\dot x_i = f_\sigma^i(x_1, \ldots, x_m), \qquad i \in \{1, \ldots, m\}
\end{equation}
with state $ x = (x_1, \ldots, x_m) \in \dR^n $ and functions $ f_p(x) = (f_p^1(x), \ldots, f_p^m(x)) $ for $ p \in \Index $. For brevity, we denote by $ J_{x_j} f_p^i(v) := J_{x_j} f_p^i(x)|_{x = v} $ the Jacobian matrix of function $ f_p^i(x) $ in mode $ p $ of the $ i $-th subsystem of \eqref{eq:sw-inter} with respect to state $ x_j $ of the $ j $-th subsystem at $ x = v $.

In this subsection, we show that the upper bound \eqref{eq:sw-ent-upper} for the topological entropy of the general switched nonlinear system \eqref{eq:sw} can be readily applied to derive upper bounds for the entropy of \eqref{eq:sw-inter} that depend only on ``network-level'' information. Following \cite{RussoBernardoSontag2013}, we assume that the following norms are given:
\begin{enumerate}
	\item a ``local'' norm $ |\cdot|_{\local i} $ on $ \dR^{n_i} $ for each $ i \in \{1, \ldots, m\} $, and
	\item a monotone ``network'' norm $ |\cdot|_\Network $ on $ \dR^m $, that is, for all $ v, w \in \dR^m $, we have\footnote{In this subsection, an inequality between two vectors or matrices of the same size, or between a vector or matrix and a scalar, is to be interpreted entrywise (e.g., $ A \geq 0 $ means that $ A $ is a nonnegative matrix).}
		\begin{equation*}
			v \geq w \geq 0 \implies |v|_\Network \geq |w|_\Network.
		\end{equation*}
		In particular, all $ p $-norms with $ p \geq 1 $ are monotone.
\end{enumerate}
For a vector $ v = (v_1, \ldots, v_m) \in \dR^n $ with $ v_i \in \dR^{n_i} $ for $ i \in \{1, \ldots, m\} $, we define a ``global'' norm $ |\cdot|_\Global $ by
\begin{equation}\label{eq:global-norm}
	|v|_\Global := |(|v_1|_{\local 1}, \ldots, |v_m|_{\local m})|_\Network.
\end{equation}
Clearly, as $ |\cdot|_\Network $ is monotone, $ |\cdot|_\Global $ is indeed a norm. We denote by $ \|\cdot\|_{\local i} $, $ \|\cdot\|_\Network $, and $ \|\cdot\|_\Global $ the corresponding induced norms on $ \dR^{n_i \times n_i} $, $ \dR^{m \times m} $, and $ \dR^{n \times n} $, respectively, and by $ \mu_{\local i} $, $ \mu_\Network $, and $ \mu_\Global $ the corresponding matrix measures. We also denote by $ \|\cdot\|_{\local i j} $ the induced norm on $ \dR^{n_i \times n_j} $ defined by
\begin{equation*}
	\|A\|_{\local i j} := \max_{|v|_{\local j} = 1} |A v|_{\local i}, \qquad i, j \in \{1, \ldots, m\}: i \neq j.
\end{equation*}

As the ``network'' norm $ |\cdot|_\Network $ is monotone, the induced norm $ \|\cdot\|_\Network $ and matrix measure $ \mu_\Network $ satisfy similar monotonicity properties for nonnegative and Metzler matrices, respectively:
\begin{lem}\label{lem:nonneg-mx-norm-meas}
\begin{enumerate}
	\item For all nonnegative matrices $ A, B \in \dR_{\geq 0}^{m \times m} $, we have
		\begin{equation}\label{eq:nneg-mx-norm}
			A \geq B \implies \|A\|_\Network \geq \|B\|_\Network.
		\end{equation}
	\item For all Metzler matrices $ A, B \in \dR^{m \times m} $, that is, matrices with nonnegative off-diagonal entries, we have
		\begin{equation}\label{eq:nneg-mx-meas}
			A \geq B \implies \mu_\Network(A) \geq \mu_\Network(B).
		\end{equation}
\end{enumerate}
\end{lem}
\begin{proof}
\begin{enumerate}[wide]
\item The implication in \eqref{eq:nneg-mx-norm} follows from the definition of induced norm and monotonicity of $ |\cdot|_\Network $. More specifically, let $ \bar v \in \argmax_{|v|_{\scriptstyle \Network}=1} |B v|_\Network $. Then $ \bar v \geq 0 $ as $ B \geq 0 $ and $ |\cdot|_\Network $ is monotone. Hence $ \|A\|_\Network \geq |A \bar v|_\Network \geq |B \bar v|_\Network = \|B\|_\Network $ as $ A \geq B \geq 0 $ and $ |\cdot|_\Network $ is monotone.
\item The implication in \eqref{eq:nneg-mx-meas} follows from \eqref{eq:nneg-mx-norm} and the definition of matrix measure \eqref{eq:mx-meas-dfn}. More specifically, we have
		\begin{equation*}
			\mu_\Network(A) = \lim_{t \searrow 0} \frac{\|I + t A\|_\Network - 1}{t} \geq \lim_{t \searrow 0} \frac{\|I + t B\|_\Network - 1}{t} = \mu_\Network(B),
		\end{equation*}
		where the inequality follows from \eqref{eq:nneg-mx-norm} as $ I + t A \geq I + t B \geq 0 $ for the Metzler matrices $ A \geq B $ and small enough $ t > 0 $. \qedhere
\end{enumerate}
\end{proof}

\begin{thm}\label{thm:sw-inter-ent-upper}
Given a switching signal $ \sigma $, consider the active rates $ \rho_p(t) $ defined by \eqref{eq:act-rate} and the sets of persistent and strongly persistent modes $ \IndexInf $ and $ \IndexPos $ defined by \eqref{eq:index-inf-pos}. Let
\begin{equation}\label{eq:sw-inter-meas-sup}
	\hat\mu_p^\Network := \limsup_{t \to \infty:\, \sigma(t) = p} \max_{v \in \xi_\sigma(t, \co(K))} \mu_\Network( A_p^\Network(v)), \qquad p \in \Index
\end{equation}
with the matrix-valued functions $ A_p^\Network(v) = [a_p^{ij}(v)] \in \dR^{m \times m} $ defined by
\begin{equation}\label{eq:sw-inter-meas}
	a_p^{ii}(v) := \mu_{\local i}(J_{x_i} f_p^i(v)), \quad a_p^{ij}(v) := \|J_{x_j} f_p^i(v)\|_{\local i j}, \qquad i, j \in \{1, \ldots, m\}: i \neq j.
\end{equation}
If the constants $ \hat\mu_p^\Network $ are finite for all $ p \in \IndexInf $, then the topological entropy of the switched interconnected system \eqref{eq:sw-inter} is upper bounded by
\begin{equation}\label{eq:sw-inter-ent-upper}
    h(f_\sigma, K) \leq \max \bigg\{ \limsup_{t \to \infty} \sum_{p \in \IndexPos} n \hat\mu_p^\Network \rho_p(t),\, 0 \bigg\}.
\end{equation}
\end{thm}

\begin{proof}
The upper bound \eqref{eq:sw-inter-ent-upper} follows from the general upper bound \eqref{eq:sw-ent-upper} as the constants $ \hat\mu_p $ defined by \eqref{eq:sw-meas-sup} with matrix measure $ \mu_\Global $ satisfy $ \hat\mu_p \leq \hat\mu_p^\Network $ for all $ p \in \IndexInf $, which can be established using Lemma~\ref{lem:block-mx-meas} below.
\end{proof}

\begin{lem}[{\cite[Th.~2]{RussoBernardoSontag2013}}]\label{lem:block-mx-meas}
Consider a block matrix $ A = [A_{ij}] \in \dR^{n \times n} $ with $ A_{ij} \in \dR^{n_i \times n_j} $ for $ i, j \in \{1, \ldots, m\} $. Let $ A_\Network = [a_{ij}] \in \dR^{m \times m} $ be a Metzler matrix such that
\begin{equation*}
	a_{ii} \geq \mu_{\local i}(A_{ii}), \quad a_{ij} \geq \|A_{ij}\|_{\local i j} \qquad \forall i, j \in \{1, \ldots, m\}: i \neq j.
\end{equation*}
Then
\begin{equation*}
	\mu_\Global(A) \leq \mu_\Network(A_\Network).
\end{equation*}
\end{lem}

The next result shows that the upper limit and maximum in \eqref{eq:sw-inter-meas-sup} can also be taken entrywise.
\begin{cor}\label{cor:sw-inter-ent-upper-entry}
Given a switching signal $ \sigma $, consider the active rates $ \rho_p(t) $ defined by \eqref{eq:act-rate} and the sets of persistent and strongly persistent modes $ \IndexInf $ and $ \IndexPos $ defined by \eqref{eq:index-inf-pos}. Let
\begin{equation}\label{eq:sw-inter-meas-sup-entry}
	\hat a_p^{ij} := \limsup_{t \to \infty:\, \sigma(t) = p} \max_{v \in \xi_\sigma(t, \co(K))} a_p^{ij}(v), \qquad i, j \in \{1, \ldots, m\},\, p \in \Index
\end{equation}
with the functions $ a_p^{ij}(v) $ defined by \eqref{eq:sw-inter-meas}. If the constants $ \hat a_p^{ij} $ are finite for all $ i, j \in \{1, \ldots, m\} $ and $ p \in \IndexInf $, then the topological entropy of the switched interconnected system \eqref{eq:sw-inter} is upper bounded by
\begin{equation}\label{eq:sw-inter-ent-upper-entry}
    h(f_\sigma, K) \leq \max \bigg\{ \limsup_{t \to \infty} \sum_{p \in \IndexPos} n \mu_\Network(\hat A_p^\Network) \rho_p(t),\, 0 \bigg\}
\end{equation}
with the matrices $ \hat A_p^\Network = [\hat a_p^{ij}] \in \dR^{m \times m} $.
\end{cor}

\begin{proof}
Note that the matrices $ A_p^\Network(v) = [a_p^{ij}(v)] $ and $ \hat A_p^\Network = [\hat a_p^{ij}] $ defined by \eqref{eq:sw-inter-meas} and \eqref{eq:sw-inter-meas-sup-entry} are all Metzler matrices. Moreover, the upper limit in \eqref{eq:sw-inter-meas-sup-entry} imply that, for an arbitrary $ \delta > 0 $, there is a large enough $ t_\delta \geq 0 $ such that
\begin{equation*}
	\max_{v \in \xi_\sigma(t, \co(K))} a_p^{ij}(v) \leq \hat a_p^{ij} + \delta \qquad \forall i, j \in \{1, \ldots, m\},\, p \in \IndexInf,\, t > t_\delta: \sigma(t) = p.
\end{equation*}
Then \eqref{eq:nneg-mx-meas} implies that
\begin{equation*}
	\max_{v \in \xi_\sigma(t, \co(K))} \mu_\Network(A_p^\Network(v)) \leq \mu_\Network(\hat A_p^\Network + \delta \mathbf{1}_{m \times m}) \leq \mu_\Network(\hat A_p^\Network) + \delta \mu_\Network(\mathbf{1}_{m \times m})
\end{equation*}
for all $ p \in \IndexInf $ and $ t > t_\delta $ such that $ \sigma(t) = p $, where the last inequality follows from \eqref{eq:mx-meas-vec}. Hence the constants $ \hat\mu_p^\Network $ defined by \eqref{eq:sw-inter-meas-sup} satisfy $ \hat\mu_p^\Network \leq \mu_\Network(\hat A_p^\Network) $ for all $ p \in \IndexInf $ as $ \delta > 0 $ is arbitrary. Then \eqref{eq:sw-inter-ent-upper-entry} follows from the upper bound \eqref{eq:sw-inter-ent-upper}.
\end{proof}

Thinking of the non-switched case as a switched system with constant switched signal, Theorem~\ref{thm:sw-inter-ent-upper} and Corollary~\ref{cor:sw-inter-ent-upper-entry} imply the following upper bounds for the entropy of a time-invariant interconnected system.
\begin{cor}\label{cor:ti-inter-ent-upper}
The topological entropy of a time-invariant interconnected system
\begin{equation}\label{eq:ti-inter}
	\dot x_i = f^i(x_1, \ldots, x_m), \qquad  i \in \{1, \ldots, m\}
\end{equation}
with states $ x_i \in \dR^{n_i} $ and $ x = (x_1, \ldots, x_m) \in \dR^n $ and functions $ f(x) = (f^1(x), \ldots, f^m(x)) $ is upper bounded by
\begin{equation}\label{eq:ti-inter-ent-upper}
	h(f, K) \leq \max \Big\{ \limsup_{t \to \infty} \max_{v \in \xi(t, \co(K))} n \mu_\Network( A^\Network(v)),\, 0 \Big\} \leq \max\{n \mu_\Network(\hat A^\Network),\, 0\} 
\end{equation}
with the matrix-valued function $ A^\Network(v) = [a^{ij}(v)] \in \dR^{m \times m} $ defined by
\begin{equation}\label{eq:ti-inter-meas-sup}
	a^{ii}(v) := \mu_{\local i}(J_{x_i} f^i(v)), \quad a^{ij}(v) := \|J_{x_j} f^i(v)\|_{\local i, j}, \qquad i, j \in \{1, \ldots, m\}: i \neq j,
\end{equation}
and the matrix $ \hat A^\Network = [\hat a^{ij}] \in \dR^{m \times m} $ defined by
\begin{equation}\label{eq:ti-inter-meas-sup-entry}
	\hat a^{ij} := \limsup_{t \to \infty} \max_{v \in \xi(t, \co(K))} a^{ij}(v), \qquad i, j \in \{1, \ldots, m\}.
\end{equation}
\end{cor}

A similar upper bound for the entropy of \eqref{eq:ti-inter} was constructed in \cite[Th.~1]{Liberzon2021}:
\begin{equation}\label{eq:ti-inter-ent-upper-dl}
	h(f, K) \leq \max\{n \lambda_{\max}(\tilde A^\Network),\, 0\},
\end{equation}
where $ \lambda_{\max}(\tilde A^\Network) $ is the largest real part of the eigenvalues of a matrix $ \tilde A^\Network \in \dR^{m \times m} $ satisfying
\begin{equation*}
	\tilde A^\Network \geq A^\Network(v) \qquad \forall v \in \dR^n
\end{equation*}
with the matrix-valued function $ A^\Network(v) = [a^{ij}(v)] $ defined by \eqref{eq:ti-inter-meas-sup}.\footnote{Note that $ \tilde A^\Network $ has to be a Metzler matrix; thus its eigenvalue with the largest real part is real \cite[Th.~10.2, p.\,167]{Bullo2022}.}
The upper bounds given by the inequalities in \eqref{eq:ti-inter-ent-upper} and \eqref{eq:ti-inter-ent-upper-dl} are both useful in the sense that neither is less conservative. On one hand, \eqref{eq:ti-inter-ent-upper-dl} may be less conservative due to the second inequality in \eqref{eq:mx-meas-norm-eig}. On the other hand, \eqref{eq:ti-inter-ent-upper-dl} may be more conservative as the matrix $ \hat A^\Network = [\hat a^{ij}] $ defined by \eqref{eq:ti-inter-meas-sup-entry} depends on the values of $ a^{ij}(v) $ over only the $ \omega $-limit set from $ \co(K) $, instead of the entire $ \dR^n $.

Based on the upper bounds \eqref{eq:sw-inter-ent-upper} and \eqref{eq:sw-inter-ent-upper-entry}, we construct additional upper bounds for the entropy of \eqref{eq:sw-inter} that require less information about the switching signal; the proof is along the lines of that of Corollary~\ref{cor:sw-ent-upper-sum-max} and thus omitted here. The relations between them are similar to those between \eqref{eq:sw-ent-upper}, \eqref{eq:sw-ent-upper-sum}, and \eqref{eq:sw-ent-upper-max} described in Remark~\ref{rmk:sw-ent-compare}.
\begin{cor}\label{cor:sw-inter-ent-upper-sum-max}
Given a switching signal $ \sigma $, consider the asymptotic active rates $ \hat\rho_p $ defined by \eqref{eq:act-rate-sup} and the sets of persistent and strongly persistent modes $ \IndexInf $ and $ \IndexPos $ defined by \eqref{eq:index-inf-pos}. If the constants $ \hat\mu_p^\Network $ and $ \hat a_p^{ij} $ defined by \eqref{eq:sw-inter-meas-sup} and \eqref{eq:sw-inter-meas-sup-entry} are finite for all $ i, j \in \{1, \ldots, m\} $ and $ p \in \IndexInf $, then the topological entropy of the switched interconnected system \eqref{eq:sw-inter} is upper bounded by
\begin{equation*}
    h(f_\sigma, K) \leq \sum_{p \in \IndexPos} \max\{n \hat\mu_p^\Network,\, 0\}\, \hat\rho_p \leq \sum_{p \in \IndexPos} \max\{n \mu_\Network(\hat A_p^\Network),\, 0\}\, \hat\rho_p
\end{equation*}
and also by
\begin{equation*}
    h(f_\sigma, K) \leq \max_{p \in \IndexPos} \max\{n \hat\mu_p^\Network,\, 0\} \leq \max_{p \in \IndexPos} \max\{n \mu_\Network(\hat A_p^\Network),\, 0\}
\end{equation*}
with the matrices $ \hat A_p^\Network = [\hat a_p^{ij}] \in \dR^{m \times m} $.
\end{cor}

\subsection{Switched block-diagonal systems}\label{ssec:ent-blk-diag}
Consider the case where the dynamics of the $ m \geq 2 $ subsystems of the switched interconnected system \eqref{eq:sw-inter} are independent. Specifically, for each $ i \in \{1, \ldots, m\} $ and $ p \in \Index $, the function $ f_p^i(x_1, \ldots, x_m) $ is independent of $ x_j $ for all $ j \neq i $. For brevity, we regard it as a function on $ \dR^{n_i} $ and rewrite it as $ f_p^i(x_i) $. Then \eqref{eq:sw-inter} can be written as a \emph{switched block-diagonal system}
\begin{equation}\label{eq:sw-blk-diag}
	\dot x_i = f_\sigma^i(x_i), \qquad i \in \{1, \ldots, m\}
\end{equation}
with functions $ f_p(x) = (f_p^1(x_1), \ldots, f_p^m(x_m)) $ for $ p \in \Index $. Clearly, the solution to the $ i $-th subsystem of \eqref{eq:sw-blk-diag} is independent of the initial states of other subsystems. For brevity, we denote it by $ \xi_\sigma^i(t, x_i) $ and the corresponding reachable set by $ \xi_\sigma^i(t, K_i) $, where $ K_i \subset \dR^{n_i} $ is the corresponding projection of the initial set $ K $.

In this subsection, we construct upper and lower bounds for the topological entropy of \eqref{eq:sw-blk-diag} that are less conservative than the results of simply applying the general bounds from Subsection~\ref{ssec:ent-inter} to \eqref{eq:sw-blk-diag}. Let the ``local'', ``network'', and ``global'' norms $ |\cdot|_{\local i} $ for $ i \in \{1, \ldots, m\} $, $ |\cdot|_\Network $, and $ |\cdot|_\Global $, and the corresponding induced norms $ \|\cdot\|_{\local i} $, $ \|\cdot\|_\Network $, and $ \|\cdot\|_\Global $ and matrix measures $ \mu_{\local i} $, $ \mu_\Network $, and $ \mu_\Global $ be given as in Subsection~\ref{ssec:ent-inter}.

The following result provides upper and lower bound for the entropy of \eqref{eq:sw-blk-diag}; the proof can be found in Subsection~\ref{ssec:proof-sw-blk-diag-ent}.
\begin{thm}\label{thm:sw-blk-diag-ent}
Given a switching signal $ \sigma $, consider the active times $ \tau_p(t) $ defined by \eqref{eq:act-time} and the sets of persistent and strongly persistent modes $ \IndexInf $ and $ \IndexPos $ defined by \eqref{eq:index-inf-pos}.
\begin{enumerate}
	\item Let
		\begin{equation}\label{eq:sw-blk-diag-meas-sup}
		    \hat\mu_p^i := \limsup_{t \to \infty:\, \sigma(t) = p} \max_{v_i \in \xi_\sigma^i(t, \co(K_i))} \mu_i(J_{x_i} f_p^i(v_i)), \qquad i \in \{1, \ldots, m\},\, p \in \Index.
		\end{equation}
		If the constants $ \hat\mu_p^i $ are finite for all $ i \in \{1, \ldots, m\} $ and $ p \in \IndexInf $, then the topological entropy of the switched block-diagonal system \eqref{eq:sw-blk-diag} is upper bounded by
		\begin{equation}\label{eq:sw-blk-diag-ent-upper}
			h(f_\sigma, K) \leq \limsup_{T \to \infty} \sum_{i=1}^{m} \frac{1}{T} \max_{t \in [0, T]} \sum_{p \in \IndexPos} n_i \hat\mu_p^i \tau_p(t).
		\end{equation}
	\item Let
		\begin{equation}\label{eq:sw-blk-diag-meas-inf}
			\check\mu_p^i := \liminf_{t \to \infty:\, \sigma(t) = p} \min_{v_i \in \co(\xi_\sigma^i(t, K_i))} -\mu_i(-J_{x_i} f_p^i(v_i)), \qquad i \in \{1, \ldots, m\},\, p \in \Index.
		\end{equation}
		If the constants $ \check\mu_p^i $ are finite for all $ i \in \{1, \ldots, m\} $ and $ p \in \IndexInf $, then the topological entropy of the switched block-diagonal system \eqref{eq:sw-blk-diag} is lower bounded by
		\begin{equation}\label{eq:sw-blk-diag-ent-lower}
			h(f_\sigma, K) \geq \limsup_{T \to \infty} \sum_{i=1}^{m} \frac{1}{T} \max_{t \in [0, T]} \sum_{p \in \IndexPos} n_i \check\mu_p^i \tau_p(t).
		\end{equation}
\end{enumerate}
\end{thm}

The following result provides an alternative lower bound for the entropy of \eqref{eq:sw-blk-diag}; the proof can be found in Subsection~\ref{ssec:proof-sw-blk-diag-ent-tr}.
\begin{thm}\label{thm:sw-blk-diag-ent-tr}
Given a switching signal $ \sigma $, consider the active rates $ \rho_p(t) $ defined by \eqref{eq:act-rate} and the sets of persistent and strongly persistent modes $ \IndexInf $ and $ \IndexPos $ defined by \eqref{eq:index-inf-pos}. Let
\begin{equation}\label{eq:sw-blk-diag-tr-inf}
	\check\chi_p^i := \liminf_{t \to \infty:\, \sigma(t) = p} \min_{v_i \in \xi_\sigma^i(t, K_i)} \tr(J_{x_i} f_p^i(v_i)), \qquad i \in \{1, \ldots, m\},\, p \in \Index.
\end{equation}
If the constants $ \check\chi_p^i $ are finite for all $ i \in \{1, \ldots, m\} $ and $ p \in \IndexInf $, then the topological entropy of the switched block-diagonal system \eqref{eq:sw-blk-diag} is lower bounded by
\begin{equation}\label{eq:sw-blk-diag-ent-lower-tr}
	h(f_\sigma, K) \geq \limsup_{t \to \infty} \sum_{i=1}^{m} \max \bigg\{ \sum_{p \in \IndexPos} \check\chi_p^i \rho_p(t),\, 0 \bigg\}.
\end{equation}
\end{thm}

Thinking of the non-switched case as a switched system with constant switched signal, Theorems~\ref{thm:sw-blk-diag-ent} and~\ref{thm:sw-blk-diag-ent-tr} imply the following upper and lower bounds for the entropy of a time-invariant block-diagonal system.
\begin{cor}\label{cor:ti-blk-diag-ent}
The topological entropy of a time-invariant block-diagonal system $ \dot x_i = f^i(x_i) $ for $ i \in \{1, \ldots, m\} $ with states $ x_i \in \dR^{n_i} $ and $ x = (x_1, \ldots, x_m) \in \dR^n $ and functions $ f(x) = (f^1(x_1), \ldots, f^m(x_m)) $ satisfies
\begin{equation*}
	\sum_{i=1}^{m} \max\{\check\chi^i,\, 0\} \leq h(f, K) \leq \sum_{i=1}^{m} \max\{n_i \hat\mu^i,\, 0\}
\end{equation*}
with
\begin{equation*}
	\begin{aligned}
		\check\chi^i &:= \liminf_{t \to \infty} \min_{v_i \in \xi^i(t, K_i)} \tr(J_{x_i} f^i(v_i)), \\
		\hat\mu^i &:= \limsup_{t \to \infty} \max_{v_i \in \xi^i(t, \co(K_i))} \mu_i(J_{x_i} f^i(v_i)),
	\end{aligned} \qquad i \in \{1, \ldots, m\},
\end{equation*}
where $ K_i \subset \dR^{n_i} $ is the corresponding projection of the initial set $ K $.
\end{cor}

Based on the upper bound \eqref{eq:sw-blk-diag-ent-upper}, we construct additional upper bounds for the entropy of \eqref{eq:sw-blk-diag} by considering each subsystem or each mode separately.
\begin{cor}\label{cor:sw-blk-diag-ent-upper-sub}
Given a switching signal $ \sigma $, consider the active rates $ \rho_p(t) $ defined by \eqref{eq:act-rate} and the sets of persistent and strongly persistent modes $ \IndexInf $ and $ \IndexPos $ defined by \eqref{eq:index-inf-pos}. If the constants $ \hat\mu_p^i $ defined by \eqref{eq:sw-blk-diag-meas-sup} are finite for all $ i \in \{1, \ldots, m\} $ and $ p \in \IndexInf $, then the topological entropy of the switched block-diagonal system \eqref{eq:sw-blk-diag} is upper bounded by
\begin{equation}\label{eq:sw-blk-diag-ent-upper-sub}
    h(f_\sigma, K) \leq \sum_{i=1}^{m} \max \bigg\{ \limsup_{t \to \infty} \sum_{p \in \IndexPos} n_i \hat\mu_p^i \rho_p(t),\, 0 \bigg\}.
\end{equation}
\end{cor}
\begin{proof}
As the upper limit is a subadditive function, the upper bound \eqref{eq:sw-blk-diag-ent-upper} implies that
\begin{equation*}
	h(f_\sigma, K) \leq \sum_{i=1}^{m} \limsup_{T \to \infty} \frac{1}{T} \max_{t \in [0, T]} \sum_{p \in \IndexPos} n_i \hat\mu_p^i \tau_p(t).
\end{equation*}
Then we obtain \eqref{eq:sw-blk-diag-ent-upper-sub} by invoking the second equality in \eqref{eq:sw-lin-avg-max-sup-new} for each $ i \in \{1, \ldots, m\} $ with the constants $ c_p = n_i \hat\mu_p^i $ for $ p \in \IndexPos $.
\end{proof}

\begin{cor}\label{cor:sw-blk-diag-ent-upper-mode}
Given a switching signal $ \sigma $, consider the active rates $ \rho_p(t) $ defined by \eqref{eq:act-rate} and the sets of persistent and strongly persistent modes $ \IndexInf $ and $ \IndexPos $ defined by \eqref{eq:index-inf-pos}. If the constants $ \hat\mu_p^i $ defined by \eqref{eq:sw-blk-diag-meas-sup} are finite for all $ i \in \{1, \ldots, m\} $ and $ p \in \IndexInf $, then the topological entropy of the switched block-diagonal system \eqref{eq:sw-blk-diag} is upper bounded by
\begin{equation}\label{eq:sw-blk-diag-ent-upper-mode}
    h(f_\sigma, K) \leq \limsup_{t \to \infty} \sum_{p \in \IndexPos} \bigg( \sum_{i=1}^{m} \max\{n_i \hat\mu_p^i,\, 0\} \bigg) \rho_p(t).
\end{equation}
\end{cor}
\begin{proof}
As the active times $ \tau_p(t) $ are nondecreasing, the upper bound \eqref{eq:sw-blk-diag-ent-upper} implies that
\begin{equation*}
	h(f_\sigma, K) \leq \limsup_{T \to \infty} \sum_{i=1}^{m} \sum_{p \in \IndexPos} \max\{n_i \hat\mu_p^i,\, 0\} \rho_p(T).
\end{equation*}
Then we obtain \eqref{eq:sw-blk-diag-ent-upper-mode} by grouping together the terms for the same mode.
\end{proof}

Based on the upper bounds \eqref{eq:sw-blk-diag-ent-upper-sub} and \eqref{eq:sw-blk-diag-ent-upper-mode}, we also construct additional upper bounds for the entropy of \eqref{eq:sw-blk-diag} that require less information about the switching signal; the proof is along the lines of that of Corollary~\ref{cor:sw-ent-upper-sum-max} and thus omitted here.
\begin{cor}\label{cor:sw-blk-diag-ent-upper-sum-max}
Given a switching signal $ \sigma $, consider the asymptotic active rates $ \hat\rho_p $ defined by \eqref{eq:act-rate-sup} and the sets of persistent and strongly persistent modes $ \IndexInf $ and $ \IndexPos $ defined by \eqref{eq:index-inf-pos}. If the constants $ \hat\mu_p^i $ defined by \eqref{eq:sw-blk-diag-meas-sup} are finite for all $ i \in \{1, \ldots, m\} $ and $ p \in \IndexInf $, then the topological entropy of the switched block-diagonal system \eqref{eq:sw-blk-diag} is upper bounded by
\begin{equation}\label{eq:sw-blk-diag-ent-upper-sum}
    h(f_\sigma, K) \leq \sum_{p \in \IndexPos} \bigg( \sum_{i=1}^{m} \max\{n_i \hat\mu_p^i,\, 0\} \bigg) \hat\rho_p
\end{equation}
and also by
\begin{equation}\label{eq:sw-blk-diag-ent-upper-max}
    h(f_\sigma, K) \leq \max_{p \in \IndexPos} \sum_{i=1}^{m} \max\{n_i \hat\mu_p^i,\, 0\}.
\end{equation}
\end{cor}

The results in Theorems~\ref{thm:sw-ent},~\ref{thm:sw-blk-diag-ent}, and~\ref{thm:sw-blk-diag-ent-tr} and Corollaries~\ref{cor:sw-blk-diag-ent-upper-sub},~\ref{cor:sw-blk-diag-ent-upper-mode}, and~\ref{cor:sw-blk-diag-ent-upper-sum-max} are compared in Fig.~\ref{fig:sw-blk-diag-ent} and Remark~\ref{rmk:sw-blk-diag-ent} below; for a numerical example, see Example~\ref{eg:sw-blk-diag} in Subsection~\ref{ssec:eg-sw-blk-diag}.
\begin{figure}[ht]
\normalsize
\centering
\renewcommand\tabcolsep{2pt}
\begin{tabular}{ccccc}
    \eqref{eq:sw-blk-diag-ent-upper} & $ \overset{\text{(C)}}{\implies} $ & \eqref{eq:sw-blk-diag-ent-upper-mode} & $ \implies $ & \eqref{eq:sw-blk-diag-ent-upper-max} \\
    \quad\,\rotatebox[origin=c]{270}{$ \,\implies $}\,{\scriptsize\text{(A)}} && \quad\,\rotatebox[origin=c]{270}{$ \,\implies $}\,{\scriptsize\text{(B)}} \\
    \eqref{eq:sw-blk-diag-ent-upper-sub} & $ \overset{\text{(D)}}{\implies} $ & \eqref{eq:sw-blk-diag-ent-upper-sum}
\end{tabular}
\caption{The relations between the upper bounds \eqref{eq:sw-blk-diag-ent-upper}, \eqref{eq:sw-blk-diag-ent-upper-sub}, \eqref{eq:sw-blk-diag-ent-upper-mode}, \eqref{eq:sw-blk-diag-ent-upper-sum}, and \eqref{eq:sw-blk-diag-ent-upper-max}. The implications (A) and (B) become equivalence if the active rates $ \rho_p(t) $ converge for all $ p \in \IndexPos $; the implication (C) becomes equivalence if the constants $ \hat\mu_p^i \geq 0 $ for all $ i \in \{1, \ldots, m\} $ and $ p \in \IndexPos $; the implication (D) becomes equivalence if both of these conditions hold.}\label{fig:sw-blk-diag-ent}
\end{figure}

\begin{rmk}\label{rmk:sw-blk-diag-ent}
\begin{enumerate}
	\item The upper bound \eqref{eq:sw-blk-diag-ent-upper} is less conservative than or equivalent to the upper bounds \eqref{eq:sw-blk-diag-ent-upper-sub} and \eqref{eq:sw-blk-diag-ent-upper-mode}, while \eqref{eq:sw-blk-diag-ent-upper-sub} is less conservative than or equivalent to the upper bound \eqref{eq:sw-blk-diag-ent-upper-sum}, and \eqref{eq:sw-blk-diag-ent-upper-mode} is less conservative than or equivalent to the upper bounds \eqref{eq:sw-blk-diag-ent-upper-sum} and \eqref{eq:sw-blk-diag-ent-upper-max}.
		The upper bounds \eqref{eq:sw-blk-diag-ent-upper-sub} and \eqref{eq:sw-blk-diag-ent-upper-mode} are both useful in the sense that neither is less conservative (cf. Fig.~\ref{fig:sw-blk-diag-ent}).
		The same holds for the upper bounds \eqref{eq:sw-blk-diag-ent-upper-sum} and \eqref{eq:sw-blk-diag-ent-upper-max} (cf. item~\ref{rmk:sw-ent-compare-order} of Remark~\ref{rmk:sw-ent-compare}), and also for the lower bounds \eqref{eq:sw-blk-diag-ent-lower} and \eqref{eq:sw-blk-diag-ent-lower-tr}.
	\item As explained in the proof of Theorem~\ref{thm:sw-ent} in Subsection~\ref{ssec:proof-sw-ent}, the upper bound \eqref{eq:sw-ent-upper} and lower bound \eqref{eq:sw-ent-lower} for the entropy of the general switched nonlinear systems can be obtained by taking $ m = 1 $ and $ n_1 = n $ in the bounds \eqref{eq:sw-blk-diag-ent-upper-sub} and \eqref{eq:sw-blk-diag-ent-lower-tr}, respectively.
		For the switched block-diagonal system \eqref{eq:sw-blk-diag}, the bounds \eqref{eq:sw-blk-diag-ent-upper-sub}, \eqref{eq:sw-blk-diag-ent-lower-tr}, \eqref{eq:sw-blk-diag-ent-upper-sum}, and \eqref{eq:sw-blk-diag-ent-upper-max} are less conservative than or equivalent to the bounds \eqref{eq:sw-ent-upper}, \eqref{eq:sw-ent-lower}, \eqref{eq:sw-ent-upper-sum}, and \eqref{eq:sw-ent-upper-max}, respectively.
	\item For a fixed family of modes, the upper bounds \eqref{eq:sw-blk-diag-ent-upper-sum} and \eqref{eq:sw-blk-diag-ent-upper-max} require less information about the switching signal than the upper bounds \eqref{eq:sw-blk-diag-ent-upper}, \eqref{eq:sw-blk-diag-ent-upper-sub}, and \eqref{eq:sw-blk-diag-ent-upper-mode}, as the former two depend on the asymptotic active rates $ \hat\rho_p $ instead of the active times $ \tau_p(t) $ and active rates $ \rho_p(t) $. Moreover, if for each $ p \in \IndexPos $, a constant $ \hat\mu_p^* $ as in the first two scenarios in Remark~\ref{rmk:ent-simp} is used in place of $ \hat\mu_p $, then \eqref{eq:sw-blk-diag-ent-upper-max} yields the same upper bound for all switching signals such that all modes are strongly persistent (i.e., $ \IndexPos = \IndexInf = \Index $).
\end{enumerate}
\end{rmk}

Consider the case where the subsystems of the switched block-diagonal system \eqref{eq:sw-blk-diag} are all scalar and their modes are all linear, that is, there is a family of diagonal matrices $ D_p = \diagMx(a_p^1, \ldots, a_p^n) \in \dR^{n \times n} $ for $ p \in \Index $ such that
\begin{equation}\label{eq:sw-diag-lin}
    f_p^i(x_i) = a_p^i x_i \qquad \forall i \in \{1, \ldots, n\},\, p \in \Index,\, (x_1, \ldots, x_n) \in \dR^n.
\end{equation}
Then the constants $ \hat\mu_p^i $, $ \check\mu_p^i $, and $ \check\chi_p^i $ defined by \eqref{eq:sw-blk-diag-meas-sup}, \eqref{eq:sw-blk-diag-meas-inf}, and \eqref{eq:sw-blk-diag-tr-inf} satisfy
\begin{equation*}
    \hat\mu_p^i = \check\mu_p^i = \check\chi_p^i = a_p^i, \qquad \forall i \in \{1, \ldots, n\},\, p \in \IndexInf.
\end{equation*}
Therefore, Theorem~\ref{thm:sw-blk-diag-ent} and Corollaries~\ref{cor:sw-blk-diag-ent-upper-sub},~\ref{cor:sw-blk-diag-ent-upper-mode}, and~\ref{cor:sw-blk-diag-ent-upper-sum-max} extend \cite[Th.~7, Prop.~8 and~9, and Cor.~10]{YangSchmidtLiberzon2018} for the case of linear modes with diagonal structure to the case of nonlinear modes with block-diagonal structure, respectively, and improve them by considering persistent and strongly persistent modes. Moreover, the lower bound \eqref{eq:sw-blk-diag-ent-lower-tr} is novel even in the former case, which is presented in the following corollary.

\begin{cor}\label{cor:sw-diag-lin-ent-lower-tr}
The topological entropy of the switched linear diagonal system \eqref{eq:sw-diag-lin} is lower bounded by
\begin{equation}\label{eq:sw-diag-lin-ent-lower-tr}
	h(D_\sigma) \geq \limsup_{t \to \infty} \sum_{i=1}^{n} \max \bigg\{ \sum_{p \in \IndexPos} a_p^i \rho_p(t),\, 0 \bigg\}.
\end{equation}
\end{cor}

Theorem~\ref{thm:sw-blk-diag-ent} and Corollaries~\ref{cor:sw-blk-diag-ent-upper-sub} and~\ref{cor:sw-blk-diag-ent-upper-sum-max} also extend \cite[Th.~4.1 and Cor~4.2]{YangLiberzonHespanha2021} for the case of diagonal structure to the case of block-diagonal structure. Moreover, due to the upper limit in \eqref{eq:sw-blk-diag-meas-sup}, the upper bound \eqref{eq:sw-blk-diag-ent-upper} depends on the values of Jacobian matrices $ J_{x_i} f_p^i(x_i) $ over only the $ \omega $-limit set from the convex hull of initial set, $ \co(K) $, whereas the upper bound \cite[eq.~(39)]{YangLiberzonHespanha2021} involves all reachable points from $ \co(K) $.

\section{Numerical examples}\label{sec:eg}
Consider the following switched nonlinear system in the nonnegative orthant $ \dR_{\geq 0}^n $ from \cite{AleksandrovChenPlatonovZhang2011}:
\begin{equation}\label{eq:eg-sw}
    \dot x = f_\sigma(x) := (r_\sigma + A_\sigma x) \HProd x
\end{equation}
with state $ x \in \dR_{\geq 0}^n $, switching signal $ \sigma: \dR_{\geq 0} \to \Index $, and finite index set $ \Index $, where $ \circ $ denotes the Hadamard (entrywise) product. Equivalently, \eqref{eq:eg-sw} can be written as
\begin{equation*}
    \dot x_i = \bigg( r_\sigma^i + \sum_{j=1}^{n} a_\sigma^{ij} x_j \bigg) x_i, \qquad i \in \{1, \ldots, n\}
\end{equation*}
with $ x = (x_1, \ldots, x_n) $ and $ r_p = (r_p^1, \ldots, r_p^n) $ and $ A_p = [a_p^{ij}] $ for $ p \in \Index $. Each mode $ p $ of \eqref{eq:eg-sw} is a Lotka--Volterra ecosystem model that describes the population dynamics of $ n $ species in a biological community \cite[Ch.~5]{HofbauerSigmund1998}, where $ x_i $ is the population density of the $ i $-th species, $ r_p^i \in \dR $ quantifies its intrinsic growth rate, $ a_p^{ii} < 0 $ is a self-interaction term justified by the limitation of resources in the environment, and $ a_p^{ij} \in \dR $ for $ j \neq i $ is an interaction term quantifying the influence of the $ j $-th species on the $ i $-th one. Switching in \eqref{eq:eg-sw} may be due to seasonal changes or environment fluctuations. Clearly, $ \dR_{\geq 0}^n $ is a positively invariant set for the switched system \eqref{eq:eg-sw}.

We construct two switching signals $ \sigma_1, \sigma_2: \dR_{\geq 0} \to \Index = \{1,\, 2\} $ as follows\footnote{We denote by $ 0 < t_1 < t_2 < \cdots $ the sequence of switches and let $ t_0 := 0 $, with $ \sigma(t) = 1 $ on $ [t_{2k}, t_{2k+1}) $ and $ \sigma(t) = 2 $ on $ [t_{2k+1}, t_{2k+2}) $.}:
\begin{itemize}
    \item $ \sigma_1 $ with periodic switching: Let $ t_k := 1000 k $ for $ k \geq 1 $. Simple computation shows that $ \hat\rho_1 = \hat\rho_2 = 0.5 $.%
    \item $ \sigma_2 $ with constant set-points: Let $ t_1 := 1 $ and $ t_{2k} := \min\{t > t_{2k - 1}: \rho_{2}(t) \geq 0.9\} $ and $ t_{2k+1} := \min\{t > t_{2k}: \rho_{1}(t) \geq 0.9\} $ for $ k \geq 1 $. Simple computation shows that $ t_k = 9^{k-1} + 9^{k-2} $ for $ k \geq 2 $ and $ \hat\rho_1 = \hat\rho_2 = 0.9 $.%
\end{itemize}
Clearly, $ \IndexPos = \IndexInf = \Index $ for both $ \sigma_1 $ and $ \sigma_2 $.

\subsection{A general case}\label{ssec:eg-sw}
Suppose that the matrices $ A_p = [a_p^{ij}] $ in the switched Lotka--Volterra system \eqref{eq:eg-sw} satisfy
\begin{equation}\label{eq:eg-sw-cond}
	a_p^{ii} + \sum_{j \neq i} |a_p^{ij}| < 0, \quad a_p^{ii} + \sum_{j \neq i} |a_p^{ji}| < 0 \qquad \forall i \in \{1, \ldots, n\},\, p \in \Index.
\end{equation}
Then \cite[Th.~3]{AleksandrovChenPlatonovZhang2011} implies that \eqref{eq:eg-sw} is \emph{uniformly ultimately bounded (UUB)} in $ \dR_{\geq 0}^n $, and its $ \omega $-limit set is a subset of\footnote{Specifically, following the proof of \cite[Th.~3]{AleksandrovChenPlatonovZhang2011}, $ S $ contains the $ \omega $-limit set of \eqref{eq:eg-sw} if for all $ p \in \Index $ and $ x \in \dR_{\geq 0}^n \backslash (S \cup \{0\}) $, we have $ \mathbf{1}_n^\top f_p(x) < 0 $, i.e., $ r_p^\top x + x^\top A_p x < 0 $. Note that $ r_p^\top x + x^\top A_p x = r_p^\top x + x^\top (A_p + A_p^\top)\, x/2 \leq (r_p + \lambda_{\max}(A_p + A_p^\top)\, x/2)^\top x $, which is negative if $ 2 r_p^i + \lambda_{\max}(A_p + A_p^\top)\, x_i < 0 $ for all $ i \in \{1, \ldots, n\} $ as $ x \in \dR_{\geq 0}^n \backslash \{0\} $.}
\begin{equation}\label{eq:eg-sw-set}
    S := \prod_{i=1}^{n} \bigg[ 0, \max_{p \in \Index} \max \bigg\{ -\frac{2 r_p^i}{\lambda_{\max}(A_p + A_p^\top)},\, 0 \bigg\} \bigg],
\end{equation}
where $ \lambda_{\max}(A_p + A_p^\top) $ denotes the largest eigenvalue of the symmetric matrix $ A_p + A_p^\top $ and satisfies $ \lambda_{\max}(A_p + A_p^\top) < 0 $ due to the condition \eqref{eq:eg-sw-cond}, the formula for matrix measure \eqref{eq:mx-meas-inf}, and the second inequality in \eqref{eq:mx-meas-norm-eig}.

\begin{eg}\label{eg:sw}
Consider the switched system \eqref{eq:eg-sw} in $ \dR_{\geq 0}^2 $ with index set $ \Index = \{1,\, 2\} $ and coefficients
\begin{equation*}
	r_1 = \begin{bmatrix}
		-1 \\
		2
	\end{bmatrix}, \qquad r_2 = \begin{bmatrix}
		3 \\
		-1
	\end{bmatrix}, \qquad A_1 = A_2 = \begin{bmatrix}
		-1 & 0.1 \\
		0.1 & -1
	\end{bmatrix}.
\end{equation*}
Simple computation shows that mode~$ 1 $ has an attractor $ (0, 2) $ and a saddle point $ (0, 0) $ with stable manifold $ \dR_{\geq 0} \times \{0\} $, and mode~$ 2 $ has an attractor $ (3, 0) $ and a saddle point $ (0, 0) $ with stable manifold $ \{0\} \times \dR_{\geq 0} $. Moreover, the condition \eqref{eq:eg-sw-cond} holds and the set defined by \eqref{eq:eg-sw-set} is given by $ S = [0, 10/3] \times [0, 20/9] $. Typical trajectories of \eqref{eq:eg-sw} for the individual modes~$ 1 $ and~$ 2 $ and the switching signals $ \sigma_1 $ and $ \sigma_2 $ are plotted in Fig.~\ref{fig:eg-sw}. In particular, $ S $ is not a positively invariant set.
\begin{figure}[!htbp]
\centering
\subfloat[Mode~$ 1 $]{\includegraphics[width=.49\columnwidth,max width=128pt,trim=2em 0ex 3.5em 5ex,clip]{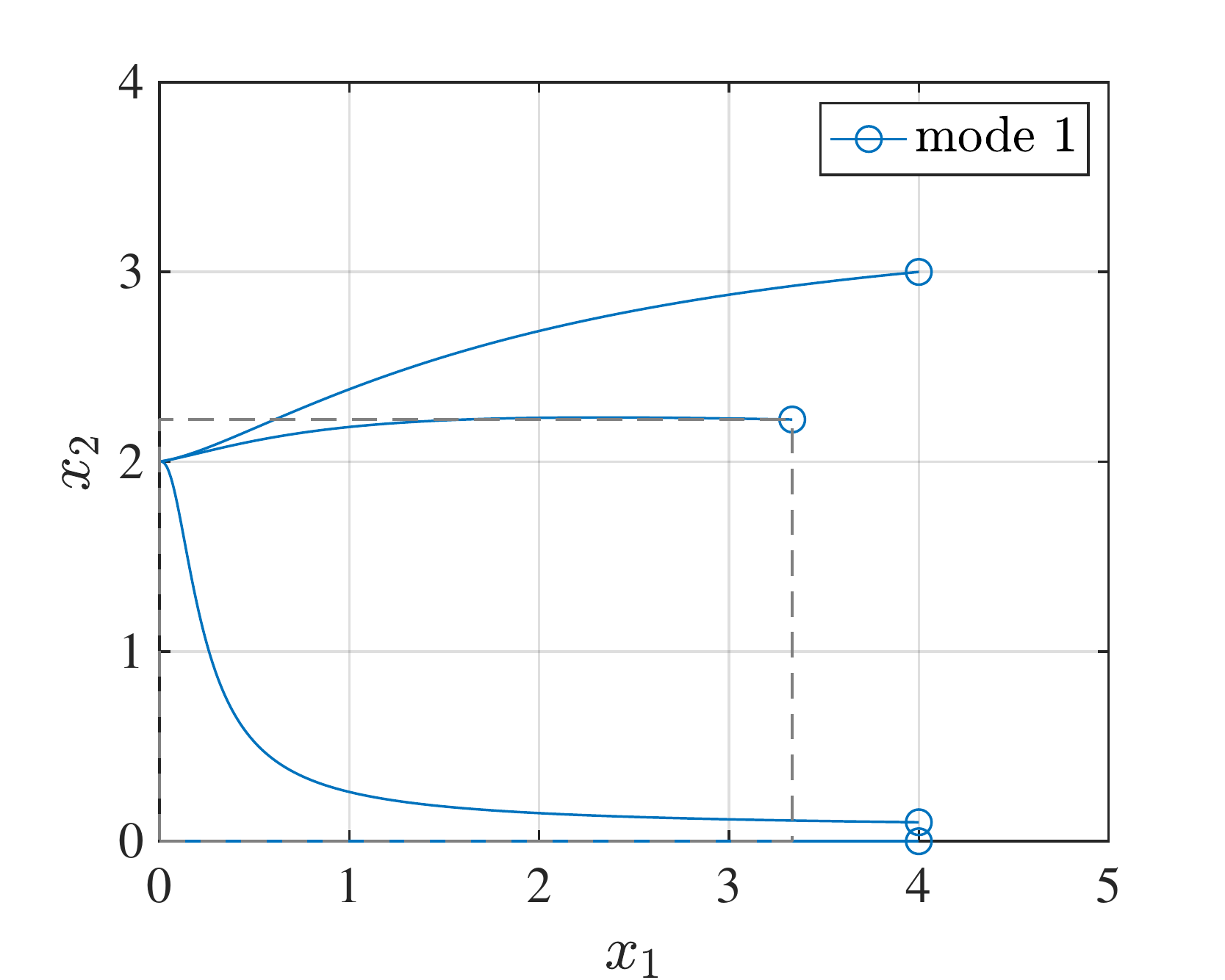}\label{fig:eg-sw-1}}%
\subfloat[Mode~$ 2 $]{\includegraphics[width=.49\columnwidth,max width=128pt,trim=2em 0ex 3.5em 5ex,clip]{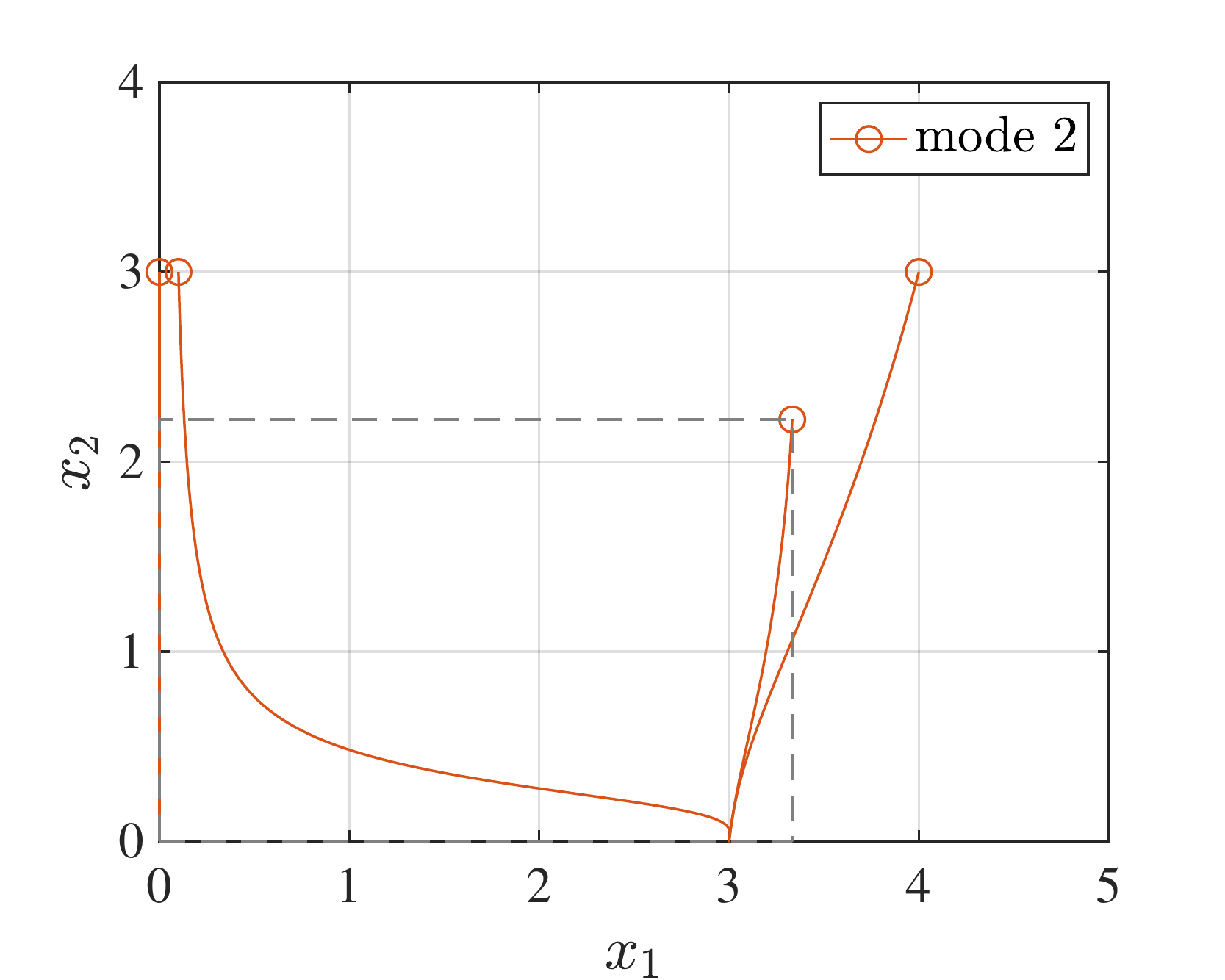}\label{fig:eg-sw-2}}%
\\%
\subfloat[Switching signal $ \sigma_1 $]{\includegraphics[width=.49\columnwidth,max width=128pt,trim=2em 0ex 3.5em 5ex,clip]{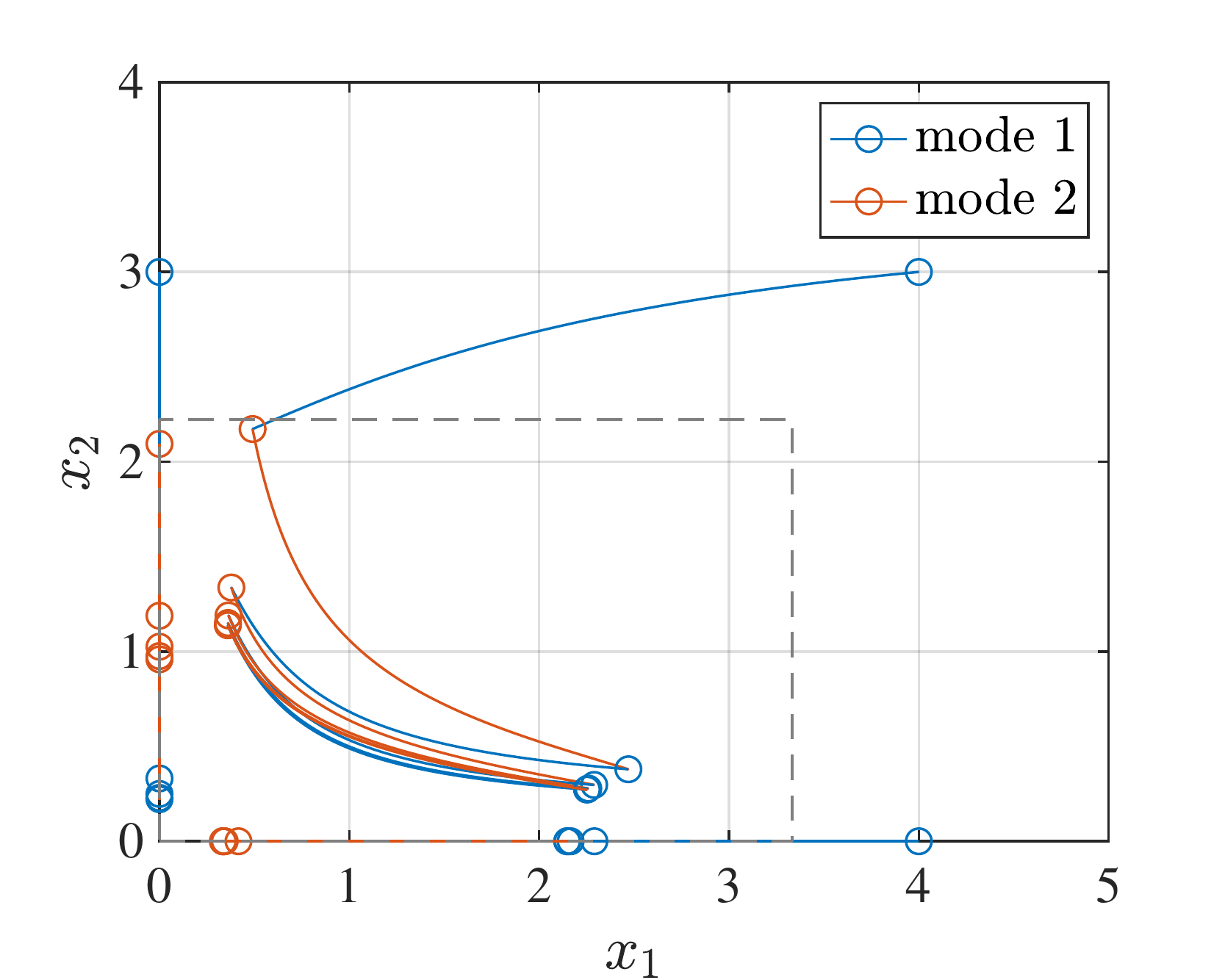}\label{fig:eg-sw-3}}%
\subfloat[Switching signal $ \sigma_2 $]{\includegraphics[width=.49\columnwidth,max width=128pt,trim=2em 0ex 3.5em 5ex,clip]{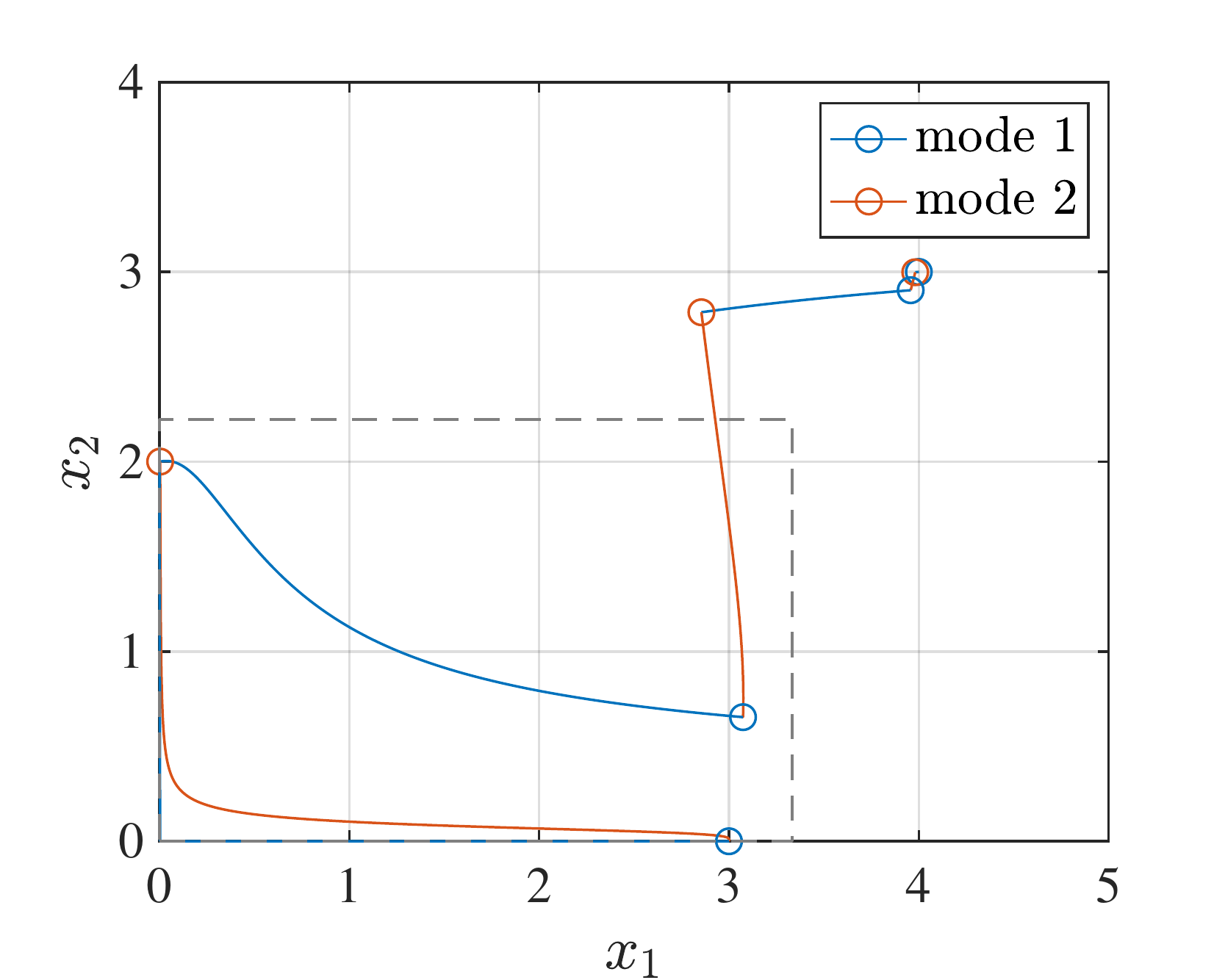}\label{fig:eg-sw-4}}%
\caption{Trajectories of the switched system in Example~\ref{eg:sw}, for (a) individual mode~$ 1 $ with initial states $ (4, 3) $, $ (4, 0.1) $, $ (4, 0) $, and $ (10/3, 20/9) $; (b) individual mode~$ 2 $ with initial states $ (4, 3) $, $ (0.1, 3) $, $ (0, 3) $, and $ (10/3, 20/9) $; (c) switching signal $ \sigma_1 $ with initial states $ (4, 3) $, $ (4, 0) $, and $ (0, 3) $; (d) switching signal $ \sigma_2 $ with initial state $ (4, 3) $. The circles mark the beginning of a segment after switching. The dashed rectangles represent $ S = [0, 10/3] \times [0, 20/9] $.}\label{fig:eg-sw}%
\end{figure}

The Jacobian matrices of individual modes are given by
\begin{equation*}
\begin{aligned}
    J_x f_1(v) &= \begin{bmatrix}
        -1 - 2 v_1 + 0.1 v_2 & 0.1 v_1 \\
        0.1 v_2 & 2 + 0.1 v_1 - 2 v_2
    \end{bmatrix}, \\
    J_x f_2(v) &= \begin{bmatrix}
        3 - 2 v_1 + 0.1 v_2 & 0.1 v_1 \\
        0.1 v_2 & -1 + 0.1 v_1 - 2 v_2
    \end{bmatrix}
\end{aligned}
\end{equation*}
with $ v = (v_1, v_2) \in \dR_{\geq 0}^2 $. As the switched system is UUB and its $ \omega $-limit set is a subset of $ S $, for all initial sets $ K \subset \dR_{\geq 0}^2 $, one can obtain upper bounds for the constants $ \hat\mu_p $ defined by \eqref{eq:sw-meas-sup} by replacing the upper limit over $ \{t \geq 0: \sigma(s) = p\} $ and maximum over $ \xi_\sigma(t, \co(K)) $ with a maximum over $ S $, that is,
\begin{equation*}
\begin{aligned}
    \hat\mu_1 &\leq \max_{v \in S} \mu(J_x f_1(v)) = \max_{v \in S} \max\{-1 - 1.9 v_1 + 0.1 v_2,\, 2 + 0.1 v_1 - 1.9 v_2\} = 7/3, \\
    \hat\mu_2 &\leq \max_{v \in S} \mu(J_x f_2(v)) = \max_{v \in S} \max\{3 - 1.9 v_1 + 0.1 v_2,\, -1 + 0.1 v_1 - 1.9 v_2\} = 29/9,
\end{aligned}
\end{equation*}
where the matrix measures are computed with the induced $ \infty $-norm using the formula \eqref{eq:mx-meas-inf}.

The upper bounds for the entropies $ h(f_{\sigma_1}, K) $ and $ h(f_{\sigma_2}, K) $ computed using \eqref{eq:sw-ent-upper}, \eqref{eq:sw-ent-upper-sum}, and \eqref{eq:sw-ent-upper-max} for all initial sets $ K \subset \dR^2_{\geq 0} $ are summarized in Table~\ref{tbl:eg-sw} below.\footnote{In this example, we use the set $ S $ defined by \eqref{eq:eg-sw-set} which contains the $ \omega $-limit set for every switched system \eqref{eq:eg-sw} satisfying \eqref{eq:eg-sw-cond}. For a given family of coefficients, one can usually construct more precise over-approximation of the $ \omega $-limit set and thus obtain less conservative upper bounds for entropy, such as those in \cite[Example~3.6]{YangLiberzonHespanha2021}.}
In particular, the upper bound \eqref{eq:sw-ent-upper} for $ h(f_{\sigma_2}, K) $ can be computed as follows:
\begin{equation*}
\begin{aligned}
    h(f_{\sigma_2}, K) &\leq \limsup_{t \to \infty} \big( 2 \hat\mu_1 \rho_1(t) + 2 \hat\mu_2 \rho_2(t) \big) \\
    &\leq \limsup_{t \to \infty}\,  2 \bigg( \frac{7}{3} (1 - \rho_2(t)) + \frac{29}{9} \rho_2(t) \bigg) = 2 \bigg( \frac{7}{3} + \bigg( \frac{29}{9} - \frac{7}{3} \bigg) \hat\rho_2 \bigg) \approx 6.27.
\end{aligned}
\end{equation*}
The values in Table~\ref{tbl:eg-sw} are consistent with the relations between these upper bounds described in Remark~\ref{rmk:sw-ent-compare}.
\begin{table}[!ht]
\begin{center}
\begin{minipage}{\textwidth}
\caption{Upper bounds for the entropy of the switched system in Example~\ref{eg:sw}.}\label{tbl:eg-sw}
\begin{tabular}{@{\extracolsep{0.4em}}ccccc@{\extracolsep{0.4em}}}
    \hline%
    & $(\hat\rho_1, \hat\rho_2)$ & \eqref{eq:sw-ent-upper} & \eqref{eq:sw-ent-upper-sum} & \eqref{eq:sw-ent-upper-max} \\
    \hline%
    $\sigma_1$ & $(0.5, 0.5)$ & $5.56$ & $5.56$ & $6.45$ \\
    $\sigma_2$ & $(0.9, 0.9)$ & $6.27$ & $10$ & $6.45$ \\
    \hline%
\end{tabular}
\end{minipage}
\end{center}
\end{table}
\end{eg}

\subsection{A block-diagonal case}\label{ssec:eg-sw-blk-diag}
Consider the case where the switched Lotka--Volterra system \eqref{eq:eg-sw} describes a biological community of $ m \geq 2 $ groups of species, where the $ i $-th group consists of $ n_i $ species, and the dynamics of species from different groups are independent. Regarding the groups as interconnected subsystems, we denote by $ \bar x_i \in \dR_{\geq 0}^{n_i} $ the state of the $ i $-th subsystem, and by $ \bar f_p^i(\bar x_i) $, $ \bar r_p^i $, and $ \bar A_p^i $ the corresponding function and coefficients of mode $ p $ to avoid confusion. Then \eqref{eq:eg-sw} can be written as the switched block-diagonal system
\begin{equation}\label{eq:eg-sw-blk-diag}
    \dot{\bar x}_i = \bar f_\sigma^i(\bar x_i) := (\bar r_\sigma^i + \bar A_\sigma^i \bar x_i) \HProd \bar x_i, \qquad i \in \{1, \ldots, m\}
\end{equation}
with state $ x = (\bar x_1, \ldots, \bar x_k) \in \dR^n $, and functions $ f_p(x) = (\bar f_p^1(\bar x_1), \ldots, \bar f_p^m(\bar x_m)) $, coefficients $ r_p = (\bar r_p^1, \ldots, \bar r_p^m) $, and matrices $ A_p = \diagMx(\bar A_p^1, \ldots, \bar A_p^m) $ for $ p \in \Index $.

For \eqref{eq:eg-sw-blk-diag}, the formula \eqref{eq:mx-meas-inf} implies that the condition \eqref{eq:eg-sw-cond} can be written as
\begin{equation}\label{eq:eg-sw-blk-diag-cond}
	\mu(\bar A_p^i), \mu((\bar A_p^i)^\top) < 0 \qquad \forall i \in \{1, \ldots, m\},\, p \in \Index,
\end{equation}
where the matrix measures are computed with the induced $ \infty $-norm. If \eqref{eq:eg-sw-blk-diag-cond} holds, then \cite[Th.~3]{AleksandrovChenPlatonovZhang2011} implies that \eqref{eq:eg-sw-blk-diag} is UUB in $ \dR_{\geq 0}^n $, and its $ \omega $-limit set is a subset of
\begin{equation}\label{eq:eg-sw-blk-diag-set}
    S := \prod_{j=1}^{n} \bigg[ 0, \max_{p \in \Index} \max \bigg\{ -\frac{2 r_p^j}{\lambda_{\max}(A_p^{i(j)} + (A_p^{i(j)})^\top)},\, 0 \bigg\} \bigg],
\end{equation}
where the $ j $-th species is in the $ i(j) $-th group, and $ \lambda_{\max}(A_p^{i(j)} + (A_p^{i(j)})^\top) $ denotes the largest eigenvalue of the symmetric matrix $ A_p^{i(j)} + (A_p^{i(j)})^\top $ and satisfies $ \lambda_{\max}(A_p^{i(j)} + (A_p^{i(j)})^\top) < 0 $ due to the condition \eqref{eq:eg-sw-blk-diag-cond}, the formula for matrix measure \eqref{eq:mx-meas-inf}, and the second inequality in \eqref{eq:mx-meas-norm-eig}.

\begin{eg}\label{eg:sw-blk-diag}
Consider the switched block-diagonal system \eqref{eq:eg-sw-blk-diag} in $ \dR_{\geq 0}^3 $ with index set $ \Index = \{1,\, 2\} $ and coefficients
\begin{equation*}
\begin{aligned}
	\bar r_1^1 &= \begin{bmatrix}
		-1 \\
		-1
	\end{bmatrix}, &\qquad \bar r_2^1 &= \begin{bmatrix}
		3 \\
		3
	\end{bmatrix}, &\qquad \bar A_1^1 &= \bar A_2^1 = \begin{bmatrix}
		-1 & 0.1 \\
		0.1 & -1
	\end{bmatrix}, \\
	\bar r_1^2 &= 2, &\qquad \bar r_2^2 &= -1, &\qquad \bar A_1^2 &= \bar A_2^2 = -1.
\end{aligned}
\end{equation*}
Simple computation shows that mode~$ 1 $ has an attractor $ (0, 0, 2) $ and a saddle point $ (0, 0, 0) $ with stable manifold $ \dR_{\geq 0}^2 \times \{0\} $, and mode~$ 2 $ has an attractor $ (10/3, 10/3, 0) $ and three saddle points $ (0, 0, 0) $, $ (3, 0, 0) $, and $ (0, 3, 0) $ with stable manifolds $ \{(0, 0)\} \times \dR_{\geq 0} $, $ \dR_{> 0} \times \{0\} \times \dR_{\geq 0} $, and $ \{0\} \times \dR_{> 0} \times \dR_{\geq 0} $, respectively. Moreover, the condition \eqref{eq:eg-sw-blk-diag-cond} holds and the set defined by \eqref{eq:eg-sw-blk-diag-set} is given by $ S = [0, 10/3]^2 \times [0, 2] $. Typical trajectories of \eqref{eq:eg-sw-blk-diag} for the individual modes~$ 1 $ and~$ 2 $ and the switching signals $ \sigma_1 $ and $ \sigma_2 $ are plotted in Fig.~\ref{fig:eg-sw-blk-diag}.
\begin{figure}[!htbp]
\centering
\subfloat[Mode~$ 1 $]{\includegraphics[width=.49\columnwidth,max width=128pt,trim=4.5em 1ex 4.5em 5.5ex,clip]{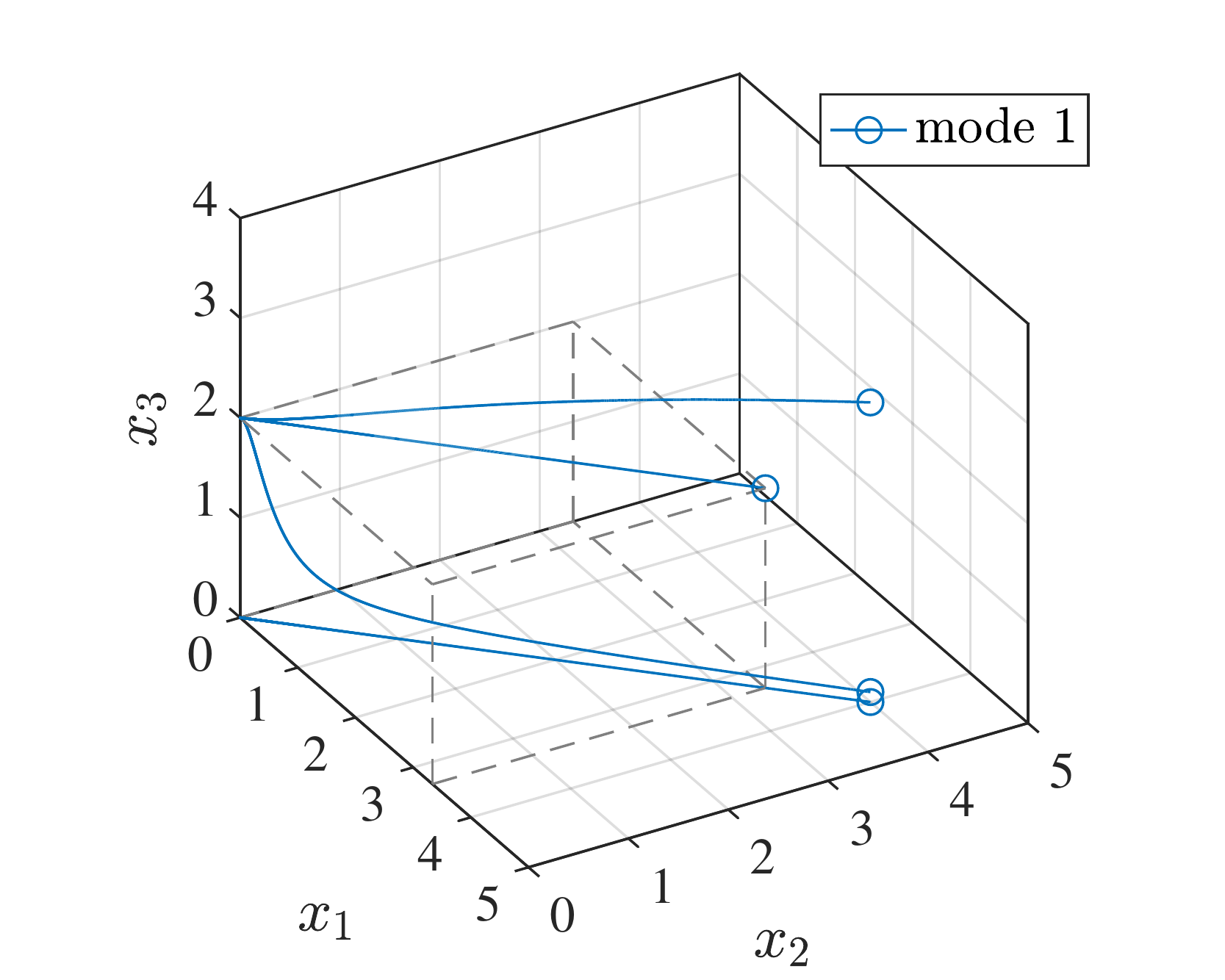}\label{fig:eg-sw-blk-diag-1}}%
\subfloat[Mode~$ 2 $]{\includegraphics[width=.49\columnwidth,max width=128pt,trim=4.5em 1ex 4.5em 5.5ex,clip]{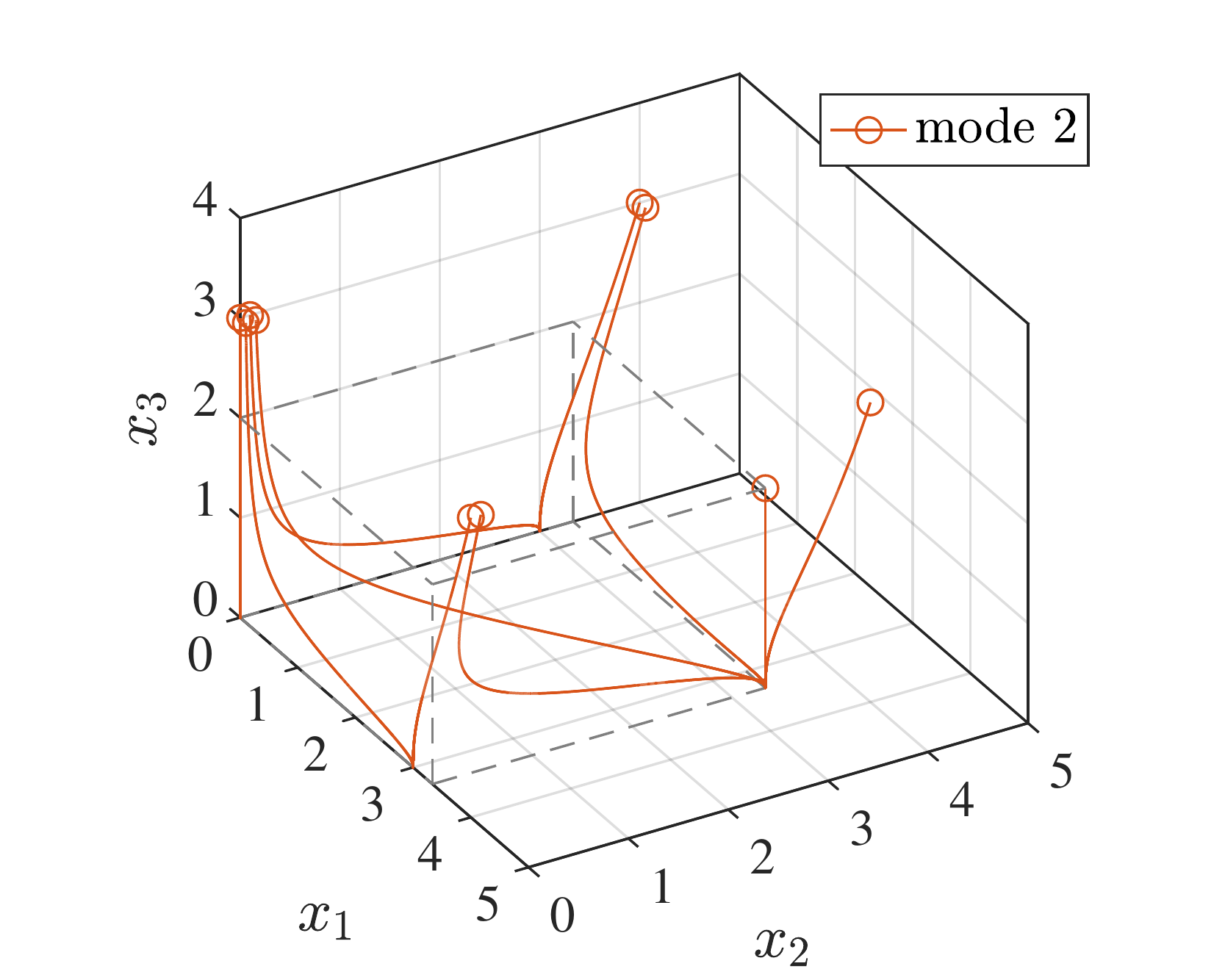}\label{fig:eg-sw-blk-diag-2}}%
\\%
\subfloat[Switching signal $ \sigma_1 $]{\includegraphics[width=.49\columnwidth,max width=128pt,trim=4.5em 1ex 4.5em 5.5ex,clip]{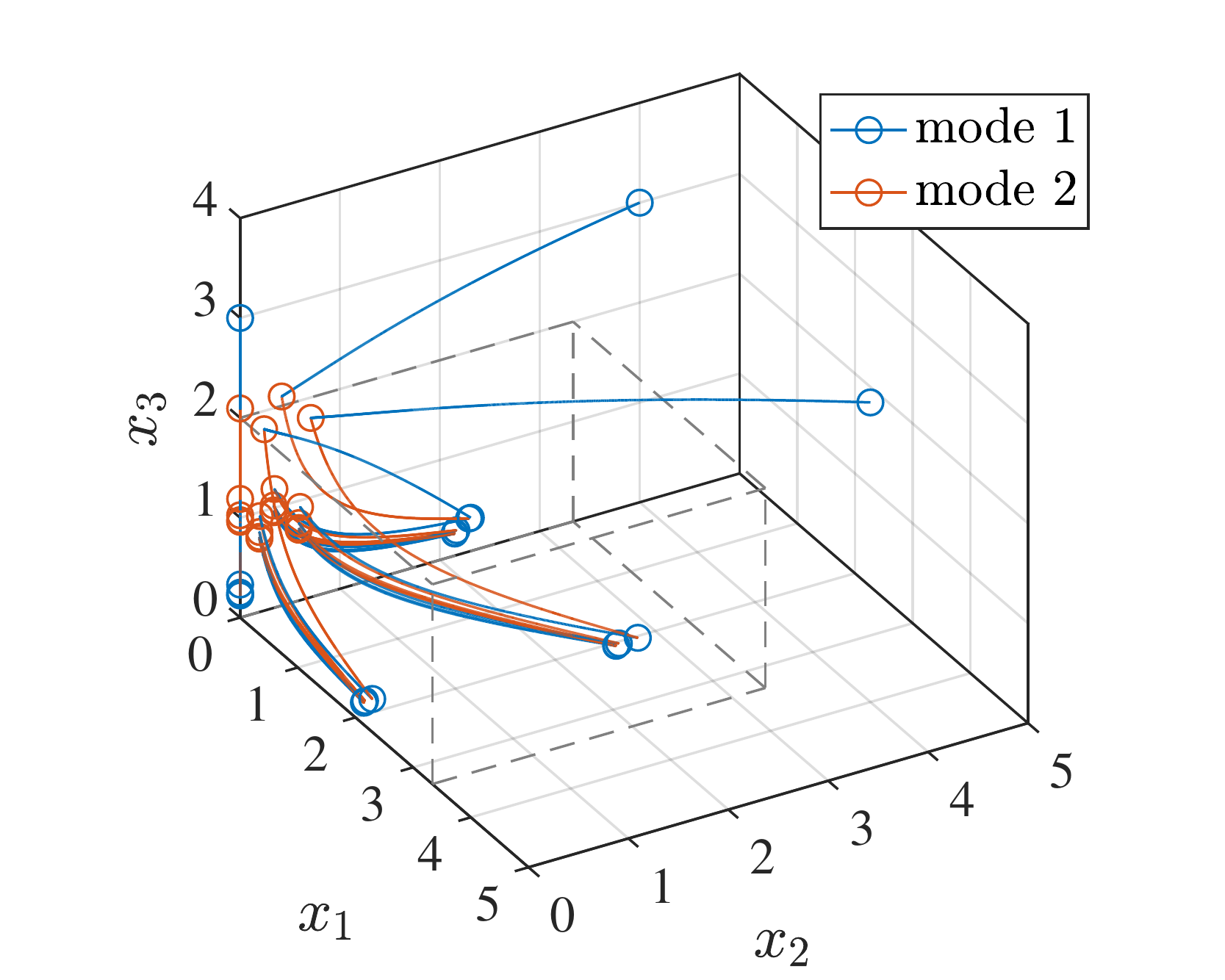}\label{fig:eg-sw-blk-diag-3}}%
\subfloat[Switching signal $ \sigma_2 $]{\includegraphics[width=.49\columnwidth,max width=128pt,trim=4.5em 1ex 4.5em 5.5ex,clip]{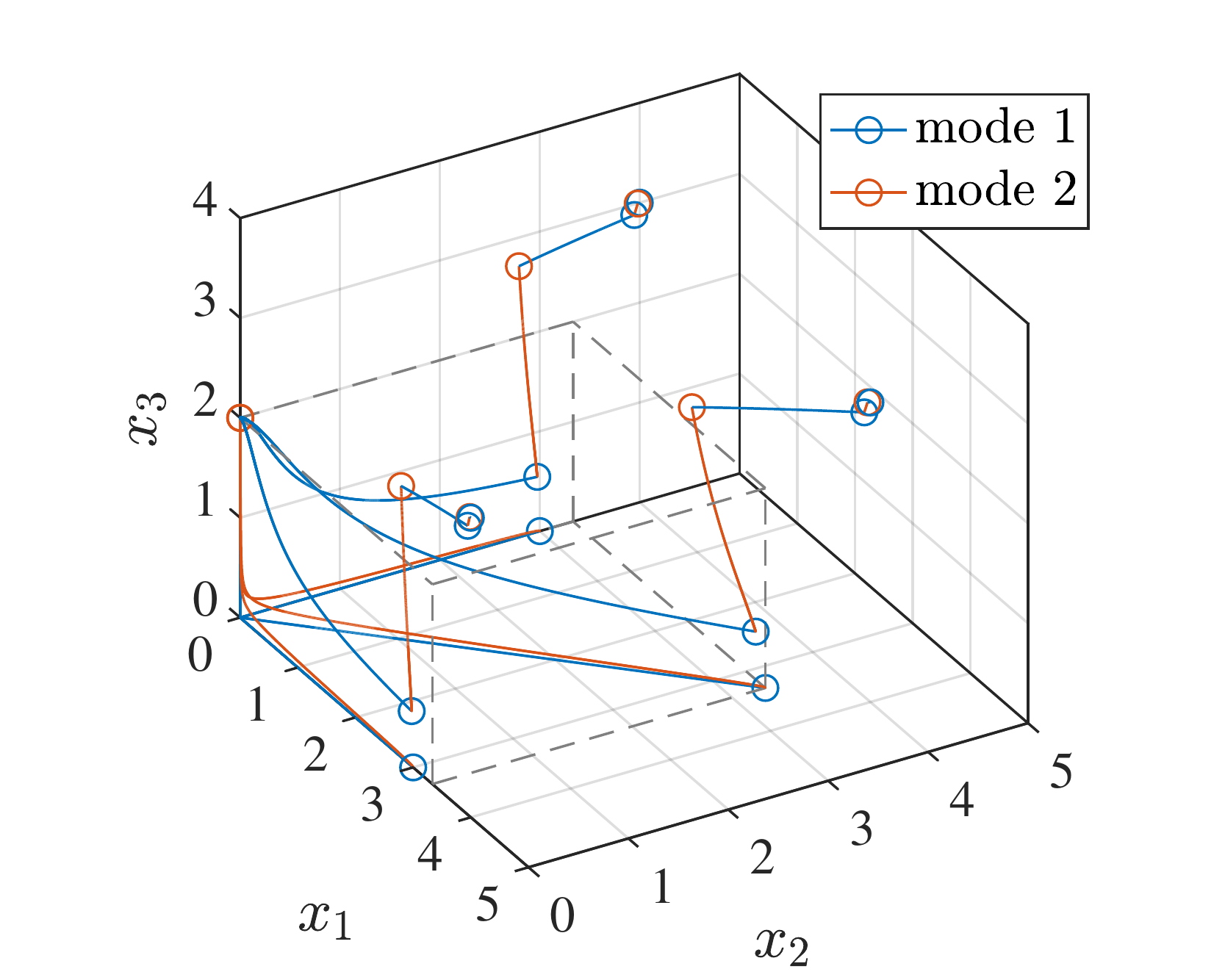}\label{fig:eg-sw-blk-diag-4}}%
\caption{Trajectories of the switched block-diagonal system in Example~\ref{eg:sw-blk-diag}, for (a) individual mode~$ 1 $ with initial states $ (4, 4, 3) $, $ (4, 4, 0.1) $, $ (4, 4, 0) $, and $ (10/3, 10/3, 2) $; (b) individual mode~$ 2 $ with initial states $ (4, 4, 3) $, $ (4, 0.1, 3) $, $ (0.1, 4, 3) $, $ (4, 0, 3) $, $ (0, 4, 3) $, $ (0.1, 0.1, 3) $, $ (0.1, 0, 3) $, $ (0, 0.1, 3) $, $ (0, 0, 3) $, and $ (10/3, 10/3, 2) $; (c) switching signal $ \sigma_1 $ with initial states $ (4, 4, 3) $, $ (4, 0, 3) $, $ (0, 4, 3) $, and $ (0, 0, 3) $; (d) switching signal $ \sigma_2 $ with initial states $ (4, 4, 3) $, $ (4, 0, 3) $, and $ (0, 4, 3) $. The circles mark the beginning of a segment after switching. The dashed hyperrectangles represent $ S = [0, 10/3]^2 \times [0, 2] $.}\label{fig:eg-sw-blk-diag}%
\end{figure}

The Jacobian matrices of individual modes are given by
\begin{equation*}
\begin{aligned}
    J_x f_1(v) &= \diagMx(J_{\bar x_1} \bar f_1^1(\bar v_1), J_{\bar x_2} \bar f_1^2(\bar v_2)), \\
    J_x f_2(v) &= \diagMx(J_{\bar x_1} \bar f_2^1(\bar v_1), J_{\bar x_2} \bar f_2^2(\bar v_2))
\end{aligned}
\end{equation*}
with
\begin{equation*}
\begin{aligned}
    J_{\bar x_1} \bar f_1^1(\bar v_1)) &= \begin{bmatrix}
        -1 - 2 v_1 + 0.1 v_2 & 0.1 v_1 \\
        0.1 v_2 & -1 + 0.1 v_1 - 2 v_2
    \end{bmatrix}, &\qquad J_{\bar x_2} \bar f_1^2(\bar v_2)) &= 2 - 2 v_3, \\
    J_{\bar x_1} \bar f_2^1(\bar v_1)) &= \begin{bmatrix}
        3 - 2 v_1 + 0.1 v_2 & 0.1 v_1 \\
        0.1 v_2 & 3 + 0.1 v_1 - 2 v_2
    \end{bmatrix}, &\qquad J_{\bar x_2} \bar f_2^2(\bar v_2)) &= -1 - 2 v_3,
\end{aligned}
\end{equation*}
where $ v = (\bar v_1, \bar v_2) $ with $ \bar v_1 = (v_1, v_2) \in \dR_{\geq 0}^2 $, and $ \bar v_2 = v_3 \in \dR_{\geq 0} $. As the switched block-diagonal system is UUB and its $ \omega $-limit set is a subset of $ S $, for all initial sets $ K \subset \dR_{\geq 0}^3 $, one can obtain upper bounds for the constants $ \hat\mu_p $ and $ \hat\mu_p^i $ defined by \eqref{eq:sw-meas-sup} and \eqref{eq:sw-blk-diag-meas-sup} by replacing the upper limit over $ \{t \geq 0: \sigma(s) = p\} $ and maximum over $ \xi_\sigma(t, \co(K)) $ and $ \xi^i_\sigma(t, \co(K_i)) $ with a maximum over $ S $, that is,
\begin{equation*}
\begin{aligned}
    \hat\mu_1^1 &\leq \max_{v \in S} \mu(J_{\bar x_1} \bar f_1^1(\bar v_1)) = \max_{v \in S} \max\{-1 - 1.9 v_1 + 0.1 v_2,\, -1 + 0.1 v_1 - 1.9 v_2\} = -2/3, \\
    \hat\mu_1^2 &\leq \max_{v \in S} \mu(J_{\bar x_2} \bar f_1^2(\bar v_2)) = \max_{v \in S} 2 - 2 v_3 = 2, \\
    \hat\mu_2^1 &\leq \max_{v \in S} \mu(J_{\bar x_1} \bar f_2^1(\bar v_1)) = \max_{v \in S} \max\{3 - 1.9 v_1 + 0.1 v_2,\, 3 + 0.1 v_1 - 1.9 v_2\} = 10/3, \\
    \hat\mu_2^2 &\leq \max_{v \in S} \mu(J_{\bar x_2} \bar f_2^2(\bar v_2)) = \max_{v \in S} -1 - 2 v_3 = -1,
\end{aligned}
\end{equation*}
and
\begin{equation*}
\begin{aligned}
    \hat\mu_1 &\leq \max_{v \in S} \mu(J_x f_1(v)) = \max\{-2/3,\, 2\} = 2, \\
    \hat\mu_2 &\leq \max_{v \in S} \mu(J_x f_2(v)) = \max\{10/3,\, -1\} = 10/3,
\end{aligned}
\end{equation*}
where the matrix measures are computed with the induced $ \infty $-norm using the formula \eqref{eq:mx-meas-inf}.

The upper bounds for the entropies $ h(f_{\sigma_1}, K) $ and $ h(f_{\sigma_2}, K) $ computed using \eqref{eq:sw-ent-upper}, \eqref{eq:sw-ent-upper-sum}, \eqref{eq:sw-ent-upper-max}, \eqref{eq:sw-blk-diag-ent-upper}, \eqref{eq:sw-blk-diag-ent-upper-sub}, \eqref{eq:sw-blk-diag-ent-upper-mode}, \eqref{eq:sw-blk-diag-ent-upper-sum}, and \eqref{eq:sw-blk-diag-ent-upper-max} for all initial sets $ K \subset \dR_{\geq 0}^3 $ are summarized in Table~\ref{tbl:eg-sw-blk-diag} below. In particular, the upper bounds \eqref{eq:sw-ent-upper}, \eqref{eq:sw-blk-diag-ent-upper-sub}, and \eqref{eq:sw-blk-diag-ent-upper-mode} for $ h(f_{\sigma_2}, K) $ are computed along the lines of the computation of \eqref{eq:sw-ent-upper} in Example~\ref{eg:sw}; the upper bound \eqref{eq:sw-blk-diag-ent-upper} is computed along the lines of the computation of $ h(D_{\sigma_2}) $ in \cite[Example~3]{YangSchmidtLiberzonHespanha2020}. The values in Table~\ref{tbl:eg-sw-blk-diag} are consistent with the relations between these upper bounds described in Fig.~\ref{fig:sw-blk-diag-ent} and Remark~\ref{rmk:sw-blk-diag-ent}.
\begin{table}[!ht]
\begin{center}
\begin{minipage}{\textwidth}
\caption{Upper bounds for the entropy of the switched block-diagonal system in Example~\ref{eg:sw-blk-diag}.}\label{tbl:eg-sw-blk-diag}
\begin{tabular}{@{\extracolsep{.4em}}cccccccccc@{\extracolsep{.4em}}}
    \hline%
    & $(\hat\rho_1, \hat\rho_2)$ & \eqref{eq:sw-ent-upper} & \eqref{eq:sw-ent-upper-sum} & \eqref{eq:sw-ent-upper-max} & \eqref{eq:sw-blk-diag-ent-upper} & \eqref{eq:sw-blk-diag-ent-upper-sub} & \eqref{eq:sw-blk-diag-ent-upper-mode} & \eqref{eq:sw-blk-diag-ent-upper-sum} & \eqref{eq:sw-blk-diag-ent-upper-max} \\
    \hline%
    $\sigma_1$ & $(0.5, 0.5)$ & $8$ & $8$ & $10$ & $3.17$ & $3.17$ & $4.34$ & $4.34$ & $6.67$ \\
    $\sigma_2$ & $(0.9, 0.9)$ & $9.6$ & $14.4$ & $10$ & $6.06$ & $7.57$ & $6.2$ & $7.8$ & $6.67$ \\
    \hline%
\end{tabular}
\end{minipage}
\end{center}
\end{table}
\end{eg}

\section{Proofs of main results}\label{sec:proof}
In this section, we present the proofs of Theorems~\ref{thm:sw-ent}, \ref{thm:sw-blk-diag-ent}, and~\ref{thm:sw-blk-diag-ent-tr}, after introducing some necessary technical preliminaries.

\subsection{Bounds for distance between solutions and volume of reachable set}\label{ssec:pre-est}
We construct several upper and lower bounds for the distance between two solutions to the switched system \eqref{eq:sw} and a lower bound for the volume of its reachable set; the proofs can be found in Appendix~\ref{apx:sw-soln-vol}.

\begin{lem}\label{lem:sw-soln-vol}
For all initial states $ x, \bar x \in K $, the corresponding solutions to the switched system \eqref{eq:sw} satisfy
\begin{equation}\label{eq:sw-soln-upper}
    |\xi_\sigma(t, \bar x) - \xi_\sigma(t, x)| \leq e^{\overline\eta_\sigma(t)} |\bar x - x| \qquad \forall t \geq 0
\end{equation}
with
\begin{equation*}
    \overline\eta_\sigma(t) := \max_{v \in \co(K)} \sum_{p \in \Index} \int_{0}^{t} \mu(J_x f_p(\xi_\sigma(s, v))) \indFcn_p(\sigma(s)) \d s.
\end{equation*}
Also, the reachable set of \eqref{eq:sw} satisfies
\begin{equation}\label{eq:sw-vol-lower}
    \vol(\xi_\sigma(t, K)) \geq e^{\gamma_\sigma(t)} \vol(K) \qquad \forall t \geq 0
\end{equation}
with
\begin{equation*}
    \gamma_\sigma(t) := \min_{v \in K} \sum_{p \in \Index} \int_{0}^{t} \tr(J_x f_p(\xi_\sigma(s, v))) \indFcn_p(\sigma(s)) \d s.
\end{equation*}
\end{lem}

Note that $ \overline\eta_\sigma(t) $ and $ \gamma_\sigma(t) $ are defined in terms of integrals of the matrix measure and trace of Jacobian matrix of the active mode over $ [0, t] $, respectively, rewritten via the following observation: For a family of functions $ a_p: \dR_{\geq 0} \to \dR $ with $ p \in \Index $, we have
\begin{equation}\label{eq:sw-int}
	\int_{0}^{t} a_{\sigma(s)}(s) \d s = \sum_{p \in \Index} \int_{0}^{t} a_p(s) \indFcn_p(\sigma(s)) \d s.
\end{equation}%
The upper bound \eqref{eq:sw-soln-upper} extends a similar upper bound in \cite[Th.~1]{Sontag2010} to the case of noncontractive switched systems, and also an upper bound in the proof of \cite[Th.~4.2]{BoichenkoLeonov1998} to the case of switched systems without a compact invariant set. Its proof is inspired by the variational constructions in the proofs of \cite[Th.~1]{Sontag2010} and \cite[Th.~4.2]{BoichenkoLeonov1998}.

The following result provides alternative upper and lower bounds for the distance between two solutions to \eqref{eq:sw}.
\begin{lem}\label{lem:sw-soln-alt}
For all initial states $ x, \bar x \in K $, the corresponding solutions to the switched system \eqref{eq:sw} satisfy
\begin{equation}\label{eq:sw-soln-bnd}
	e^{\underline\eta_\sigma(t)} |\bar x - x| \leq |\xi_\sigma(t, \bar x) - \xi_\sigma(t, x)| \leq e^{\overline\eta_\sigma^\alt(t)} |\bar x - x| \qquad \forall t \geq 0
\end{equation}
with
\begin{equation*}
\begin{aligned}
    \underline\eta_\sigma(t) &:= \sum_{p \in \Index} \int_{0}^{t} \Big( \min_{v \in \co(\xi_\sigma(s, K))} -\mu(-J_x f_p(v)) \Big) \indFcn_p(\sigma(s)) \d s, \\
	\overline\eta_\sigma^\alt(t) &:= \sum_{p \in \Index} \int_{0}^{t} \Big( \max_{v \in \co(\xi_\sigma(s, K))} \mu(J_x f_p(v)) \Big) \indFcn_p(\sigma(s)) \d s.
\end{aligned}
\end{equation*}
\end{lem}

Lemma~\ref{lem:sw-soln-alt} extends similar upper and lower bounds in \cite[Th.~2.5.22, p.\,52]{Vidyasagar2002} by taking the maximum and minimum over only the convex hull of reachable set instead of the entire state space in $ \underline\eta_\sigma(t) $ and $ \overline\eta_\sigma^\alt(t) $, respectively. Its proof is inspired by the variational construction in \cite[Sec.~2.5]{Vidyasagar2002} and is similar to the proof of the upper bound \eqref{eq:sw-soln-upper} in Lemma~\ref{lem:sw-soln-vol}.\footnote{As shown in Appendix~\ref{apx:sw-soln-vol}, their main difference is that, in the proof of Lemma~\ref{lem:sw-soln-alt}, the variational arguments are applied to the line segment connecting two solutions instead of the one connecting two initial states, which results in the different locations of convex hulls in, e.g., $ \overline\eta_\sigma(t) $ and $ \overline\eta_\sigma^\alt(t) $.}

The results in Lemmas~\ref{lem:sw-soln-vol} and~\ref{lem:sw-soln-alt} are compared in Remark~\ref{rmk:sw-soln-vol} below.
\begin{rmk}\label{rmk:sw-soln-vol}
\begin{enumerate}
	\item\label{rmk:sw-soln-upper} The upper bound \eqref{eq:sw-soln-upper} and the one given by the second inequality in \eqref{eq:sw-soln-bnd} are both useful in the sense that neither is less conservative, in part because the relation between the sets $ \xi_\sigma(t, \co(K)) $ and $ \co(\xi_\sigma(t, K)) $ is undetermined for nonlinear systems. However, \eqref{eq:sw-soln-upper} is less conservative or equivalent if the initial set $ K $ is convex, or if all modes of \eqref{eq:sw} are linear, as $ \xi_\sigma(t, \co(K)) \subset \co(\xi_\sigma(t, K)) $ for the former case and $ \xi_\sigma(t, \co(K)) = \co(\xi_\sigma(t, K)) $ for the latter, and the maximum is a subadditive function. All upper bounds for topological entropy in this paper are constructed using \eqref{eq:sw-soln-upper}, and alternative results can be obtained if the second inequality in \eqref{eq:sw-soln-bnd} is used instead, that is, if terms in the form of $ \xi_\sigma(t, \co(K)) $ are replaced with the corresponding ones in the form of $ \co(\xi_\sigma(t, K)) $.
	\item\label{rmk:sw-soln-vol-lower} The first inequality in \eqref{eq:mx-meas-norm-eig} implies that $ \gamma_\sigma(t) $ in \eqref{eq:sw-vol-lower} and $ \underline\eta_\sigma(t) $ in \eqref{eq:sw-soln-bnd} satisfy $ \gamma_\sigma(t) \geq n \underline\eta_\sigma(t) $ for all $ t \geq 0 $. For this reason, volume-based arguments using \eqref{eq:sw-vol-lower} usually lead to less conservative lower bounds for entropy. However, the first inequality in \eqref{eq:sw-soln-bnd} is useful for constructing a possibly less conservative lower bound for entropy in the case of block-diagonal structure in Subsection~\ref{ssec:ent-blk-diag}.
\end{enumerate}
\end{rmk}

\begin{rmk}\label{rmk:local-int-able}
When analyzing the exponential growth rate of the distance between two solutions to a nonlinear system, a measurable upper bound is often used in place of the matrix measure of Jacobian matrix in integrations over time. For example, in \cite[Th.~2.5.22, p.\,52]{Vidyasagar2002}, the author used a measurable function $ \alpha(t) \geq \mu(J_x f(t, x)) $ for all $ t \geq 0 $ and $ x \in \dR^n $, and in \cite[Th.~1]{LuBernardo2016}, the authors used a measurable function $ \alpha(t) \geq \mu_{\chi(t)}(J_x f(x, \sigma(t))) $ for all $ t $ that are not switches and $ x $ in a set $ C $. In this paper, the matrix measures and traces of Jacobian matrices are used directly in integrations over time, as they can be shown to be locally integrable in time under the assumptions in Section~\ref{sec:pre}.
Even for switched systems in which the Jacobian matrices are not locally integrable in time, one can still follow the analysis in this paper with trivial modifications (i.e., replacing the matrix measures and traces of Jacobian matrices with suitable integrable bounds) to obtain similar bounds for exponential growth rates and thus similar bounds for topological entropy.
\end{rmk}

\subsection{Universal spanning and separated sets}\label{ssec:pre-span-sep}
Given a time horizon $ T \geq 0 $ and a radius $ \varepsilon > 0 $, we provide a universal construction of $ (T, \varepsilon) $-spanning and $ (T, \varepsilon) $-separated sets by extending a notion of grid from \cite{YangSchmidtLiberzonHespanha2020}. For a vector $ \theta = (\theta_1, \ldots, \theta_n) \in \dR_{>0}^n $ which may depend on $ T $ and $ \varepsilon $, we define the following grid on the initial set $ K $:
\begin{equation}\label{eq:grid-dfn}
    G(\theta) := \{(k_1 \theta_1, \ldots, k_n \theta_n) \in K: k_1, \ldots, k_n \in \dZ\}.
\end{equation}
As $ K $ is a compact set with nonempty interior, there exist closed hypercubes $ B_1 $ with radius $ r_1 > 0 $ and $ B_2 $ with radius $ r_2 > 0 $ such that $ B_1 \subset K \subset B_2 $. Then the cardinality of the grid $ G(\theta) $ satisfies
\begin{equation*}
	 \prod_{i=1}^{n} \bigg\lfloor \frac{2 r_1}{\theta_i} \bigg\rfloor \leq \#G(\theta) \leq \prod_{i=1}^{n} \bigg( \bigg\lfloor \frac{2 r_2}{\theta_i} \bigg\rfloor + 1 \bigg).
\end{equation*}
For a point $ x = (x_1, \ldots, x_n) \in G(\theta) $, let $ R(x) $ be the open hyperrectangle in $ \dR^n $ with center $ x $ and sides $ 2 \theta_1, \ldots, 2 \theta_n $, that is,
\begin{equation}\label{eq:grid-rect}
    R(x) := \{\bar x_1, \ldots, \bar x_n) \in \dR^n: |\bar x_1 - x_1| < \theta_1, \ldots, |\bar x_n - x_n| < \theta_n\}.
\end{equation}
Then the points in $ G(\theta) $ adjacent to $ x $ are on the boundary of $ R(x) $, and the union of all $ R(x) $ covers $ K $, that is,
\begin{equation*}
    K \subset \bigcup_{x \in G(\theta)} R(x).
\end{equation*}
Comparing the hyperrectangle $ R(x) $ to the open ball $ B_{f_\sigma}(x, \varepsilon, T) $ defined by \eqref{eq:ball-dfn}, we obtain the following result; the proof can be found in Appendix~\ref{apx:grid}.
\begin{lem}\label{lem:grid}
\begin{enumerate}
	\item \label{lem:grid-upper} If the vector $ \theta $ is selected so that
		\begin{equation}\label{eq:grid-span-cond}
			R(x) \subset B_{f_\sigma}(x, \varepsilon, T) \qquad \forall x \in G(\theta),
		\end{equation}
		then the grid $ G(\theta) $ is $ (T, \varepsilon) $-spanning. Moreover, if \eqref{eq:grid-span-cond} holds for all $ T \geq 0 $ and $ \varepsilon > 0 $, and
		\begin{equation}\label{eq:grid-span-conv-cond}
            \limsup_{\varepsilon \searrow 0} \limsup_{T \to \infty} \frac{\log\theta_i}{T} \leq 0 \qquad \forall i \in \{1, \ldots, n\},
        \end{equation}
        then the topological entropy of the switched system \eqref{eq:sw} is upper bounded by
		\begin{equation}\label{eq:grid-span-ent}
            h(f_\sigma, K) \leq \limsup_{\varepsilon \searrow 0} \limsup_{T \to \infty} \sum_{i=1}^{n} \frac{\log(1/\theta_i)}{T}.
        \end{equation}
	\item \label{lem:grid-lower} If the vector $ \theta $ is selected so that
		\begin{equation}\label{eq:grid-sep-cond}
			B_{f_\sigma}(x, \varepsilon, T) \subset R(x) \qquad \forall x \in G(\theta),
		\end{equation}		
		then the grid $ G(\theta) $ is $ (T, \varepsilon) $-separated. Moreover, if \eqref{eq:grid-sep-cond} holds for all $ T \geq 0 $ and $ \varepsilon > 0 $, then the topological entropy of the switched system \eqref{eq:sw} is lower bounded by
		\begin{equation}\label{eq:grid-sep-ent}
            h(f_\sigma, K) \geq \liminf_{\varepsilon \searrow 0} \limsup_{T \to \infty} \sum_{i=1}^{n} \frac{\log(1/\theta_i)}{T}.
        \end{equation}
\end{enumerate}
\end{lem}

Lemma~\ref{lem:grid} extends \cite[Lemma~2]{YangSchmidtLiberzonHespanha2020} to the case of nonlinear modes. Note that \eqref{eq:grid-span-conv-cond} is satisfied if all $ \theta_i $ are nonincreasing in $ T $.\footnote{Also, the lower bound \eqref{eq:grid-sep-ent} does not require a condition similar to \eqref{eq:grid-span-conv-cond}. For this reason, in \cite[Lemma~2.3]{YangLiberzonHespanha2021}, the requirement that all $ \theta_i $ are nonincreasing in $ T $ is not actually needed for the lower bound \cite[eq.~(13)]{YangLiberzonHespanha2021}.}

\subsection{Asymptotic weighted averages}\label{ssec:pre-avg}
Given a switching signal $ \sigma $, consider the active times $ \tau_p(t) $ defined by \eqref{eq:act-time}, active rates $ \rho_p(t) $ defined by \eqref{eq:act-rate}, and sets of persistent and strongly persistent modes $ \IndexInf $ and $ \IndexPos $ defined by \eqref{eq:index-inf-pos}. It was shown in \cite[Lemma~1]{YangSchmidtLiberzonHespanha2020} that, for a family of constants $ c_p \in \dR $ with $ p \in \Index $, we have
\begin{equation}\label{eq:sw-lin-avg-max-sup}
    \limsup_{T \to \infty} \frac{1}{T} \max_{t \in [0, T]} \sum_{p \in \Index} c_p \tau_p(t) = \max \bigg\{ \limsup_{t \to \infty} \sum_{p \in \Index} c_p \rho_p(t),\, 0 \bigg\}.
\end{equation}
In this subsection, we present similar results for a case where the constants are replaced by locally (Lebesgue) integrable functions; the proof can be found in Appendix~\ref{apx:sw-diag-avg-max-sup}.
\begin{lem}\label{lem:sw-diag-avg-max-sup}
Consider a family of locally integrable functions $ a_p^i: \dR_{\geq 0} \to \dR $ with $ i \in \{1, \ldots, m\} $ and $ p \in \Index $.
\begin{enumerate}
	\item Let
		\begin{equation*}
		    \hat a_p^i := \limsup_{t \to \infty:\, \sigma(t) = p} a_p^i(t), \qquad i \in \{1, \ldots, m\},\, p \in \Index.
		\end{equation*}
		If the constants $ \hat a_p^i $ are finite for all $ i \in \{1, \ldots, m\} $ and $ p \in \IndexInf $, then
		\begin{equation}\label{eq:sw-diag-avg-max-sup-upper}
			\limsup_{T \to \infty} \sum_{i=1}^{m} \frac{1}{T} \max_{t \in [0, T]} \sum_{p \in \Index} \int_{0}^{t} a_p^i(s) \indFcn_p(\sigma(s)) \d s \leq \limsup_{T \to \infty} \sum_{i=1}^{m} \frac{1}{T} \max_{t \in [0, T]} \sum_{p \in \IndexPos} \hat a_p^i \tau_p(t).
		\end{equation}
	\item Let
		\begin{equation*}
		    \check a_p^i := \liminf_{t \to \infty:\, \sigma(t) = p} a_p^i(t), \qquad i \in \{1, \ldots, m\},\, p \in \Index.
		\end{equation*}
		If the constants $ \check a_p^i $ are finite for all $ i \in \{1, \ldots, m\} $ and $ p \in \IndexInf $, then
		\begin{equation}\label{eq:sw-diag-avg-max-sup-lower}
			\limsup_{T \to \infty} \sum_{i=1}^{m} \frac{1}{T} \max_{t \in [0, T]} \sum_{p \in \Index} \int_{0}^{t} a_p^i(s) \indFcn_p(\sigma(s)) \d s \geq \limsup_{T \to \infty} \sum_{i=1}^{m} \frac{1}{T} \max_{t \in [0, T]} \sum_{p \in \IndexPos} \check a_p^i \tau_p(t)
		\end{equation}
		and
		\begin{equation}\label{eq:sw-diag-avg-max-sup-lower-alt}
			\limsup_{t \to \infty} \sum_{i=1}^{m} \max \bigg\{ \frac{1}{t} \sum_{p \in \Index} \int_{0}^{t} a_p^i(s) \indFcn_p(\sigma(s)) \d s,\, 0 \bigg\} \geq \limsup_{t \to \infty} \sum_{i=1}^{m} \max \bigg\{ \sum_{p \in \IndexPos} \check a_p^i \rho_p(t),\, 0 \bigg\}.
		\end{equation}
\end{enumerate}
\end{lem}

\begin{rmk}\label{rmk:sw-diag-avg-max-sup}
The inequalities \eqref{eq:sw-diag-avg-max-sup-upper}--\eqref{eq:sw-diag-avg-max-sup-lower-alt} can be extended to some cases where the functions $ a_p^i $ are locally integrable but possibly unbounded. For example, \eqref{eq:sw-diag-avg-max-sup-lower} still holds if for all $ i \in \{1, \ldots, m\} $, we have $ \check a_p^i \in \dR \cup \{+\infty\} $ for all $ p \in \IndexInf\backslash\IndexPos $ and $ \check a_p^i \in \dR $ for all $ p \in \IndexPos $. Moreover, if $ \check a_p^i \in \dR \cup \{+\infty\} $ for all $ i \in \{1, \ldots, m\} $ and $ p \in \IndexInf $, and there exist $ j \in \{1, \ldots, m\} $ and $ q \in \IndexPos $ such that $ \check a_q^j = +\infty $, then
\begin{equation*}
	\limsup_{T \to \infty} \sum_{i=1}^{m} \frac{1}{T} \max_{t \in [0, T]} \sum_{p \in \Index} \int_{0}^{t} a_p^i(s) \indFcn_p(\sigma(s)) \d s = +\infty.
\end{equation*}
The proofs are along the lines of that of Lemma~\ref{lem:sw-diag-avg-max-sup} and thus omitted here. Similar extensions can be made for the inequalities \eqref{eq:sw-diag-avg-max-sup-upper} and \eqref{eq:sw-diag-avg-max-sup-lower-alt}, and consequently all bounds for entropy of switched systems in this paper.
\end{rmk}

For a family of constants $ c_p \in \dR $ with $ p \in \Index $, the equality \eqref{eq:sw-lin-avg-max-sup} can be improved using arguments similar to those in the first two steps of the proof of Lemma~\ref{lem:sw-diag-avg-max-sup}, which leads to the following result:
\begin{equation}\label{eq:sw-lin-avg-max-sup-new}
    \limsup_{T \to \infty} \frac{1}{T} \max_{t \in [0, T]} \sum_{p \in \Index} c_p \tau_p(t) = \limsup_{T \to \infty} \frac{1}{T} \max_{t \in [0, T]} \sum_{p \in \IndexPos} c_p \tau_p(t) = \max \bigg\{ \limsup_{t \to \infty} \sum_{p \in \IndexPos} c_p \rho_p(t),\, 0 \bigg\}.
\end{equation}

\subsection{Proof of Theorem~\ref{thm:sw-blk-diag-ent}}\label{ssec:proof-sw-blk-diag-ent}
We prove only the upper bound \eqref{eq:sw-blk-diag-ent-upper} here; the lower bound \eqref{eq:sw-blk-diag-ent-lower} can be established using similar arguments (see
Appendix~\ref{supp:sw-blk-diag-ent}
for the omitted proof). By applying the upper bound for the distance between two solutions \eqref{eq:sw-soln-upper} to each subsystem of \eqref{eq:sw-blk-diag} with the corresponding ``local'' norm and matrix measure, we obtain that, for all initial states $ x = (x_1, \ldots, x_m), \bar x = (\bar x_1, \ldots, \bar x_m) \in K $ with $ x_i, \bar x_i \in \dR^{n_i} $ for $ i \in \{1, \ldots, m\} $, the corresponding solutions satisfy
\begin{equation*}
    |\xi_\sigma^i(t, \bar x_i) - \xi_\sigma^i(t, x_i)|_{\local i} \leq e^{\overline\eta_\sigma^i(t)} |\bar x_i - x_i|_{\local i} \qquad \forall i \in \{1, \ldots, m\},\, t \geq 0
\end{equation*}
with
\begin{equation*}
    \overline\eta_\sigma^i(t) := \max_{v_i \in \co(K_i)} \sum_{p \in \Index} \int_{0}^{t} \mu_{\local i}(J_{x_i} f_p^i(\xi_\sigma^i(s, v_i)) \indFcn_p(\sigma(s)) \d s.
\end{equation*}
Due to the equivalence of norms on a finite-dimensional vector space, there exist constants $ r_{\local 1}, \ldots, r_{\local m}, r_\Network > 0 $ such that
\begin{equation*}
	|v_i|_{\local i} \leq r_{\local i} |v_i|_\infty \qquad \forall i \in \{1, \ldots, m\},\, v_i \in \dR^{n_i}
\end{equation*}
and
\begin{equation*}
	|v|_\Network \leq r_\Network |v|_\infty \qquad \forall v \in \dR^m.
\end{equation*}
Then the definition of the ``global'' norm \eqref{eq:global-norm} implies that
\begin{equation*}
\begin{aligned}
    |\xi_\sigma(t, \bar x) - \xi_\sigma(t, x)|_\Global &\leq 
    \max_{1 \leq i \leq m} r_\Network |\xi_\sigma^i(t, \bar x_i) - \xi_\sigma^i(t, x_i)|_{\local i} \\
    &\leq \max_{1 \leq i \leq m} e^{\overline\eta_\sigma^i(t)} r_\Network |\bar x_i - x_i|_{\local i} \leq \max_{1 \leq i \leq m} e^{\overline\eta_\sigma^i(t)} r_\Network r_{\local i} |\bar x_i - x_i|_\infty 
\end{aligned}
\end{equation*}
for all $ t \geq 0 $. Consequently, given arbitrary time horizon $ T \geq 0 $ and radius $ \varepsilon > 0 $, we have
\begin{equation}\label{eq:sw-blk-diag-soln-max-upper}
	\max_{t \in [0, T]} |\xi_\sigma(t, \bar x) - \xi_\sigma(t, x)|_\Global \leq \max_{1 \leq i \leq m} e^{\max_{t \in [0, T]} \overline\eta_\sigma^i(t)} r_\Network r_{\local i} |\bar x_i - x_i|_\infty.
\end{equation}
Consider the grid $ G(\theta) $ defined by \eqref{eq:grid-dfn} with the vector $ \theta = (\theta_1 \mathbf{1}_{n_1}, \ldots, \theta_m \mathbf{1}_{n_m}) \in \dR_{> 0}^n $ defined by
\begin{equation*}
    \theta_i := e^{-\max_{t \in [0, T]} \overline\eta_\sigma^i(t)} \varepsilon/(r_\Network r_{\local i}), \qquad i \in \{1, \ldots, m\}.
\end{equation*}
Comparing the corresponding hyperrectangles $ R(x) $ defined by \eqref{eq:grid-rect} to the open balls $ B_{f_\sigma}(x, \varepsilon, T) $ defined by \eqref{eq:ball-dfn} with the ``global'' norm $ |\cdot|_\Global $, we see that \eqref{eq:sw-blk-diag-soln-max-upper} implies $ R(x) \subset B_{f_\sigma}(x, \varepsilon, T) $ for all $ x \in G(\theta) $. Then part~\ref{lem:grid-upper} of Lemma~\ref{lem:grid} implies that $ G(\theta) $ is $ (T, \varepsilon) $-spanning. As $ T \geq 0 $ and $ \varepsilon > 0 $ are arbitrary, and all $ \theta_i $ are nonincreasing in $ T $, the upper bound \eqref{eq:grid-span-ent} implies that
\begin{equation*}
\begin{aligned}
    &h(f_\sigma, K) \leq \limsup_{\varepsilon \searrow 0} \limsup_{T \to \infty} \sum_{i=1}^{m} \frac{n_i \log(1/\theta_i)}{T} \\
    &\qquad = \limsup_{T \to \infty} \sum_{i=1}^{m} \frac{1}{T} \max_{t \in [0, T]} n_i \overline\eta_\sigma^i(t) + \lim_{\varepsilon \searrow 0} \lim_{T \to \infty} \sum_{i=1}^{m} \frac{n_i \log(r_\Network r_{\local i}/\varepsilon)}{T} \\
    &\qquad \leq \limsup_{T \to \infty} \sum_{i=1}^{m} \frac{1}{T} \max_{t \in [0, T]} \sum_{p \in \Index} \int_{0}^{t} \Big( \max_{\bar v_i \in \xi_\sigma^i(s, \co(K_i))} n_i \mu_i(J_{x_i} f_p^i(\bar v_i)) \Big) \indFcn_p(\sigma(s)) \d s,
\end{aligned}
\end{equation*}
where the last inequality holds in part because the maximum is a subadditive function. Finally, we obtain \eqref{eq:sw-blk-diag-ent-upper} by invoking the upper bound \eqref{eq:sw-diag-avg-max-sup-upper} with the functions
\begin{equation*}
\pushQED{\qed}
	a_p^i(t) = \max_{\bar v_i \in \xi_\sigma^i(t, \co(K_i))} n_i \mu_i(J_{x_i} f_p^i(\bar v_i)), \qquad i \in \{1, \ldots, m\},\, p \in \Index. \qedhere
\popQED
\end{equation*}

\subsection{Proof of Theorem~\ref{thm:sw-blk-diag-ent-tr}}\label{ssec:proof-sw-blk-diag-ent-tr}
We prove the lower bound \eqref{eq:sw-blk-diag-ent-lower-tr} using volume-based arguments. As $ K $ is a compact set with nonempty interior, there exists a hyperrectangle $ R = R_1 \times \cdots \times R_m $ with $ R_i \subset K_i $ for $ i \in \{1, \ldots, m\} $. Then $ R \subset K $ and
\begin{equation*}
	\xi_\sigma(t, R) = \xi_\sigma^1(t, R_1) \times \cdots \times \xi_\sigma^m(t, R_m) \qquad \forall t \geq 0
\end{equation*}
for the switched block-diagonal system \eqref{eq:sw-blk-diag}.

Given arbitrary time horizon $ T \geq 0 $ and radius $ \varepsilon > 0 $, by applying the lower bound for the volume of reachable set \eqref{eq:sw-vol-lower} to each subsystem of \eqref{eq:sw-blk-diag}, we obtain that
\begin{equation*}
    \vol(\xi_\sigma^i(T, R_i)) \geq e^{\gamma_\sigma^i(T)} \vol(R_i) \qquad \forall i \in \{1, \ldots, m\}
\end{equation*}
with (recall that $ R_i \subset K_i $)
\begin{equation*}
    \gamma_\sigma^i(t) := \min_{v_i \in K_i} \sum_{p \in \Index} \int_{0}^{t} \tr(J_{x_i} f_p^i(\xi_\sigma^i(s, v_i))) \indFcn_p(\sigma(s)) \d s.
\end{equation*}
Let $ \cI = \{i_1, \ldots, i_l\} := \{i: \gamma_\sigma^i(T) > 0\} $ which depends on $ T $, and define
\begin{equation*}
	\xi_\sigma^\cI(T, R_\cI) := \xi_\sigma^{i_1}(T, R_{i_1}) \times \cdots \times \xi_\sigma^{i_l}(T, R_{i_l}).
\end{equation*}
Then
\begin{align}
	\vol(\xi_\sigma^\cI(T, R_\cI)) &= \prod_{i \in \cI} \vol(\xi_\sigma^i(T, R_i)) \geq e^{\sum_{i \in \cI} \gamma_\sigma^i(T)} \prod_{i \in \cI} \vol(R_i) \notag\\
	&\geq e^{\sum_{i=1}^{m} \max\{\gamma_\sigma^i(T),\, 0\}} \prod_{i=1}^{m} \min\{\vol(R_i),\, 1\}. \label{eq:vol-lower}
\end{align}

Let $ E $ be a minimal $ (T, \varepsilon) $-spanning set defined by \eqref{eq:ball-dfn} and \eqref{eq:span-dfn} with the ``global'' norm $ |\cdot|_\Global $. Due to the equivalence of norms on a finite-dimensional vector space, there exist constants $ \bar r_{\local 1}, \ldots, \bar r_{\local m}, \bar r_\Network > 0 $ such that
\begin{equation*}
	|v_i|_\infty \leq \bar r_{\local i} |v_i|_{\local i} \qquad \forall i \in \{1, \ldots, m\},\, v_i \in \dR^{n_i}
\end{equation*}
and
\begin{equation*}
	|v|_\infty \leq \bar r_\Network |v|_\Network \qquad \forall v \in \dR^m.
\end{equation*}
Then for all initial states $ x = (x_1, \ldots, x_m), \bar x = (\bar x_1, \ldots, \bar x_m) \in\dR^n $ with $ x_i, \bar x_i \in \dR^{n_i} $ for $ i \in \{1, \ldots, m\} $, the definition of the ``global'' norm \eqref{eq:global-norm} implies that
\begin{equation*}
\begin{aligned}
    |\xi_\sigma(T, \bar x) - \xi_\sigma(T, x)|_\Global &\geq \max_{1 \leq i \leq m} |\xi_\sigma^i(T, \bar x_i) - \xi_\sigma^i(T, x_i)|_{\local i}/\bar r_\Network \\
    &\geq \max_{1 \leq i \leq m} |\xi_\sigma^i(T, \bar x_i) - \xi_\sigma^i(T, x_i)|_\infty/(\bar r_\Network \bar r_{\local i}).
\end{aligned}
\end{equation*}
Hence for the set $ \cI = \{i_1, \ldots, i_l\} $, we have (recall that $ R \subset K $)
\begin{equation*}
\begin{aligned}
    \xi_\sigma^\cI(T, R_\cI) &\subset \{(\xi_\sigma^{i_1}(T, \bar x_{i_1}), \ldots, \xi_\sigma^{i_l}(T, \bar x_{i_l})): (\bar x_1, \ldots, \bar x_m) \in K\} \\
    &\subset \bigcup_{x \in E} \{(\xi_\sigma^{i_1}(T, \bar x_{i_1}), \ldots, \xi_\sigma^{i_l}(T, \bar x_{i_l})): |\xi_\sigma(T, (\bar x_1, \ldots, \bar x_m)) - \xi_\sigma(T, x)|_\Global < \varepsilon\} \\
    &\subset \bigcup_{(x_1, \ldots, x_m) \in E} \{(v_{i_1}, \ldots, v_{i_l}): |v_i - \xi_\sigma^i(T, x_i)|_\infty < \bar r_\Network \bar r_{\local i} \varepsilon \text{ for all } i \in \cI\},
\end{aligned}
\end{equation*}
and thus
\begin{equation}\label{eq:vol-upper}
    \vol(\xi_\sigma^\cI(T, R_\cI)) \leq \#E \prod_{i \in \cI} (2 \bar r_\Network \bar r_{\local i} \varepsilon)^{n_i} \leq \#E \prod_{i=1}^{m} \max\{(2 \bar r_\Network \bar r_{\local i} \varepsilon)^{n_i},\, 1\}.
\end{equation}

Combining \eqref{eq:vol-lower} and \eqref{eq:vol-upper}, we obtain that the minimal cardinality of a $ (T, \varepsilon) $-spanning set satisfies
\begin{equation*}
\begin{aligned}
    S(f_\sigma, \varepsilon, T, K) = \#E \geq e^{\sum_{i=1}^{m} \max\{\gamma_\sigma^j(T),\, 0\}} \prod_{i=1}^{m} \frac{\min\{\vol(R_i),\, 1\}}{\max\{(2 \bar r_\Network \bar r_{\local i} \varepsilon)^{n_i},\, 1\}}.
\end{aligned}
\end{equation*}
Hence the definition of entropy \eqref{eq:ent-dfn} implies that
\begin{equation*}
\begin{aligned}
    h(f_\sigma, K) &\geq \liminf_{\varepsilon \searrow 0} \limsup_{T \to \infty} \frac{1}{T} \log \bigg( e^{\sum_{i=1}^{m} \max\{\gamma_\sigma^i(T),\, 0\}} \prod_{i=1}^{m} \frac{\min\{\vol(R_i),\, 1\}}{\max\{(2 r_\Network r_{\local i} \varepsilon)^{n_i},\, 1\}} \bigg) \\
    &= \limsup_{T \to \infty} \sum_{i=1}^{m} \max \bigg\{ \frac{\gamma_\sigma^i(T)}{T},\, 0 \bigg\} + \lim_{\varepsilon \searrow 0} \lim_{T \to \infty} \frac{1}{T} \sum_{i=1}^{m} \log \bigg( \frac{\min\{\vol(R_i),\, 1\}}{\max\{(2 r_\Network r_{\local i} \varepsilon)^{n_i},\, 1\}} \bigg) \\
    &\geq \limsup_{T \to \infty} \sum_{i=1}^{m} \max \bigg\{ \frac{1}{T} \sum_{p \in \Index} \int_{0}^{T} \Big( \min_{\bar v_i \in \xi_\sigma^i(s, K_i)} \tr(J_{x_i} f_p^i(\bar v_i)) \Big) \indFcn_p(\sigma(s)) \d s,\, 0 \bigg\},
\end{aligned}
\end{equation*}
where the last inequality holds in part because the minimum is a superadditive function. Finally, we obtain \eqref{eq:sw-blk-diag-ent-lower-tr} by invoking the lower bound \eqref{eq:sw-diag-avg-max-sup-lower-alt} with the functions
\begin{equation*}
\pushQED{\qed}
	a_p^i(t) = \min_{\bar v_i \in \xi_\sigma^i(t, K_i)} \tr(J_{x_i} f_p^i(\bar v_i)), \qquad i \in \{1, \ldots, m\},\, p \in \Index. \qedhere
\popQED
\end{equation*}

\subsection{Proof of Theorem~\ref{thm:sw-ent}}\label{ssec:proof-sw-ent}
Note that the switched system \eqref{eq:sw} can be seen as a switched block-diagonal system \eqref{eq:sw-blk-diag} with a single block. Consequently, the upper bound \eqref{eq:sw-ent-upper} and lower bound \eqref{eq:sw-ent-lower} for the entropy of \eqref{eq:sw} can be obtained by taking $ m = 1 $ and $ n_1 = n $ in the upper bound \eqref{eq:sw-blk-diag-ent-upper-sub} and lower bound \eqref{eq:sw-blk-diag-ent-lower-tr} for the entropy of \eqref{eq:sw-blk-diag}, respectively. For a complete proof of slightly weaker results (without considering persistent and strongly persistent modes), see the proof of \cite[Th.~3.1]{YangLiberzonHespanha2021}. \qed

\section{Conclusion}\label{sec:end}
Upper and lower bounds for topological entropy of switched nonlinear and interconnected systems were constructed, which generalized and extended previous results on switched linear systems and advanced our understanding of the effects of switching on topological entropy. A feature of these bounds is that they depend on the values of Jacobian matrices of only strongly persistent modes and over only an $ \omega $-limit set. As a result, they are easier to compute and are able to yield finite bounds in the case of unbounded Jacobian matrices but a compact global attractor.

Future research directions include analyzing the computation complexities of the bounds for topological entropy in this paper, investigating the relations between these bounds and stability conditions for switched nonlinear and interconnected systems, and constructing bounds for topological entropy of switched nonlinear systems with triangular structure and general time-varying systems.

\appendix[Proofs of technical lemmas]
\subsection{Proofs of Lemmas~\ref{lem:sw-soln-vol} and~\ref{lem:sw-soln-alt}}\label{apx:sw-soln-vol}
Lemmas~\ref{lem:sw-soln-vol} and~\ref{lem:sw-soln-alt} are established based on similar properties of a linear time-varying (LTV) system
\begin{equation}\label{eq:ltv}
    \dot x = A(t)\, x
\end{equation}
with a piecewise-continuous, matrix-valued function $ A: \dR_{\geq 0} \to \dR^{n \times n} $. The solution to \eqref{eq:ltv} at time $ t \geq 0 $ with initial state $ v $ is given by $ \xi(t, v) = \Phi_A(t, 0)\, v $, where $ \Phi_A(t, 0) $ is the state-transition matrix and satisfies Coppel's inequalities \cite[Th.~2.8.27, p.\,34]{DesoerVidyasagar2009}
\begin{equation}\label{eq:ltv-tmx-norm}
    e^{\int_{0}^{t} -\mu(-A(s)) \d s} |v| \leq |\Phi_A(t, 0)\, v| \leq e^{\int_{0}^{t} \mu(A(s)) \d s} |v| \qquad \forall t \geq 0,\, v \in \dR^n
\end{equation}
and Liouville's formula \cite[Th.~4.1, p.\,28]{Brockett1970}
\begin{equation}\label{eq:ltv-tmx-det}
    \det(\Phi_A(t, 0)) = e^{\int_{0}^{t} \tr(A(s)) \d s} \qquad \forall t \geq 0.
\end{equation}

\begin{proof}[Proof of Lemma~\ref{lem:sw-soln-vol}]
We prove Lemma~\ref{lem:sw-soln-vol} by writing the Jacobian matrix of a solution to the switched nonlinear system \eqref{eq:sw} with respect to initial state $ J_x \xi_\sigma(t, x) $ as a matrix solution to the LTV system \eqref{eq:ltv} with a suitable function $ A(t) $, using variational arguments from nonlinear systems analysis (see, e.g., \cite[Sec.~4.2.4]{Liberzon2012}). For all $ v \in \dR^n $, we have $ J_x \xi_\sigma(0, v) = I $ and
\begin{equation*}
    \partial_t J_x \xi_\sigma(t, v) = J_x \dot\xi_\sigma(t, v) = J_x f_{\sigma(t)}(\xi_\sigma(t, v))\, J_x \xi_\sigma(t, v)
\end{equation*}
for all $ t \geq 0 $ that are not switches. Hence for each fixed $ v \in \dR^n $, the matrix $ J_x \xi_\sigma(t, v) $ is equal to the state-transition matrix $ \Phi_A(t, 0) $ of \eqref{eq:ltv} with $ A(t) = J_x f_{\sigma(t)}(\xi_\sigma(t, v)) $.

First, given arbitrary initial states $ x, \bar x \in K $, let
\begin{equation*}
	\nu(\rho) := \rho \bar x + (1 - \rho)\, x, \qquad \rho \in [0, 1].
\end{equation*}
Then $ \nu(\rho) \in \co(K) $ and $ \nu'(\rho) = \bar x - x $ for all $ \rho \in [0, 1] $. Hence
\begin{equation*}
\begin{aligned}
    &|\xi_\sigma(t, \bar x) - \xi_\sigma(t, x)| = |\xi_\sigma(t, \nu(1)) - \xi_\sigma(t, \nu(0))| = \bigg| \int_{0}^{1} J_x \xi_\sigma(t, \nu(\rho)) (\bar x - x) \d\rho \bigg| \\
    &\qquad \leq \int_{0}^{1} |J_x \xi_\sigma(t, \nu(\rho)) (\bar x - x)| \d\rho \leq \int_{0}^{1} e^{\int_{0}^{t} \mu(J_x f_{\sigma(s)}(\xi_\sigma(s, \nu(\rho)))) \d s} |\bar x - x| \d\rho \\
    &\qquad \leq \Big( \max_{v \in \co(K)} e^{\int_{0}^{t} \mu(J_x f_{\sigma(s)}(\xi_\sigma(s, v))) \d s} \Big) |\bar x - x| = e^{\overline\eta_\sigma(t)} |\bar x - x|
\end{aligned}
\end{equation*}
for all $ t \geq 0 $, where the second inequality follows from the second inequality in \eqref{eq:ltv-tmx-norm} with $ A(t) = J_x f_{\sigma(t)}(\xi_\sigma(t, \nu(\rho))) $ and $ \Phi_A(t, 0) = J_x \xi_\sigma(t, \nu(\rho)) $, and the last equality follows from the transformation \eqref{eq:sw-int}. Hence \eqref{eq:sw-soln-upper} holds.

Second, for each fixed $ v \in \dR^n $, Liouville's formula \eqref{eq:ltv-tmx-det} with $ A(t) = J_x f_{\sigma(t)}(\xi_\sigma(t, v)) $ and $ \Phi_A(t, 0) = J_x \xi_\sigma(t, v) $ implies that
\begin{equation*}
    \det(J_x \xi_\sigma(t, v)) = e^{\int_{0}^{t} \tr(J_x f_{\sigma(s)}(\xi_\sigma(s, v))) \d s} \qquad \forall t \geq 0.
\end{equation*}
Then
\begin{equation*}
\begin{aligned}
    \vol(\xi_\sigma(t, K)) &= \int_{K} |\det(J_x \xi_\sigma(t, v))| \d v \geq \Big( \min_{v \in K} |\det(J_x \xi_\sigma(t, v))| \Big) \vol(K) \\
    &= \Big( \min_{v \in K} e^{\int_{0}^{t} \tr(J_x f_{\sigma(s)}(\xi_\sigma(s, v))) \d s} \Big) \vol(K) = e^{\gamma_\sigma(t)} \vol(K)
\end{aligned}
\end{equation*}
for all $ t \geq 0 $, where the last equality follows from the transformation \eqref{eq:sw-int}. Hence \eqref{eq:sw-vol-lower} holds.
\end{proof}

\begin{proof}[Proof of Lemma~\ref{lem:sw-soln-alt}]
We prove Lemma~\ref{lem:sw-soln-alt} by writing the difference between two solutions to the switched nonlinear system \eqref{eq:sw} as a solution to the LTV system \eqref{eq:ltv} with a suitable function $ A(t) $, using variational arguments similar to those in the proof of Lemma~\ref{lem:sw-soln-vol}. Given arbitrary initial states $ x, \bar x \in K $, let
\begin{equation*}
	\nu(t, \rho) := \rho \xi_\sigma(t, \bar x) + (1 - \rho) \xi_\sigma(t, x), \qquad \rho \in [0, 1],\, t \geq 0.
\end{equation*}
Then $ \nu(t, \rho) \in \co(\xi_\sigma(t, K)) $ and $ \partial_\rho \nu(t, \rho) = \xi_\sigma(t, \bar x) - \xi_\sigma(t, x) $ for all $ \rho \in [0, 1] $ and $ t \geq 0 $. Hence
\begin{equation*}
\begin{aligned}
    \partial_t (\xi_\sigma(t, \bar x) - \xi_\sigma(t, x)) &= f_{\sigma(t)}(\xi_\sigma(t, \bar x)) - f_{\sigma(t)}(\xi_\sigma(t, x)) \\
    &= f_{\sigma(t)}(\nu(t, 1)) - f_{\sigma(t)}(\nu(t, 0)) \\
    &= \bigg( \int_{0}^{1} J_x f_{\sigma(t)}(\nu(t, \rho)) \d\rho \bigg) (\xi_\sigma(t, \bar x) - \xi_\sigma(t, x))
\end{aligned}
\end{equation*}
for all $ t \geq 0 $ that are not switches. Therefore, $ \xi_\sigma(t, \bar x) - \xi_\sigma(t, x) $ is the solution to \eqref{eq:ltv} with $ A(t) = \int_{0}^{1} J_x f_{\sigma(t)}(\nu(t, \rho)) \d\rho $ at time $ t $ with initial state $ \bar x - x $. Consequently, \eqref{eq:ltv-tmx-norm} implies that
\begin{equation*}
\begin{aligned}
	e^{\int_{0}^{t} -\mu \big( -\int_{0}^{1} J_x f_{\sigma(s)}(\nu(s, \rho)) \d\rho \big) \d s} |\bar x - x| &\leq |\xi_\sigma(t, \bar x) - \xi_\sigma(t, x)| \\
	&\leq e^{\int_{0}^{t} \mu \big( \int_{0}^{1} J_x f_{\sigma(s)}(\nu(s, \rho)) \d\rho \big) \d s} |\bar x - x|
\end{aligned}
\end{equation*}
for all $ t \geq 0 $. Moreover, as the matrix measure $ \mu $ is a convex function, Jensen's inequality \cite[Th.~11.24, p.\,417]{AliprantisBorder2006} implies that
\begin{equation*}
\begin{aligned}
    \mu \bigg( \int_{0}^{1} J_x f_{\sigma(t)}(\nu(t, \rho)) \d\rho \bigg) &\leq \int_{0}^{1} \mu(J_x f_{\sigma(t)}(\nu(t, \rho)) \d\rho \\
    &\leq \max_{v \in \co(\xi_\sigma(t, K))} \mu(J_x f_{\sigma(t)}(v))
\end{aligned}
\end{equation*}
and
\begin{equation*}
\begin{aligned}
    -\mu \bigg( {-\int_{0}^{1}} J_x f_{\sigma(t)}(\nu(t, \rho)) \d\rho \bigg) &\geq \int_{0}^{1} -\mu(-J_x f_{\sigma(t)}(\nu(t, \rho)) \d\rho \\
    &\geq \min_{v \in \co(\xi_\sigma(t, K))} -\mu(-J_x f_{\sigma(t)}(v))
\end{aligned}
\end{equation*}
for all $ t \geq 0 $. Then \eqref{eq:sw-soln-bnd} follows from the transformation \eqref{eq:sw-int}.
\end{proof}

\subsection{Proof of Lemma~\ref{lem:grid}}\label{apx:grid}
\begin{enumerate}[wide]
\item If \eqref{eq:grid-span-cond} holds, then
\begin{equation*}
    K \subset \bigcup_{x \in G(\theta)} R(x) \subset \bigcup_{x \in G(\theta)} B_{f_\sigma}(x, \varepsilon, T).
\end{equation*}
Hence $ G(\theta) $ is $ (T, \varepsilon) $-spanning following the definition \eqref{eq:span-dfn}, and thus
\begin{equation*}
    S(f_\sigma, \varepsilon, T, K) \leq \#G(\theta) \leq \prod_{i=1}^{n} \bigg( \bigg\lfloor \frac{2 r_2}{\theta_i} \bigg\rfloor + 1 \bigg) \leq \prod_{i=1}^{n} \bigg( \frac{2 r_2}{\theta_i} + 1 \bigg).
\end{equation*}
If \eqref{eq:grid-span-cond} holds for all $ T \geq 0 $ and $ \varepsilon > 0 $, then the definition of entropy \eqref{eq:ent-dfn} implies that
\begin{align}
    h(f_\sigma, K) &\leq \limsup_{\varepsilon \searrow 0} \limsup_{T \to \infty} \sum_{i=1}^{n} \frac{\log(2 r_2/\theta_i + 1)}{T} \notag\\
    &\leq \limsup_{\varepsilon \searrow 0} \limsup_{T \to \infty} \sum_{i=1}^{n} \frac{\log(1/\theta_i)}{T} + \sum_{i=1}^{n} \limsup_{\varepsilon \searrow 0} \limsup_{T \to \infty} \frac{\log(\theta_i + 2 r_2)}{T}, \label{eq:grid-span-ent-temp}
\end{align}
where the last inequality holds as the upper limit is a subadditive function. Moreover, for all $ i \in \{1, \ldots, n\} $, the summands in the last term in \eqref{eq:grid-span-ent-temp} satisfy
\begin{equation*}
\begin{aligned}
    &\limsup_{\varepsilon \searrow 0} \limsup_{T \to \infty} \frac{\log(\theta_i + 2 r_2)}{T} \leq \limsup_{\varepsilon \searrow 0} \limsup_{T \to \infty} \frac{\max\{\log(2\theta_i),\, \log(4 r_2)\}}{T} \\
    &\qquad = \max \bigg\{ \limsup_{\varepsilon \searrow 0} \limsup_{T \to \infty} \frac{\log(2 \theta_i)}{T},\, \lim_{T \to \infty} \frac{\log(4 r_2)}{T} \bigg\} \\
    &\qquad = \max \bigg\{ \limsup_{\varepsilon \searrow 0} \limsup_{T \to \infty} \frac{\log\theta_i}{T},\, 0 \bigg\},
\end{aligned}
\end{equation*}
where the inequality holds as the logarithm is an increasing function and $ \theta_i, r_2 > 0 $, and the last equality holds in part because $ r_2 $ is independent of $ T $.
Then \eqref{eq:grid-span-conv-cond} implies \eqref{eq:grid-span-ent}.\footnote{There was a typo in \cite[Lemma~2]{YangSchmidtLiberzonHespanha2020}, where \cite[eq.~(17)]{YangSchmidtLiberzonHespanha2020} should be replaced with \eqref{eq:grid-span-conv-cond}, as the inequality in \eqref{eq:grid-span-conv-cond} needs to hold for each $ i \in \{1, \ldots, n\} $ instead of the upper limits of the sum.}
\item If \eqref{eq:grid-sep-cond} holds, then for all distinct points $ x, \bar x \in G(\theta) $, we have $ \bar x \notin B_{f_\sigma}(x, \varepsilon, T) $ as $ \bar x \notin R(x) $. Hence $ G(\theta) $ is $ (T, \varepsilon) $-separated following the definition \eqref{eq:sep-dfn}, and thus
\begin{equation*}
    N(f_\sigma, \varepsilon, T, K) \geq \#G(\theta) \geq \prod_{i=1}^{n} \bigg\lfloor \frac{2 r_1}{\theta_i} \bigg\rfloor \geq \prod_{i=1}^{n} \max \bigg\{ \frac{2 r_1}{\theta_i} - 1,\, 1 \bigg\}.
\end{equation*}
If \eqref{eq:grid-sep-cond} holds for all $ T \geq 0 $ and $ \varepsilon > 0 $, then the alternative definition of entropy \eqref{eq:ent-dfn-alt} implies that
\begin{align}
    h(f_\sigma, K) &\geq \liminf_{\varepsilon \searrow 0} \limsup_{T \to \infty} \sum_{i=1}^{n} \frac{\log(\max\{2 r_1/\theta_i - 1,\, 1\})}{T} \notag\\
    &\geq \liminf_{\varepsilon \searrow 0} \limsup_{T \to \infty} \sum_{i=1}^{n} \frac{\log(1/\theta_i)}{T} \notag\\
    &\quad\, + \sum_{i=1}^{n} \liminf_{\varepsilon \searrow 0} \liminf_{T \to \infty} \frac{\log(\max\{2 r_1 - \theta_i,\, \theta_i\})}{T}, \label{eq:grid-sep-ent-temp}
\end{align}
where the last inequality holds as the lower limit is a superadditive function, and for two functions $ g, \bar g: \dR_{\geq 0} \to \dR $, we have
\begin{equation*}
	\limsup_{T \to \infty} (g(T) + \bar g(T)) \geq \limsup_{T \to \infty} g(T) + \liminf_{T \to \infty} \bar g(T).
\end{equation*}
Moreover, for all $ i \in \{1, \ldots, n\} $, the summands in the last term in \eqref{eq:grid-sep-ent-temp} satisfy
\begin{equation*}
\begin{aligned}
    \liminf_{\varepsilon \searrow 0} \liminf_{T \to \infty} \frac{\log(\max\{2 r_1 - \theta_i,\, \theta_i\})}{T} &\geq \liminf_{\varepsilon \searrow 0} \liminf_{T \to \infty} \frac{\log(\max\{r_1,\, \theta_i\})}{T} \\
    &\geq \liminf_{\varepsilon \searrow 0} \liminf_{T \to \infty} \frac{\log r_1}{T} = 0,
\end{aligned}
\end{equation*}
where the first inequality holds as the logarithm is an increasing function and $ r_1, \theta_i > 0 $, and the equality holds as $ r_1 $ is independent of $ T $. Hence \eqref{eq:grid-sep-ent} holds. \qed
\end{enumerate}

\subsection{Proof of Lemma~\ref{lem:sw-diag-avg-max-sup}}\label{apx:sw-diag-avg-max-sup}
We prove only the lower bound \eqref{eq:sw-diag-avg-max-sup-lower} here; the upper bound \eqref{eq:sw-diag-avg-max-sup-upper} and lower bound \eqref{eq:sw-diag-avg-max-sup-lower-alt} can be established using similar arguments (see
Appendix~\ref{supp:sw-diag-avg-max-sup}
for the omitted proof). For brevity, we define the following functions on $ \dR_{\geq 0} $:
\begin{equation*}
	\begin{aligned}
		\eta_p^i(t) &:= \int_{0}^{t} a_p^i(s) \indFcn_p(\sigma(s)) \d s, \quad p \in \Index, \\
		\bar a^i(T) &:= \frac{1}{T} \max_{t \in [0, T]} \sum_{p \in \Index} \eta_p^i(t),
	\end{aligned} \qquad i \in \{1, \ldots, m\}
\end{equation*}
with $ \bar a^i(0) := \max\{a_{\sigma(0)}^i(0),\, 0\} $. Also, note that for two continuous functions $ g, \bar g: \dR_{\geq 0} \to \dR $, we have
\begin{equation}\label{eq:max-sum-lower}
	\max_{t \in [0, T]} (g(t) + \bar g(t)) \geq \max_{t \in [0, T]} g(t) + \min_{t \in [0, T]} \bar g(t) \qquad \forall T \geq 0.
\end{equation}

First, we eliminate modes that are not persistent. The definition of $ \IndexInf $ in \eqref{eq:index-inf-pos} implies that $ t_\infty := \sup\{t \geq 0: \sigma(t) \in \Index\backslash\IndexInf\} $ is finite. Then
\begin{equation*}
\begin{aligned}
	\bar a^i(T) &\geq \frac{1}{T} \bigg( \max_{t \in [0, T]} \sum_{p \in \IndexInf} \eta_p^i(t) + \min_{t \in [0, T]} \sum_{p \in \Index\backslash\IndexInf} \eta_p^i(t) \bigg) \\
	&= \frac{1}{T} \bigg( \max_{t \in [0, T]} \sum_{p \in \IndexInf} \eta_p^i(t) + \min_{t \in [0, t_\infty]} \sum_{p \in \Index\backslash\IndexInf} \eta_p^i(t) \bigg)
\end{aligned}
\end{equation*}
for all $ i \in \{1, \ldots, m\} $ and $ T > t_\infty  $, where the inequality follows from \eqref{eq:max-sum-lower}. Hence
\begin{equation*}
	\limsup_{T \to \infty} \sum_{i=1}^{m} \bar a^i(T) \geq \limsup_{T \to \infty} \sum_{i=1}^{m} \frac{1}{T} \max_{t \in [0, T]} \sum_{p \in \IndexInf} \eta_p^i(t).
\end{equation*}

Second, we eliminate modes that are persistent but not strongly persistent. For all $ i \in \{1, \ldots, m\} $ and $ p \in \IndexInf $, as $ \check a_p^i = \liminf_{t \to \infty:\, \sigma(t) = p} a_p^i(t) $ are finite, $ \check\alpha_p^i := \inf_{t \geq 0} a_p^i(t) \indFcn_p(\sigma(t)) $ are finite as well. Then
\begin{equation*}
\begin{aligned}
	\frac{1}{T} \max_{t \in [0, T]} \sum_{p \in \IndexInf} \eta_p^i(t) &\geq \frac{1}{T} \max_{t \in [0, T]} \sum_{p \in \IndexPos} \eta_p^i(t) + \frac{1}{T} \min_{t \in [0, T]} \sum_{p \in \IndexInf\backslash\IndexPos} \eta_p^i(t) \\
	&\geq \frac{1}{T} \max_{t \in [0, T]} \sum_{p \in \IndexPos} \eta_p^i(t) + \sum_{p \in \IndexInf\backslash\IndexPos} \min\{\check\alpha_p^i,\, 0\} \rho_p(T)
\end{aligned}
\end{equation*}
for all $ i \in \{1, \ldots, m\} $ and $ T > 0 $, where the first inequality follows from \eqref{eq:max-sum-lower}. Moreover, for all $ p \notin \IndexPos $, as $ \rho_p(t) \geq 0 $ for all $ t \geq 0 $ and $ \hat\rho_p = \limsup_{t \to \infty} \rho_p(t) = 0 $, we have $ \hat\rho_p = \lim_{t \to \infty} \rho_p(t) = 0 $. Then
\begin{equation*}
	\lim_{T \to \infty} \sum_{i=1}^{m} \sum_{p \in \IndexInf\backslash\IndexPos} \min\{\check\alpha_p^i,\, 0\} \rho_p(T) = \sum_{i=1}^{m} \sum_{p \in \IndexInf\backslash\IndexPos} \min\{\check\alpha_p^i,\, 0\} \hat\rho_p = 0.
\end{equation*}
Hence
\begin{equation*}
	\limsup_{T \to \infty} \sum_{i=1}^{m} \bar a^i(T) \geq \limsup_{T \to \infty} \sum_{i=1}^{m} \frac{1}{T} \max_{t \in [0, T]} \sum_{p \in \IndexPos} \eta_p^i(t).
\end{equation*}

Finally, as $ \IndexPos $ is a finite set, the lower limit in the definition of $ \check a_p^i $ implies that, for each $ \delta > 0 $, there exists a large enough $ t_\delta \geq 0 $ such that
\begin{equation*}
	a_p^i(t) > \check a_p^i - \delta \qquad \forall i \in \{1, \ldots, m\},\, p \in \IndexPos,\, t > t_\delta: \sigma(t) = p.
\end{equation*}
For all $ i \in \{1, \ldots, m\} $ and $ t \geq 0 $, consider
\begin{equation*}
\begin{aligned}
	\sum_{p \in \IndexPos} \eta_p^i(t) &= \sum_{p \in \IndexPos} \int_{0}^{t} (a_p^i(s) - \check a_p^i + \delta + \check a_p^i - \delta) \indFcn_p(\sigma(s)) \d s \\
	&= \sum_{p \in \IndexPos} \check a_p^i \tau_p(t) - \delta t + \sum_{p \in \IndexPos} \int_{0}^{t} (a_p^i(s) - \check a_p^i + \delta) \indFcn_p(\sigma(s)) \d s.
\end{aligned}
\end{equation*}
If $ t > t_\delta $, then
\begin{equation*}
	\sum_{p \in \IndexPos} \int_{0}^{t} (a_p^i(s) - \check a_p^i + \delta) \indFcn_p(\sigma(s)) \d s \geq \sum_{p \in \IndexPos} \int_{0}^{t_\delta} (a_p^i(s) - \check a_p^i + \delta) \indFcn_p(\sigma(s)) \d s.
\end{equation*}
Otherwise $ t \in [0, t_\delta] $, and hence
\begin{equation*}
	\sum_{p \in \IndexPos} \int_{0}^{t} (a_p^i(s) - \check a_p^i + \delta) \indFcn_p(\sigma(s)) \d s \geq \min_{\bar t \in [0, t_\delta]} \sum_{p \in \IndexPos} \int_{0}^{\bar t} (a_p^i(s) - \check a_p^i + \delta) \indFcn_p(\sigma(s)) \d s.
\end{equation*}
Combining the two cases, we obtain
\begin{equation*}
	\sum_{p \in \IndexPos} \eta_p^i(t) \geq \sum_{p \in \IndexPos} \check a_p^i \tau_p(t) - \delta t + \min_{\bar t \in [0, t_\delta]} \sum_{p \in \IndexPos} \int_{0}^{\bar t} (a_p^i(s) - \check a_p^i + \delta) \indFcn_p(\sigma(s)) \d s.
\end{equation*}
Then
\begin{equation*}
	\frac{1}{T} \max_{t \in [0, T]} \sum_{p \in \IndexPos} \eta_p^i(t) \geq \frac{1}{T} \max_{t \in [0, T]} \sum_{p \in \IndexPos} \check a_p^i \tau_p(t) - \delta + \frac{1}{T} \min_{\bar t \in [0, t_\delta]} \sum_{p \in \IndexPos} \int_{0}^{\bar t} (a_p^i(s) - \check a_p^i + \delta) \indFcn_p(\sigma(s)) \d s
\end{equation*}
for all $ T > 0 $, where the inequality follows in part from \eqref{eq:max-sum-lower}. Hence
\begin{equation*}
	\limsup_{T \to \infty} \sum_{i=1}^{m} \bar a_i(T) \geq \limsup_{T \to \infty} \sum_{i=1}^{m} \frac{1}{T} \max_{t \in [0, T]} \sum_{p \in \IndexPos} \check a_p^i \tau_p(t) - m \delta.
\end{equation*}
Then \eqref{eq:sw-diag-avg-max-sup-lower} holds as $ \delta > 0 $ is arbitrary. \qed

\subsection{Omitted proof of Theorem~\ref{thm:sw-blk-diag-ent}}\label{supp:sw-blk-diag-ent}
Here we prove the lower bound \eqref{eq:sw-blk-diag-ent-lower}. By applying the lower bound for the distance between two solutions in \eqref{eq:sw-soln-bnd} to each subsystem of \eqref{eq:sw-blk-diag} with the corresponding ``local'' norm and matrix measure, we obtain that, for all initial states $ x = (x_1, \ldots, x_m), \bar x = (\bar x_1, \ldots, \bar x_m) \in K $ with $ x_i, \bar x_i \in \dR^{n_i} $ for $ i \in \{1, \ldots, m\} $, the corresponding solutions satisfy
\begin{equation*}
    |\xi_\sigma^i(t, \bar x_i) - \xi_\sigma^i(t, x_i)|_{\local i} \geq e^{\underline\eta_\sigma^i(t)} |\bar x_i - x_i|_{\local i} \qquad \forall i \in \{1, \ldots, m\},\, t \geq 0
\end{equation*}
with
\begin{equation*}
    \underline\eta_\sigma^i(t) := \sum_{p \in \Index} \int_{0}^{t} \Big( \min_{v_i \in \co(\xi_\sigma^i(s, K_i))} -\mu_{\local i}(-J_{x_i} f_p^i(v_i) \Big) \indFcn_p(\sigma(s)) \d s.
\end{equation*}
Due to the equivalence of norms on a finite-dimensional vector space, there exist constants $ \tilde r_{\local 1}, \ldots, \tilde r_{\local m}, \tilde r_\Network > 0 $ such that
\begin{equation*}
	|v_i|_{\local i} \geq \tilde r_{\local i} |v_i|_\infty \qquad \forall i \in \{1, \ldots, m\},\, v_i \in \dR^{n_i}
\end{equation*}
and
\begin{equation*}
	|v|_\Network \geq \tilde r_\Network |v|_\infty \qquad \forall v \in \dR^m.
\end{equation*}
Then the definition of the ``global'' norm \eqref{eq:global-norm} implies that
\begin{equation*}
\begin{aligned}
    &|\xi_\sigma(t, \bar x) - \xi_\sigma(t, x)|_\Global \geq \tilde r_\Network \big| \big( |\xi_\sigma^1(t, \bar x_1) - \xi_\sigma^1(t, x_1)|_{\local 1}, \ldots, |\xi_\sigma^m(t, \bar x_m) - \xi_\sigma^m(t, x_m)|_{\local m} \big) \big|_\infty \\
    &\qquad = \max_{1 \leq i \leq m} \tilde r_\Network |\xi_\sigma^i(t, \bar x_i) - \xi_\sigma^i(t, x_i)|_{\local i} \geq \max_{1 \leq i \leq m} e^{\underline\eta_\sigma^i(t)} \tilde r_\Network |\bar x_i - x_i|_{\local i} \geq \max_{1 \leq i \leq m} e^{\underline\eta_\sigma^i(t)} \tilde r_\Network \tilde r_{\local i} |\bar x_i - x_i|_\infty
\end{aligned}
\end{equation*}
for all $ t \geq 0 $. Consequently, given arbitrary time horizon $ T \geq 0 $ and radius $ \varepsilon > 0 $, we have
\begin{equation}\label{eq:sw-blk-diag-soln-max-lower}
    \max_{t \in [0, T]} |\xi_\sigma(t, \bar x) - \xi_\sigma(t, x)|_\Global \geq \max_{1 \leq i \leq m} e^{\max_{t \in [0, T]} \underline\eta_\sigma^i(t)} \tilde r_\Network \tilde r_{\local i} |\bar x_i - x_i|_\infty. 
\end{equation}
Consider the grid $ G(\theta) $ defined by \eqref{eq:grid-dfn} with the vector $ \theta = (\theta_1 \mathbf{1}_{n_1}, \ldots, \theta_m \mathbf{1}_{n_m}) \in \dR_{> 0}^n $ defined by
\begin{equation*}
    \theta_i := e^{-\max_{t \in [0, T]} \underline\eta_\sigma^i(t)} \varepsilon/(\tilde r_\Network \tilde r_{\local i}), \qquad i \in \{1, \ldots, m\}.
\end{equation*}
Comparing the corresponding hyperrectangles $ R(x) $ defined by \eqref{eq:grid-rect} to the open balls $ B_{f_\sigma}(x, \varepsilon, T) $ defined by \eqref{eq:ball-dfn} with the ``global'' norm $ |\cdot|_\Global $, we see that \eqref{eq:sw-blk-diag-soln-max-lower} implies $ B_{f_\sigma}(x, \varepsilon, T) \subset R(x) $ for all $ x \in G(\theta) $. Then part~\ref{lem:grid-lower} of Lemma~\ref{lem:grid} implies that $ G(\theta) $ is $ (T, \varepsilon) $-separated. As $ T \geq 0 $ and $ \varepsilon > 0 $ are arbitrary, the lower bound \eqref{eq:grid-sep-ent} implies that
\begin{equation*}
\begin{aligned}
    &h(f_\sigma, K) \geq \liminf_{\varepsilon \searrow 0} \limsup_{T \to \infty} \sum_{i=1}^{m} \frac{n_i \log(1/\theta_i)}{T} \\
    &\qquad = \limsup_{T \to \infty} \sum_{i=1}^{m} \frac{1}{T} \max_{t \in [0, T]} n_i \underline\eta_\sigma^i(t) + \lim_{\varepsilon \searrow 0} \lim_{T \to \infty} \sum_{i=1}^{m} \frac{n_i \log(\tilde r_\Network \tilde r_{\local i}/\varepsilon)}{T} \\
    &\qquad = \limsup_{T \to \infty} \sum_{i=1}^{m} \frac{1}{T} \max_{t \in [0, T]} \sum_{p \in \Index} \int_{0}^{t} \Big( \min_{v_i \in \co(\xi_\sigma^i(s, K_i))} -n_i \mu_i(-J_{x_i} f_p^i(v_i)) \Big) \indFcn_p(\sigma(s)) \d s.
\end{aligned}
\end{equation*}
Finally, we obtain \eqref{eq:sw-blk-diag-ent-lower} by invoking the lower bound \eqref{eq:sw-diag-avg-max-sup-lower} with the functions
\begin{equation*}
\pushQED{\qed}
	a_p^i(t) = \min_{v_i \in \co(\xi_\sigma^i(s, K_i))} -n_i \mu_i(-J_{x_i} f_p^i(v_i)), \qquad i \in \{1, \ldots, m\},\, p \in \Index. \qedhere
\popQED
\end{equation*}

\subsection{Omitted proof of Lemma~\ref{lem:sw-diag-avg-max-sup}}\label{supp:sw-diag-avg-max-sup}
For brevity, we define the following functions on $ \dR_{\geq 0} $:
\begin{equation*}
	\begin{aligned}
		\eta_p^i(t) &:= \int_{0}^{t} a_p^i(s) \indFcn_p(\sigma(s)) \d s, \quad p \in \Index, \\
		\bar a^i(T) &:= \frac{1}{T} \max_{t \in [0, T]} \sum_{p \in \Index} \eta_p^i(t),
	\end{aligned} \qquad i \in \{1, \ldots, m\}
\end{equation*}
with $ \bar a^i(0) := \max\{a_{\sigma(0)}^i(0),\, 0\} $.

\begin{enumerate}[wide]
\item
Here we prove the upper bound \eqref{eq:sw-diag-avg-max-sup-upper}.

First, we eliminate modes that are not persistent. The definition of $ \IndexInf $ in \eqref{eq:index-inf-pos} implies that $ t_\infty := \sup\{t \geq 0: \sigma(t) \in \Index\backslash\IndexInf\} $ is finite. Then
\begin{equation*}
\begin{aligned}
	\bar a^i(T) &\leq \frac{1}{T} \bigg( \max_{t \in [0, T]} \sum_{p \in \IndexInf} \eta_p^i(t) + \max_{t \in [0, T]} \sum_{p \in \Index\backslash\IndexInf} \eta_p^i(t) \bigg) \\
	&= \frac{1}{T} \bigg( \max_{t \in [0, T]} \sum_{p \in \IndexInf} \eta_p^i(t) + \max_{t \in [0, t_\infty]} \sum_{p \in \Index\backslash\IndexInf} \eta_p^i(t) \bigg)
\end{aligned}
\end{equation*}
for all $ i \in \{1, \ldots, m\} $ and $ T > t_\infty $, where the inequality holds as the maximum is a subadditive function. Hence
\begin{equation*}
	\limsup_{T \to \infty} \sum_{i=1}^{m} \bar a^i(T) \leq \limsup_{T \to \infty} \sum_{i=1}^{m} \frac{1}{T} \max_{t \in [0, T]} \sum_{p \in \IndexInf} \eta_p^i(t).
\end{equation*}

Second, we eliminate modes that are persistent but not strongly persistent. For all $ i \in \{1, \ldots, m\} $ and $ p \in \IndexInf $, as $ \hat a_p^i = \limsup_{t \to \infty:\, \sigma(t) = p} a_p^i(t) $ are finite, $ \hat\alpha_p^i := \sup_{t \geq 0} a_p^i(t) \indFcn_p(\sigma(t)) $ are finite as well. Then
\begin{equation*}
\begin{aligned}
	\frac{1}{T} \max_{t \in [0, T]} \sum_{p \in \IndexInf} \eta_p^i(t) &\leq \frac{1}{T} \max_{t \in [0, T]} \sum_{p \in \IndexPos} \eta_p^i(t) + \frac{1}{T} \max_{t \in [0, T]} \sum_{p \in \IndexInf\backslash\IndexPos} \eta_p^i(t) \\
	&\leq \frac{1}{T} \max_{t \in [0, T]} \sum_{p \in \IndexPos} \eta_p^i(t) + \sum_{p \in \IndexInf\backslash\IndexPos} \max\{\hat\alpha_p^i,\, 0\} \rho_p(T)
\end{aligned}
\end{equation*}
for all $ i \in \{1, \ldots, m\} $ and $ T > 0 $, where the first inequality holds as the maximum is a subadditive function. Moreover, for all $ p \notin \IndexPos $, as $ \rho_p(t) \geq 0 $ for all $ t \geq 0 $ and $ \hat\rho_p = \limsup_{t \to \infty} \rho_p(t) = 0 $, we have $ \hat\rho_p = \lim_{t \to \infty} \rho_p(t) = 0 $. Then
\begin{equation*}
	\lim_{T \to \infty} \sum_{i=1}^{m} \sum_{p \in \IndexInf\backslash\IndexPos} \max\{\hat\alpha_p^i,\, 0\} \rho_p(T) = \sum_{i=1}^{m} \sum_{p \in \IndexInf\backslash\IndexPos} \max\{\hat\alpha_p^i,\, 0\} \hat\rho_p = 0.
\end{equation*}
Hence
\begin{equation*}
	\limsup_{T \to \infty} \sum_{i=1}^{m} \bar a^i(T) \leq \limsup_{T \to \infty} \sum_{i=1}^{m} \frac{1}{T} \max_{t \in [0, T]} \sum_{p \in \IndexPos} \eta_p^i(t).
\end{equation*}

Finally, as $ \IndexPos $ is a finite set, the upper limit in the definition of $ \hat a_p^i $ implies that, for each $ \delta > 0 $, there exists a large enough $ t_\delta \geq 0 $ such that
\begin{equation*}
	a_p^i(t) < \hat a_p^i + \delta \qquad \forall i \in \{1, \ldots, m\},\, p \in \IndexPos,\, t > t_\delta: \sigma(t) = p.
\end{equation*}
For all $ i \in \{1, \ldots, m\} $ and $ t \geq 0 $, consider
\begin{equation*}
\begin{aligned}
	\sum_{p \in \IndexPos} \eta_p^i(t) &= \sum_{p \in \IndexPos} \int_{0}^{t} (a_p^i(s) - \hat a_p^i - \delta + \hat a_p^i + \delta) \indFcn_p(\sigma(s)) \d s \\
	&= \sum_{p \in \IndexPos} \hat a_p^i \tau_p(t) + \delta t + \sum_{p \in \IndexPos} \int_{0}^{t} (a_p^i(s) - \hat a_p^i - \delta) \indFcn_p(\sigma(s)) \d s.
\end{aligned}
\end{equation*}
If $ t > t_\delta $, then
\begin{equation*}
	\sum_{p \in \IndexPos} \int_{0}^{t} (a_p^i(s) - \hat a_p^i - \delta) \indFcn_p(\sigma(s)) \d s \leq \sum_{p \in \IndexPos} \int_{0}^{t_\delta} (a_p^i(s) - \hat a_p^i - \delta) \indFcn_p(\sigma(s)) \d s.
\end{equation*}
Otherwise $ t \in [0, t_\delta] $, and hence
\begin{equation*}
	\sum_{p \in \IndexPos} \int_{0}^{t} (a_p^i(s) - \hat a_p^i - \delta) \indFcn_p(\sigma(s)) \d s \leq \max_{\bar t \in [0, t_\delta]} \sum_{p \in \IndexPos} \int_{0}^{\bar t} (a_p^i(s) - \hat a_p^i - \delta) \indFcn_p(\sigma(s)) \d s.
\end{equation*}
Combining the cases, we obtain
\begin{equation*}
	\sum_{p \in \IndexPos} \eta_p^i(t) \leq \sum_{p \in \IndexPos} \hat a_p^i \tau_p(t) + \delta t + \max_{\bar t \in [0, t_\delta]} \sum_{p \in \IndexPos} \int_{0}^{\bar t} (a_p^i(s) - \hat a_p^i - \delta) \indFcn_p(\sigma(s)) \d s.
\end{equation*}
Then
\begin{equation*}
	\frac{1}{T} \max_{t \in [0, T]} \sum_{p \in \IndexPos} \eta_p^i(t) \leq \frac{1}{T} \max_{t \in [0, T]} \sum_{p \in \IndexPos} \hat a_p^i \tau_p(t) + \delta + \frac{1}{T} \max_{\bar t \in [0, t_\delta]} \sum_{p \in \IndexPos} \int_{0}^{\bar t} (a_p^i(s) - \hat a_p^i - \delta) \indFcn_p(\sigma(s)) \d s
\end{equation*}
for all $ T > 0 $, where the inequality holds in part because the maximum is a subadditive function. Hence
\begin{equation*}
	\limsup_{T \to \infty} \sum_{i=1}^{m} \bar a_i(T) \leq \limsup_{T \to \infty} \sum_{i=1}^{m} \frac{1}{T} \max_{t \in [0, T]} \sum_{p \in \IndexPos} \hat a_p^i \tau_p(t) + m \delta.
\end{equation*}
Then \eqref{eq:sw-diag-avg-max-sup-upper} holds as $ \delta > 0 $ is arbitrary.
\item
Here we prove the lower bound \eqref{eq:sw-diag-avg-max-sup-lower-alt}. Note that
\begin{equation}\label{eq:max-sum-lower-const}
	\max\{a + b,\, 0\} \geq \max\{a,\, 0\} + \min\{b,\, 0\} \qquad \forall a, b \in \dR.
\end{equation}

First, we eliminate modes that are not persistent. The definition of $ \IndexInf $ in \eqref{eq:index-inf-pos} implies that $ t_\infty := \sup\{t \geq 0: \sigma(t) \in \Index\backslash\IndexInf\} $ is finite. Then
\begin{equation*}
\begin{aligned}
	\max \bigg\{ \frac{1}{t} \sum_{p \in \Index} \eta_p^i(t),\, 0 \bigg\} &\geq \max \bigg\{ \frac{1}{t} \sum_{p \in \IndexInf} \eta_p^i(t),\, 0 \bigg\} + \min \bigg\{ \frac{1}{t} \sum_{p \in \Index\backslash\IndexInf} \eta_p^i(t),\, 0 \bigg\} \\
	&= \max \bigg\{ \frac{1}{t} \sum_{p \in \IndexInf} \eta_p^i(t),\, 0 \bigg\} + \min \bigg\{ \frac{1}{t} \sum_{p \in \Index\backslash\IndexInf} \eta_p^i(t_\infty),\, 0 \bigg\}
\end{aligned}
\end{equation*}
for all $ i \in \{1, \ldots, m\} $ and $ t > t_\infty $, where the inequality follows from \eqref{eq:max-sum-lower-const}. Hence
\begin{equation*}
	\limsup_{t \to \infty} \sum_{i=1}^{m} \max \bigg\{ \frac{1}{t} \sum_{p \in \Index} \eta_p^i(t),\, 0 \bigg\} \geq \limsup_{t \to \infty} \sum_{i=1}^{m} \max \bigg\{ \frac{1}{t} \sum_{p \in \IndexInf} \eta_p^i(t),\, 0 \bigg\}.
\end{equation*}

Second, we eliminate modes that are persistent but not strongly persistent. For all $ i \in \{1, \ldots, m\} $ and $ p \in \IndexInf $, as $ \check a_p^i = \liminf_{t \to \infty:\, \sigma(t) = p} a_p^i(t) $ are finite, $ \check\alpha_p^i := \inf_{t \geq 0} a_p^i(t) \indFcn_p(\sigma(t)) $ are finite as well. Then
\begin{equation*}
\begin{aligned}
	\max \bigg\{ \frac{1}{t} \sum_{p \in \IndexInf} \eta_p^i(t),\, 0 \bigg\} &\geq \max \bigg\{ \frac{1}{t} \sum_{p \in \IndexPos} \eta_p^i(t),\, 0 \bigg\} + \min \bigg\{ \frac{1}{t} \sum_{p \in \IndexInf\backslash\IndexPos} \eta_p^i(t),\, 0 \bigg\} \\
	&\geq \max \bigg\{ \frac{1}{t} \sum_{p \in \IndexPos} \eta_p^i(t),\, 0 \bigg\} + \sum_{p \in \IndexInf\backslash\IndexPos} \min\{\check\alpha_p^i,\, 0\} \rho_p(t)
\end{aligned}
\end{equation*}
for all $ i \in \{1, \ldots, m\} $ and $ t > 0 $, where the first inequality follows from \eqref{eq:max-sum-lower-const}. Moreover, for all $ p \notin \IndexPos $, as $ \rho_p(t) \geq 0 $ for all $ t \geq 0 $ and $ \hat\rho_p = \limsup_{t \to \infty} \rho_p(t) = 0 $, we have $ \hat\rho_p = \lim_{t \to \infty} \rho_p(t) = 0 $. Then
\begin{equation*}
	\lim_{t \to \infty} \sum_{i=1}^{m} \sum_{p \in \IndexInf\backslash\IndexPos} \min\{\check\alpha_p^i,\, 0\} \rho_p(t) = \sum_{i=1}^{m} \sum_{p \in \IndexInf\backslash\IndexPos} \min\{\check\alpha_p^i,\, 0\} \hat\rho_p = 0.
\end{equation*}
Hence
\begin{equation*}
	\limsup_{t \to \infty} \sum_{i=1}^{m} \max \bigg\{ \frac{1}{t} \sum_{p \in \Index} \eta_p^i(t),\, 0 \bigg\} \geq \limsup_{t \to \infty} \sum_{i=1}^{m} \max \bigg\{ \frac{1}{t} \sum_{p \in \IndexPos} \eta_p^i(t),\, 0 \bigg\}.
\end{equation*}

Finally, as $ \IndexPos $ is a finite set, the lower limit in the definition of $ \check a_p^i $ implies that, for each $ \delta > 0 $, there exists a large enough $ t_\delta \geq 0 $ such that
\begin{equation*}
	a_p^i(t) > \check a_p^i - \delta \qquad \forall i \in \{1, \ldots, m\},\, p \in \IndexPos,\, t > t_\delta: \sigma(t) = p.
\end{equation*}
Then
\begin{equation*}
\begin{aligned}
	\sum_{p \in \IndexPos} \eta_p^i(t) &= \sum_{p \in \IndexPos} \int_{0}^{t} (a_p^i(s) - \check a_p^i + \delta + \check a_p^i - \delta) \indFcn_p(\sigma(s)) \d s \\
	&= \sum_{p \in \IndexPos} \check a_p^i \tau_p(t) - \delta t + \sum_{p \in \IndexPos} \int_{0}^{t} (a_p^i(s) - \check a_p^i + \delta) \indFcn_p(\sigma(s)) \d s \\
	&\geq \sum_{p \in \IndexPos} \check a_p^i \tau_p(t) - \delta t + \sum_{p \in \IndexPos} \int_{0}^{t_\delta} (a_p^i(s) - \check a_p^i + \delta) \indFcn_p(\sigma(s)) \d s
\end{aligned}
\end{equation*}
for all $ i \in \{1, \ldots, m\} $ and $ t > t_\delta $. Hence
\begin{equation*}
	\limsup_{t \to \infty} \sum_{i=1}^{m} \max \bigg\{ \frac{1}{t} \sum_{p \in \Index} \eta_p^i(t),\, 0 \bigg\} \geq \limsup_{t \to \infty} \sum_{i=1}^{m} \max \bigg\{ \sum_{p \in \IndexPos} \check a_p^i \rho_p(t) - \delta,\, 0 \bigg\}.
\end{equation*}
Then \eqref{eq:sw-diag-avg-max-sup-lower-alt} holds as $ \delta > 0 $ is arbitrary. \qed
\end{enumerate}


\bibliographystyle{IEEEtran}
\bibliography{reference-abbr,reference-temp}
\end{document}